\documentclass[a4paper,11pt]{article}
\pdfoutput=1
\usepackage{jheppub}
\usepackage{tikz}
\usetikzlibrary{decorations.pathmorphing,decorations.markings}
\usepackage{booktabs,longtable,array}
\usepackage{url}
\usepackage{orcidlink}
\newcommand{\HVTP}{\textsf{HVTP}}
\newcommand{\HVTPeta}{\textsf{HVTP\_eta}}
\title{When heavy vectors meet top partners: HVTP, a unified framework for composite Higgs searches}
\author{Natascia Vignaroli\,\orcidlink{0000-0002-6711-522X}}
\affiliation{Dipartimento di Matematica e Fisica, Universit\`a del Salento,\\ I-73100 Lecce, Italy}
\emailAdd{natascia.vignaroli@unisalento.it}
\abstract{Vector resonances and top partners are characteristic new states of composite Higgs models: they belong to the same strongly coupled sector, and partial compositeness ties their interactions to the mixings that generate the top mass. Experimental searches, however, usually treat them separately. We provide a common framework for both and release \HVTP{} (Heavy Vectors and Top Partners), a public UFO model for MadGraph in which an electroweak triplet, a hypercharge vector and a colour octet couple to a custodial top-partner multiplet, with all couplings obtained from the exact diagonalisation of the mass matrices. The models and the inputs of our results are available at \url{https://doi.org/10.5281/zenodo.24345218}. We translate the latest LHC searches at 13 and 13.6 TeV into the parameter space of the model. In natural scenarios the top partners, which are directly tied to the Higgs mass and to the residual tuning, are lighter than the vector resonances; the vectors then decay mostly into partners and become broad: the standard searches lose much of their sensitivity, while the vectors produce partner pairs at rates that can exceed QCD production. For example, at our benchmark points, with partners near their present limits and an electroweak triplet at about 4 TeV, 73--83\% of the triplet decays contain partners, and vector-mediated production reaches up to three times the QCD rate. At the HL-LHC this region can be probed by resonance searches extended to large widths and by partner-pair searches that include resonant production; the latter also reveal the partner spectrum. Composite Higgs models with a non-minimal coset generally also contain a pseudoscalar, which can alter the standard decays of the top partners; we include it in the extension \HVTPeta{}.}
\begin{document}
\maketitle
\section{Introduction}\label{sec:intro}
The discovery of the Higgs boson confirmed the Brout--Englert--Higgs mechanism: a scalar doublet with a non-zero vacuum expectation value gives mass to the $W$ and $Z$ bosons while preserving gauge invariance and renormalisability. The Standard Model (SM) does not, however, explain why electroweak symmetry is broken. The potential $V(H)=-\mu^2H^\dagger H+\lambda_H(H^\dagger H)^2$ is written down rather than derived: the negative mass term that triggers the breaking, and with it the weak scale $v^2=\mu^2/\lambda_H$, is an input parameter, with no dynamics that fixes its sign or its size. Its size is also unstable. The mass term is the only relevant operator of the SM, and any heavy state of mass $M$ coupled to the Higgs, for instance through $aH^\dagger HS^2$, shifts it by an amount of order $aM^2/(16\pi^2)$; the top-quark loop alone gives $\delta\mu^2\simeq3y_t^2\Lambda^2/(8\pi^2)$ for a cutoff $\Lambda$. If the SM remains valid up to scales far above $v$, the observed $m_h=125$ GeV requires a cancellation between unrelated contributions of order $(m_h/\Lambda)^2$. The origin of electroweak symmetry breaking (EWSB) and the stability of the weak scale are therefore two aspects of the same open question: what dynamics generates the Higgs potential?

Composite Higgs models answer it with a new strongly interacting sector at the TeV scale~\cite{Kaplan:1983fs,mchm,Contino:2010rs,Panico:2015jxa}, whose lightest resonances, spin-one states and vector-like top partners, are the main targets of the direct searches at the LHC, with production and decay patterns correlated by the same dynamics~\cite{gluon,wp,Vignaroli:2012nf,Vignaroli:2012sf,Vignaroli:2015ama,Mohan:2015doa}. QCD shows how light spin-zero states can emerge from such a sector. The pions are much lighter than the other hadrons, whose masses of order $\Lambda_{\rm QCD}$ are generated by dimensional transmutation, because they are the pseudo-Nambu--Goldstone bosons (pNGBs) of the chiral symmetry $SU(2)_L\times SU(2)_R$, spontaneously broken to $SU(2)_V$ by the quark condensate. Their masses arise only from the small explicit breaking of the chiral symmetry by the quark masses and by electromagnetism, which generates radiatively the mass difference between charged and neutral pions. In the same way, the Higgs is a bound state of the new sector, so that its mass is screened above the compositeness scale, and it is naturally lighter than the other resonances, whose masses $m_*\simeq g_*f$, with $g_*$ the coupling of the strong sector, are set by dimensional transmutation, if it is a pNGB of a global symmetry $\mathcal G$ spontaneously broken to $\mathcal H$ at a scale $f$; the minimal coset $SO(5)/SO(4)$ contains exactly one Higgs doublet. As for the pions, the potential arises from explicit breaking: the interactions of the strong sector with the SM fields break $\mathcal G$ and generate the Higgs potential radiatively, and with it EWSB. The vacuum is misaligned, $v=f\sin(\langle h\rangle/f)$, and the degree of misalignment, $\xi=v^2/f^2$, measures both the deviations of the Higgs couplings from the SM, which limit $\xi\lesssim0.1$~\cite{ATLAS:2022vpf,CMS:2022dwd}, and the residual tuning of the potential. The strong sector can be realised by a confining gauge theory of new fermions~\cite{Ferretti:2013kya,Cacciapaglia:2014uja,Cacciapaglia:2020kgq}; the global symmetries of such theories typically lead to cosets larger than the minimal one, with additional pNGBs besides the Higgs.

The dominant explicit breaking comes from the top quark. In the paradigm of partial compositeness~\cite{Kaplan:1991dc} each chiral component of the top mixes linearly with vector-like fermionic resonances of the strong sector carrying the same SM charges, the top partners, and acquires a mass proportional to its degree of compositeness. The same mixings generate the Higgs potential, and a light Higgs requires light partners: up to factors of order one that depend on the representations, $m_h^2\sim(N_c/\pi^2)\,(m_t^2/f^2)\,m_Tm_{\widetilde T}$, with $N_c=3$ the number of colours, where $T$ and $\widetilde T$ are the partners of $q_L$ and $t_R$~\cite{Matsedonskyi:2012ym,Redi:2012ha,Marzocca:2012zn,Pomarol:2012qf}. For $f\simeq800$ GeV ($\xi\simeq0.1$) this points to partners at about 1 TeV, so that the current bounds of about 1.5 TeV (section~\ref{sec:vlqlimits}) already probe the natural region. A custodial symmetry, required to protect the $T$ parameter and the $Zb\bar b$ coupling, places the top partners in multiplets that also contain the exotic charge-$5/3$ state $X_{5/3}$ and its partner $X_{2/3}$, the so-called custodians~\cite{protect,custodians}. Top partners are therefore the smoking gun of a natural composite Higgs, and their direct search is a central part of the LHC programme.

The strong sector also contains spin-one resonances: colour-singlet partners of the electroweak gauge bosons and a colour-octet partner of the gluon. The LHC has excluded such states up to 4--6 TeV in the benchmarks used by the experiments, the sequential SM and the heavy vector triplet models (section~\ref{sec:existing}), and electroweak precision tests require masses above about 2.5--3 TeV in minimal models through the $S$ parameter~\cite{Contino:2010rs,Grojean:2013qca}; together with the resonance searches, which start from a few hundred GeV (e.g.\ from 250 GeV for dileptons~\cite{ATLAS:2019erb}), these constraints cover the range below the TeV for vectors coupled to the light quarks with the strength expected in composite models. Further indirect constraints come from contact interactions, from the $T$ parameter and the $Zb\bar b$ vertex, to which both vectors and partners contribute, and from flavour observables~\cite{Vignaroli:2012si,Konig:2014iva,Glioti:2024hye}; we discuss the first two in section~\ref{sec:vectorsearches} and concentrate on the direct searches. The direct bounds on the vectors rest, however, on two assumptions: that the resonances decay only into SM final states such as dileptons, dijets or dibosons, and, for most results, that they are narrow. In a composite Higgs model neither need hold. The vector resonances couple weakly to the light quarks through which they are produced, and strongly to the other composite states, the longitudinal gauge bosons, the Higgs and the top partners. When the partner channels are open they dominate the decays~\cite{gluon,hunt,Chala:2014mma,Azatov:2015xqa}, and three things change together: the vectors become broad, their branching fractions into SM final states drop, and they become a source of top partners, in heavy--light and partner-pair final states such as $W'\to T\bar b$, $G'\to T\bar t$ and $W'\to X_{5/3}\bar X_{2/3}$, the latter with a characteristic same-sign dilepton signature~\cite{gluon,wp,Vignaroli:2015ama}. Partner and vector searches are therefore two views of a single spectrum, connected by the same mixings that generate the top mass.

In this paper we develop and make public a unified description of this spectrum. \HVTP{} (Heavy Vectors and Top Partners) is a two-site model that contains an electroweak triplet $W',Z'$, a hypercharge vector $B'$ and a colour octet $G'$, coupled to the custodial top-partner multiplet, and that brings together in a single framework the models used in our earlier studies of heavy gluons, $W'$ resonances and top partners~\cite{gluon,wp,Vignaroli:2012nf,Vignaroli:2012sf,Vignaroli:2012si,Vignaroli:2015ama}. All production and decay vertices follow from the exact diagonalisation of the gauge and fermion mass matrices at each parameter point, and the model is distributed as a generator of UFO models~\cite{Degrande:2011ua,Darme:2023jdn} for MadGraph5\_aMC@NLO~\cite{Alwall:2014hca}. We also release \HVTPeta{}, an extension that adds a composite pseudoscalar $\eta$ coupled to the top partners, as predicted by non-minimal cosets.

The paper has two aims. The first is practical: to provide a single, validated signal model in which vector and partner searches can be interpreted consistently, with physical masses as inputs and a guide that defines the inputs and the particle content for users who are not familiar with the model. With the vectors decoupled, the same model provides consistent single-production couplings and branching fractions for the top partners alone, obtained from the exact diagonalisation of the mass matrices. The second is to identify the regions of parameter space that the present search strategy may miss. In natural scenarios the top partners, which control the Higgs potential and the residual tuning, are the lightest composite states and lie below the vector resonances. With $m_\psi\simeq g_\psi f$ and $m_\rho\simeq g_\rho f$, this hierarchy corresponds to a coupling of the fermionic resonances smaller than that of the vectors, as required by a light Higgs when $m_\rho$ is above the electroweak precision bound~\cite{Matsedonskyi:2012ym,Redi:2012ha,Marzocca:2012zn,Panico:2015jxa}; it is not a general property of composite models, and \HVTP{} also describes the opposite ordering (section~\ref{sec:inverted}). The vectors then decay mostly into partners and become broad, their branching fractions into SM final states drop, and vector-mediated partner production becomes comparable to or larger than QCD production, as we quantify for representative benchmarks. The present resonance limits then rely on narrow-width templates, and partner-pair searches that include resonant production and resonance searches extended to large widths are complementary probes of this region at the HL-LHC.

Section~\ref{sec:composite} describes the composite sector and partial compositeness, section~\ref{sec:model} the model and its couplings, and section~\ref{sec:usage} the public models, their inputs and their validation. Section~\ref{sec:currents} discusses the vector currents, thresholds and widths and relates \HVTP{} to existing models. Sections~\ref{sec:vlq} and~\ref{sec:vlqlimits} discuss the phenomenology of the top partners and the current limits on them, and section~\ref{sec:vectorsearches} derives the present limits on the vector resonances from the searches at 13 and 13.6 TeV, together with the indirect constraints. Section~\ref{sec:broad} studies the region in which the vectors are broad and decay mainly into partners, including finite-width effects, the neutral vectors and projections for the HL-LHC. Section~\ref{sec:nonstandard} discusses nonstandard partner decays, including decays into $\eta$, and further extensions of the scalar sector. The appendices collect the decay widths, the details of the use of the models, the inventory of experimental searches and the inputs of the limit calculations.
\section{The composite sector and partial compositeness}\label{sec:composite}
\begin{figure}[htbp]\centering
\resizebox{\linewidth}{!}{
\begin{tikzpicture}[font=\footnotesize,
  elem/.style={rounded corners=6pt,draw=blue!55!black,fill=blue!6,thick},
  comp/.style={rounded corners=18pt,draw=red!60!black,fill=red!5,thick},
  mix/.style={<->,>=stealth,thick,draw=black!70},
  lvl/.style={thick},
  lab/.style={font=\scriptsize,text=black!75}]
\node[font=\bfseries] at (-2.0,3.25) {(a)};
\node[elem,minimum width=4.0cm,minimum height=5.8cm] at (0,0) {};
\node[font=\bfseries,text=blue!55!black] at (0,2.5) {Elementary sector};
\node[lab] at (0,2.05) {weak couplings $g_{0i}$};
\node at (0,1.3) {$W^a_{0\mu},\ B_{0\mu},\ G^a_{0\mu}$};
\node at (0,0.45) {$q_L=(t_L,b_L),\ t_R$};
\node at (0,-0.4) {light quarks, leptons};
\node[lab,align=center] at (0,-1.5) {chiral fields,\\massless before mixing};

\node[comp,minimum width=6.4cm,minimum height=5.8cm] at (7.9,0) {};
\node[font=\bfseries,text=red!60!black] at (7.9,2.5) {Composite sector};
\node[lab] at (7.9,2.05) {strong dynamics: $g_{1i}\gg g_{0i}$, scale $f\sim$ TeV};
\node at (7.9,1.3) {vectors $W',Z'\,(\rho_L)$, $B'\,(\rho_Y)$, $G'\,(\rho_c)$};
\node[lab] at (7.9,0.8) {top partners, custodial $(\mathbf2,\mathbf2)_{2/3}\oplus(\mathbf1,\mathbf1)_{2/3}$:};
\node at (7.9,0.4) {$Q=(T,B),\ X=(X_{5/3},X_{2/3}),\ \widetilde T$};
\node at (7.9,-0.4) {Higgs doublet $H\ni h,\ \pi^\pm,\pi^0$: pNGBs of $\mathcal G/\mathcal H$};
\node[lab,text=red!60!black] at (7.9,-0.82) {$\pi^\pm,\pi^0$ eaten $\Rightarrow$ $W_L^\pm,Z_L$ are composite};
\node[text=black!60] at (7.9,-1.25) {[\,$\eta$: additional pNGB in \HVTPeta\,]};
\node[lab] at (7.9,-1.85) {composite Yukawa $Y_*$ couples $H$ to $Q,X$ and $\widetilde T$};

\draw[mix] (2.1,1.3) -- (4.6,1.3);
\node[lab,align=center] at (3.35,1.75) {gauge mixing\\$\tan\theta_i=1/k_i$};
\draw[mix] (2.1,0.45) -- (4.6,0.45);
\node[lab,align=center] at (3.35,-0.05) {fermion mixing\\$\Delta_{L,R}\Rightarrow s_L,s_R$};
\draw[->,>=stealth,thick,dashed,draw=black!55] (4.6,-1.25) -- (2.1,-1.25);
\node[lab,align=center] at (3.35,-1.8) {explicit breaking\\of $\mathcal G$ $\Rightarrow V(H)$};

\node[lab,align=center] at (3.95,-3.55) {%
light states: $g_i=g_{0i}\cos\theta_i$,\quad $y_t\simeq Y_*s_Ls_R$;\quad heavy vectors couple with $g_i/k_i$ to light quarks\\
and with $g_ik_i$ to the Higgs, $W_L,Z_L$ and the partners, which fixes both their production and their decays};

\begin{scope}[shift={(11.9,-2.7)}]
\node[font=\bfseries] at (-0.7,5.95) {(b)};
\draw[->,>=stealth,thick] (0,0) -- (0,5.7) node[above,lab]{mass};
\draw[lvl,blue!55!black] (0.2,0.4) -- (0.72,0.4);
\draw[lvl,red!60!black] (0.78,0.4) -- (1.3,0.4);
\node[anchor=west,font=\footnotesize,blue!55!black] at (1.3,0.62) {$q,\ell,\ W_T,Z_T$\ \ ($\propto1/k^2$)};
\node[anchor=west,font=\footnotesize,red!60!black] at (1.3,0.2) {$h,\ W_L,Z_L$\ \ ($\propto k^2$)};
\draw[lvl,black!45,dashed] (0.2,1.6) -- (1.3,1.6) node[right,black,font=\footnotesize]{$f$};
\draw[lvl,red!60!black] (0.2,2.1) -- (1.3,2.1);
\draw[lvl,red!60!black] (0.2,2.35) -- (1.3,2.35);
\draw[lvl,red!60!black] (0.2,2.6) -- (1.3,2.6) node[right,black,font=\footnotesize]{partners};
\draw[lvl,red!60!black,very thick] (0.2,4.0) -- (1.3,4.0) node[right,black,font=\footnotesize]{$W',Z',B',G'$};
\draw[lvl,black!45,dashed] (0.2,5.1) -- (1.3,5.1) node[right,black,font=\footnotesize]{$\Lambda\sim4\pi f$};
\draw[->,>=stealth,red!60!black,thick] (0.35,3.93) -- (0.35,2.68);
\draw[->,>=stealth,blue!55!black,thick] (0.55,3.93) -- (0.55,0.48);
\draw[->,>=stealth,red!60!black,thick] (1.05,3.93) -- (1.05,0.48);
\node[font=\scriptsize,red!60!black,anchor=east] at (0.27,3.3) {$\propto k^2$};
\end{scope}
\end{tikzpicture}}
\caption{(a) Elementary and composite sectors of \HVTP{} and the mixings that connect them. The explicit breaking of the global symmetry $\mathcal G$ by these mixings, dominated by the top sector, generates the Higgs potential and EWSB. (b) Schematic spectrum: the vector resonances decay into SM fermions and transverse gauge bosons with widths proportional to $1/k^2$, and with widths proportional to $k^2$ into the partners and into the Higgs and the longitudinal gauge bosons, whose couplings at high energy are those of the eaten Goldstones $\pi^\pm,\pi^0$ of the composite Higgs doublet. Transverse and longitudinal polarisations of the same bosons are drawn at the same mass, with different colours.}\label{fig:sectors}
\end{figure}
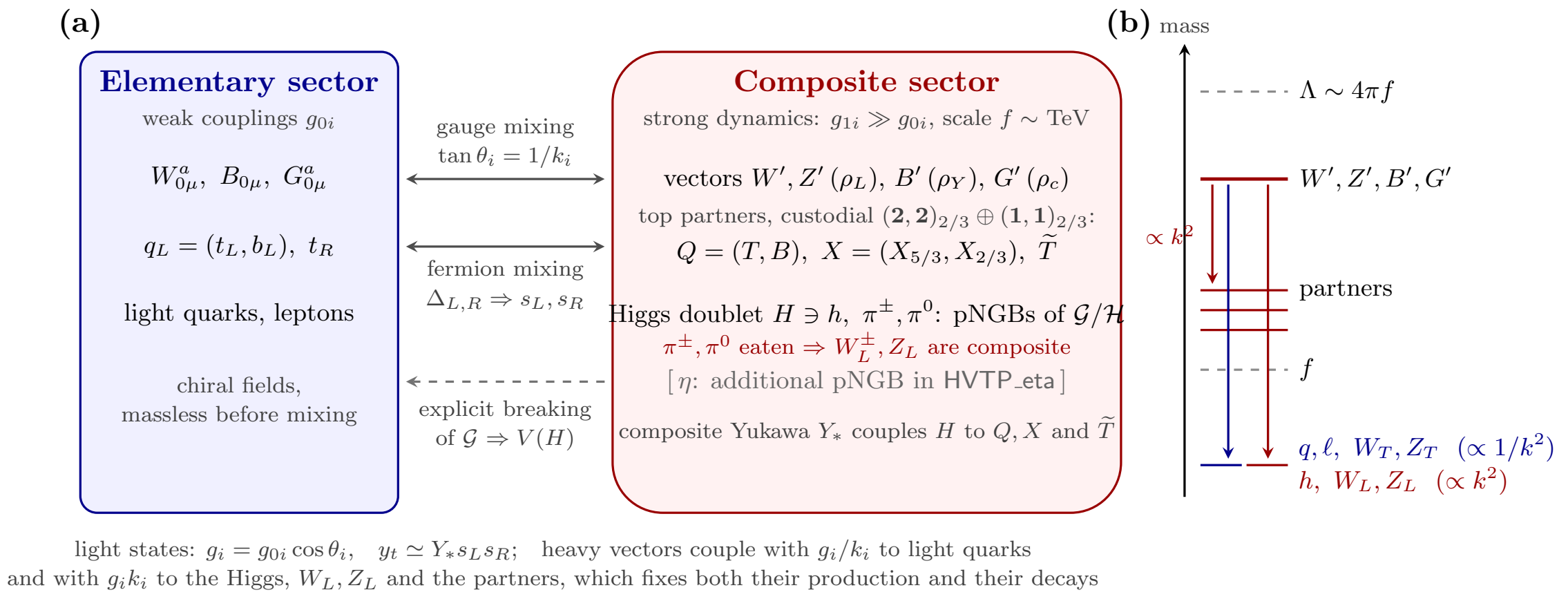

The framework implemented in \HVTP{} is summarised in figure~\ref{fig:sectors}. An elementary sector, which contains the SM gauge fields and chiral fermions with weak couplings $g_{0i}$, interacts with a composite sector characterised by large couplings $g_{1i}$, generically denoted $g_*$, and by a scale $f$ of order TeV. The composite sector contains the Higgs doublet as a pNGB, vector resonances with the quantum numbers of the SM gauge bosons and vector-like fermions with the quantum numbers of the top quark; in non-minimal cosets it also contains additional pNGBs, such as the singlet $\eta$ that we include in \HVTPeta{}. The two sectors communicate only through linear mixings: each elementary gauge field mixes with the composite vector of the same quantum numbers, and the elementary top components mix with the fermionic resonances through mass terms $\Delta_{L,R}$.

The physical states are therefore superpositions of elementary and composite fields~\cite{two}. For the gauge fields,
\begin{equation}
 A_i=\cos\theta_i\,A_i^{\rm el}+\sin\theta_i\,\rho_i^{\rm comp},\qquad
 \rho_i=-\sin\theta_i\,A_i^{\rm el}+\cos\theta_i\,\rho_i^{\rm comp},
 \label{eq:gaugesuperposition}
\end{equation}
with $\tan\theta_i=g_{0i}/g_{1i}\equiv1/k_i$, where $A_i$ is the massless SM gauge field, with coupling $g_i=g_{0i}\cos\theta_i=g_{1i}\sin\theta_i$, and $\rho_i$ the heavy vector. This mixing is the analogue of the $\rho$--photon mixing of QCD, which underlies vector-meson dominance~\cite{Sakurai:1960ju,Kroll:1967it,Contino:2010rs}. For $k_i>1$ the SM states are mostly elementary and the resonances mostly composite, and this single fact determines the phenomenology. At tree level a heavy vector couples to an elementary field only through the elementary component of its own mass eigenstate, with strength $g_i/k_i$, while it couples to a composite field with strength $g_ik_i$. For $k_i$ of a few, the vectors are produced weakly from the light quarks of the proton but decay strongly into the Higgs, the longitudinal gauge bosons and the top partners. The longitudinal $W$ and $Z$ count as composite because they originate from the Goldstone components $\pi^\pm,\pi^0$ of the composite Higgs doublet; by the equivalence theorem, their couplings at energies well above $m_W$ are those of the Goldstones. The same mixings make the top quark partly composite and generate its mass, so that the couplings that control the production and the decays of all new states are correlated.

\subsection{Partial compositeness and the top mass}
The same mechanism operates for the fermions, following the paradigm of partial compositeness~\cite{Kaplan:1991dc}. The choice of linear mixings, rather than the bilinear couplings of extended technicolour, is motivated by flavour: the hierarchy of the fermion masses can then arise from different degrees of compositeness, even with an anarchic strong sector, and the flavour-changing neutral currents of the light generations are suppressed by their small composite components~\cite{Agashe:2004cp,Csaki:2008zd,Redi:2011zi,Panico:2015jxa}. Consider an elementary chiral fermion $\psi^{\rm el}$ that mixes, through a mass parameter $\Delta$, with a composite Dirac fermion $\chi$ of mass $m_*$ carrying the same gauge quantum numbers. Before EWSB the mass eigenstates are
\begin{equation}
 \psi=\cos\phi\,\psi^{\rm el}+\sin\phi\,\chi,\qquad
 \chi'=-\sin\phi\,\psi^{\rm el}+\cos\phi\,\chi,\qquad
 \tan\phi=\frac{\Delta}{m_*} .
\end{equation}
The massless state $\psi$ is identified with the SM fermion\footnote{The relative sign of the two components depends on the phase convention of $\Delta$ and $m_*$; only $\sin^2\phi$ enters the observables discussed here, while the model uses the signs of its own mass matrices.}, and $\sin\phi$ is its degree of compositeness, while the heavy state $\chi'$, the partner, has mass $\sqrt{m_*^2+\Delta^2}=m_*/\cos\phi$. The top quark involves two such rotations, one for $q_L$, which mixes with the doublet $(T,B)$, and one for $t_R$, which mixes with the singlet $\widetilde T$,
\begin{equation}
 t_L=c_L\,t_L^{\rm el}+s_L\,T_L,\qquad
 t_R=c_R\,t_R^{\rm el}+s_R\,\widetilde T_R,
 \label{eq:superposition}
\end{equation}
and analogously $b_L=c_Lb_L^{\rm el}+s_LB_L$, where $s_{L,R}\equiv\sin\phi_{L,R}$ are the degrees of compositeness of the two chiralities and $c_{L,R}\equiv\cos\phi_{L,R}$. Since the Higgs is composite, the top acquires its mass only through its composite components,
\begin{equation}
 m_t\simeq Y_*\,s_Ls_R\,\frac{v}{\sqrt2} ,
\end{equation}
where $Y_*$ is the Yukawa coupling of the strong sector. A large top mass thus requires sizeable degrees of compositeness of both chiralities, whereas the light quarks remain mostly elementary~\cite{Kaplan:1991dc,two,mchm}.

The same rotations generate the couplings between heavy and light states. A gauge field couples to the elementary and composite fields with different strengths, $g_{\rm el}$ and $g_{\rm comp}$, and the current is diagonal only in the original basis. After the rotation, the light and heavy fermions are connected by an off-diagonal coupling $(g_{\rm comp}-g_{\rm el})\sin\phi\cos\phi$. For the massless SM gauge bosons, which couple with the same strength $g$ to both sectors, it vanishes, as required by gauge invariance. For a heavy vector, whose couplings to the two sectors have opposite signs, $-g/k$ and $gk$, the two contributions add to $g(k+1/k)\sin\phi\cos\phi$: the transition is enhanced by $k$ and mediates the decays $V'\to Q\bar q$ into one partner and one SM quark. For the SM gauge bosons, transitions between light and heavy states arise only after EWSB, from the mixing of fields with different electroweak quantum numbers, and control the single production of partners and their decays into $Wb$, $Zt$ and $ht$. Both processes therefore probe the degree of compositeness of the top quark.

\subsection{Custodial symmetry and light custodians}
The strong sector must respect a custodial $SU(2)_L\times SU(2)_R$ symmetry, which protects the $\rho$ parameter and is contained in the minimal coset $SO(5)/SO(4)$. The embedding of the top quark is then constrained by the $Zb_L\bar b_L$ coupling, measured at the per-mille level. If $q_L$ mixes with a bidoublet $(\mathbf2,\mathbf2)_{2/3}$ of $SU(2)_L\times SU(2)_R\times U(1)_X$, with $Y=T^3_R+X$, a discrete left--right parity forbids tree-level corrections to this coupling, since $b_L$ has $T^3_L=T^3_R$~\cite{protect}, provided that the whole strong sector, including its vector resonances, respects the parity. In \HVTP{}, which retains only the hypercharge component of the $SU(2)_R$ vectors, the mixing of the $Z$ with the heavy vectors leaves a small residual shift (section~\ref{sec:bprime}). The bidoublet contains two $SU(2)_L$ doublets: $Q=(T,B)$, with hypercharge $1/6$ and the quantum numbers of $q_L$, and $X=(X_{5/3},X_{2/3})$, with hypercharge $7/6$. Only $Q$ mixes with $q_L$. Before EWSB the doublet $X$ therefore remains at the Dirac mass $m_Q$, whereas $Q$ is pushed to $m_Q/c_L$, with $c_L=\sqrt{1-s_L^2}$. For the large left mixing required by the top mass the custodians $X_{5/3}$ and $X_{2/3}$ are thus among the lightest partners~\cite{custodians}, and in explicit models their mass is tied to that of the Higgs~\cite{Matsedonskyi:2012ym,Pomarol:2012qf}.

The exotic charge $5/3$ gives the custodians a distinctive signature: the decay $X_{5/3}\to W^+t\to W^+W^+b$ produces two same-sign leptons, both in pair production~\cite{Contino:2008hi,Mrazek:2009yu} and in the decays of a charged vector resonance~\cite{wp}. \HVTP{} includes the bidoublet together with the singlet $\widetilde T$ that mixes with $t_R$; together they form a $\mathbf5_{2/3}$ of $SO(5)$ (section~\ref{sec:multiplets} and figure~\ref{fig:multiplets}). Larger representations, such as the $\mathbf{10}$ or the $\mathbf{14}$, would contain further multiplets.

\subsection{A warped interpretation}
The two-site model can be viewed as a truncation of a five-dimensional warped theory. In the Randall--Sundrum geometry, $ds^2=e^{-2ky}\eta_{\mu\nu}dx^\mu dx^\nu-dy^2$ with $0\le y\le L$, mass scales near the infrared boundary are redshifted by $e^{-kL}$, so that the first Kaluza--Klein excitations of the gauge and fermion fields appear at the TeV scale. In holographic composite Higgs models the Higgs is the fifth component of a bulk gauge field, and its potential, generated by Wilson lines, is the five-dimensional counterpart of the radiative pNGB potential~\cite{rs,holo,mchm}. In this picture the elementary--composite mixing angles measure the overlap of the zero-mode profiles with the infrared brane: light quarks, localised towards the ultraviolet, couple weakly to the resonances, whereas the top quark, localised towards the infrared to obtain its large mass, couples strongly to them and to its partners. The two-site truncation retains the lightest resonances and the correlations between Drell--Yan production, vector-boson fusion and heavy-fermion decays that follow from these overlaps~\cite{two,Mohan:2015doa}, while it discards the higher resonances and the details of the potential.

\section{Symmetries, spectrum and couplings of \HVTP{}}\label{sec:model}
\subsection{Sites, symmetries and multiplets}\label{sec:multiplets}
\HVTP{} is a two-site realisation of the elementary and composite sectors of section~\ref{sec:composite}~\cite{two}, which follows the construction of our earlier studies of heavy gluons, $W'$ resonances and top partners~\cite{gluon,wp,Vignaroli:2012nf,Vignaroli:2012sf} and combines their gauge and fermion sectors in a single model. The elementary site (site 0) contains the SM gauge fields $A_{0i}$, with couplings $g_{0i}$, and the chiral SM fermions, including $t$ and $b$; the index $i=1,2,3$ refers to hypercharge, weak isospin and colour. The composite site (site 1) contains a second copy of the SM gauge group, with fields $A_{1i}$ and couplings $g_{1i}$, the Higgs doublet and the composite fermions. Link fields, which we treat in the unitary gauge, break the product of the two gauge groups to the diagonal SM group: for each $i$ one combination of $A_{0i}$ and $A_{1i}$ remains massless and becomes the SM gauge field, and the orthogonal one, the heavy vector $\rho_i$, acquires the mass $M_i$ (section~\ref{sec:gauge}).

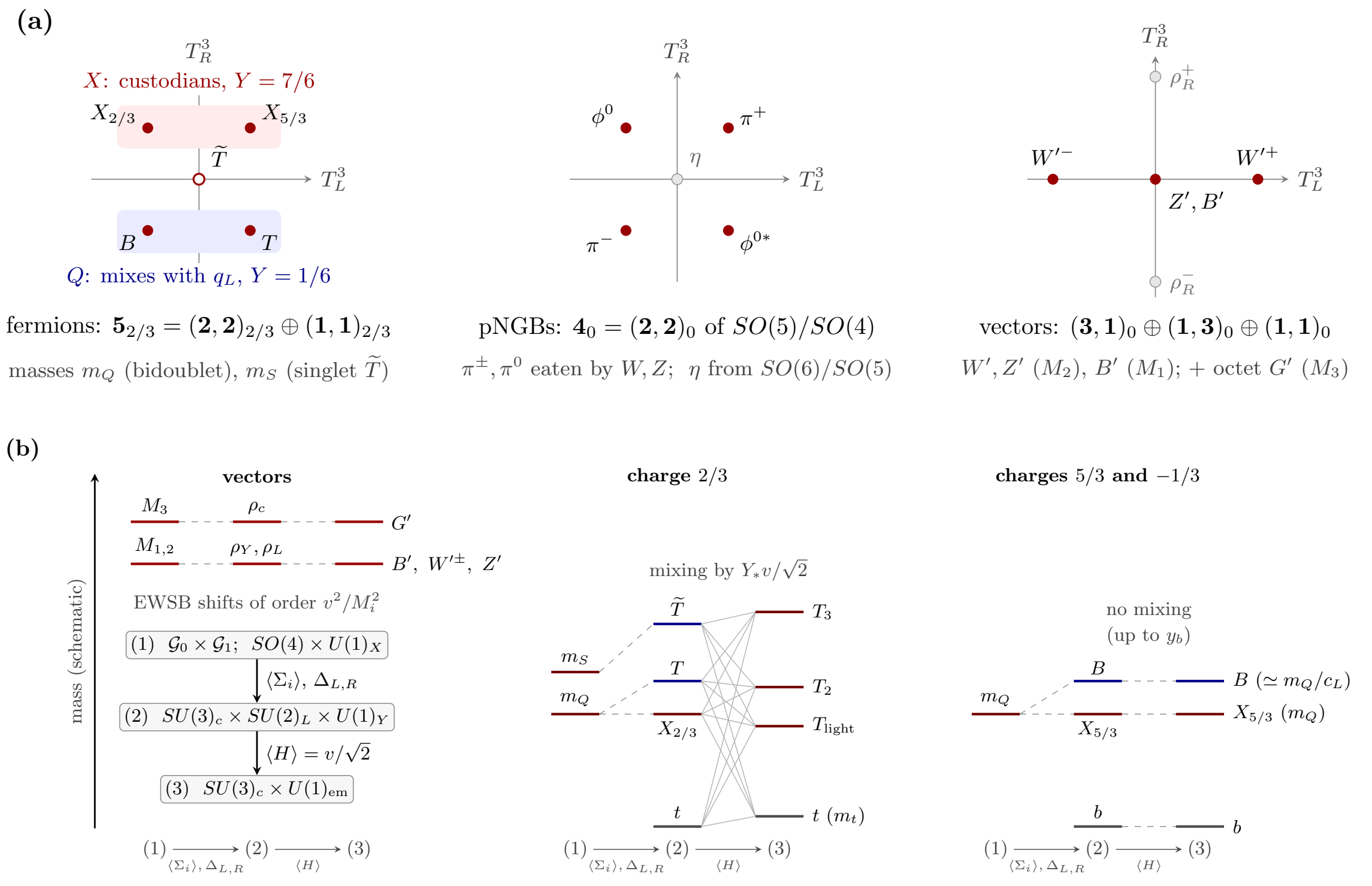
\begin{figure}[htbp]\centering
\resizebox{\linewidth}{!}{
\begin{tikzpicture}[font=\small,
  ax/.style={->,>=stealth,black!50},
  st/.style={circle,fill=red!60!black,inner sep=1.6pt},
  sth/.style={circle,draw=red!60!black,fill=white,thick,inner sep=1.6pt},
  sto/.style={circle,draw=black!40,fill=black!10,inner sep=1.6pt},
  lab/.style={font=\footnotesize,text=black!75},
  lvl/.style={very thick},
  con/.style={black!45,dashed}]
\node[font=\bfseries] at (-2.4,2.3) {(a)};
\begin{scope}[shift={(0,0)},x=1.5cm,y=1.5cm]
\draw[ax] (-1.05,0) -- (1.1,0) node[right,lab]{$T^3_L$};
\draw[ax] (0,-1.0) -- (0,1.05) node[above,lab]{$T^3_R$};
\fill[red!8,rounded corners=3pt] (-0.8,0.3) rectangle (0.8,0.72);
\fill[blue!8,rounded corners=3pt] (-0.8,-0.72) rectangle (0.8,-0.3);
\node[st] at (0.5,0.5) {}; \node[right,font=\footnotesize] at (0.52,0.62) {$X_{5/3}$};
\node[st] at (-0.5,0.5) {}; \node[left,font=\footnotesize] at (-0.52,0.62) {$X_{2/3}$};
\node[st] at (0.5,-0.5) {}; \node[right,font=\footnotesize] at (0.52,-0.62) {$T$};
\node[st] at (-0.5,-0.5) {}; \node[left,font=\footnotesize] at (-0.52,-0.62) {$B$};
\node[sth] at (0,0) {}; \node[font=\footnotesize,anchor=south west] at (0.03,0.03) {$\widetilde T$};
\node[lab,text=red!60!black,fill=white,inner sep=1pt] at (0,0.95) {$X$: custodians, $Y=7/6$};
\node[lab,text=blue!55!black,fill=white,inner sep=1pt] at (0,-0.95) {$Q$: mixes with $q_L$, $Y=1/6$};
\node[align=center] at (0,-1.45) {fermions: $\mathbf5_{2/3}=(\mathbf2,\mathbf2)_{2/3}\oplus(\mathbf1,\mathbf1)_{2/3}$};
\node[lab,align=center] at (0,-1.85) {masses $m_Q$ (bidoublet), $m_S$ (singlet $\widetilde T$)};
\end{scope}
\begin{scope}[shift={(7.0,0)},x=1.5cm,y=1.5cm]
\draw[ax] (-1.05,0) -- (1.1,0) node[right,lab]{$T^3_L$};
\draw[ax] (0,-1.0) -- (0,1.05) node[above,lab]{$T^3_R$};
\node[st] at (0.5,0.5) {}; \node[right,font=\footnotesize] at (0.52,0.62) {$\pi^+$};
\node[st] at (-0.5,-0.5) {}; \node[left,font=\footnotesize] at (-0.52,-0.62) {$\pi^-$};
\node[st] at (0.5,-0.5) {}; \node[right,font=\footnotesize] at (0.52,-0.62) {$\phi^{0*}$};
\node[st] at (-0.5,0.5) {}; \node[left,font=\footnotesize] at (-0.52,0.62) {$\phi^0$};
\node[sto] at (0,0) {}; \node[font=\footnotesize,anchor=south west,text=black!60] at (0.03,0.03) {$\eta$};
\node[align=center] at (0,-1.45) {pNGBs: $\mathbf4_0=(\mathbf2,\mathbf2)_0$ of $SO(5)/SO(4)$};
\node[lab,align=center] at (0,-1.85) {$\pi^\pm,\pi^0$ eaten by $W,Z$;\ \ $\eta$ from $SO(6)/SO(5)$};
\end{scope}
\begin{scope}[shift={(14.0,0)},x=1.5cm,y=1.5cm]
\draw[ax] (-1.25,0) -- (1.3,0) node[right,lab]{$T^3_L$};
\draw[ax] (0,-1.15) -- (0,1.2) node[above,lab]{$T^3_R$};
\node[st] at (1,0) {}; \node[above,font=\footnotesize] at (1,0.05) {$W'^+$};
\node[st] at (-1,0) {}; \node[above,font=\footnotesize] at (-1,0.05) {$W'^-$};
\node[st] at (0,0) {}; \node[font=\footnotesize,anchor=north west] at (0.03,-0.03) {$Z',B'$};
\node[sto] at (0,1) {}; \node[right,font=\footnotesize,text=black!55] at (0.05,1) {$\rho_R^+$};
\node[sto] at (0,-1) {}; \node[right,font=\footnotesize,text=black!55] at (0.05,-1) {$\rho_R^-$};
\node[align=center] at (0,-1.45) {vectors: $(\mathbf3,\mathbf1)_0\oplus(\mathbf1,\mathbf3)_0\oplus(\mathbf1,\mathbf1)_0$};
\node[lab,align=center] at (0,-1.85) {$W',Z'$ ($M_2$), $B'$ ($M_1$); $+$ octet $G'$ ($M_3$)};
\end{scope}
\end{tikzpicture}}\\[4mm]
\resizebox{\linewidth}{!}{
\begin{tikzpicture}[font=\small,
  ax/.style={->,>=stealth,black!50},
  st/.style={circle,fill=red!60!black,inner sep=1.6pt},
  sth/.style={circle,draw=red!60!black,fill=white,thick,inner sep=1.6pt},
  sto/.style={circle,draw=black!40,fill=black!10,inner sep=1.6pt},
  lab/.style={font=\footnotesize,text=black!75},
  lvl/.style={very thick},
  con/.style={black!45,dashed}]
\begin{scope}[y=1.0cm]
\node[font=\bfseries] at (-1.2,6.3) {(b)};
\draw[->,>=stealth,thick] (0,0) -- (0,5.9) node[pos=0.5,above,rotate=90,lab]{mass (schematic)};
\begin{scope}[shift={(0.6,0)}]
\node[font=\footnotesize\bfseries] at (2.1,5.85) {vectors};
\foreach \x in {0,1.7,3.4} {
 \draw[lvl,red!60!black] (\x,4.4) -- ++(0.8,0);
 \draw[lvl,red!60!black] (\x,5.1) -- ++(0.8,0);}
\foreach \x in {0.8,2.5} {\draw[con] (\x,4.4) -- ++(0.9,0); \draw[con] (\x,5.1) -- ++(0.9,0);}
\node[font=\footnotesize,above] at (0.4,4.4) {$M_{1,2}$};
\node[font=\footnotesize,above] at (0.4,5.1) {$M_3$};
\node[font=\footnotesize,above] at (2.1,4.4) {$\rho_Y,\rho_L$};
\node[font=\footnotesize,above] at (2.1,5.1) {$\rho_c$};
\node[font=\footnotesize,anchor=west] at (4.2,4.4) {$B',\ W'^\pm,\ Z'$};
\node[font=\footnotesize,anchor=west] at (4.2,5.1) {$G'$};
\node[lab,align=center] at (2.1,3.75) {EWSB shifts of order $v^2/M_i^2$};
\node[draw=black!50,rounded corners=2pt,fill=black!3,font=\footnotesize,inner sep=2pt] (s1) at (2.1,3.05) {(1)\ \ $\mathcal G_0\times\mathcal G_1$;\ \ $SO(4)\times U(1)_X$};
\node[draw=black!50,rounded corners=2pt,fill=black!3,font=\footnotesize,inner sep=2pt] (s2) at (2.1,1.85) {(2)\ \ $SU(3)_c\times SU(2)_L\times U(1)_Y$};
\node[draw=black!50,rounded corners=2pt,fill=black!3,font=\footnotesize,inner sep=2pt] (s3) at (2.1,0.65) {(3)\ \ $SU(3)_c\times U(1)_{\rm em}$};
\draw[->,>=stealth,thick] (s1) -- node[right,font=\footnotesize]{$\langle\Sigma_i\rangle$, $\Delta_{L,R}$} (s2);
\draw[->,>=stealth,thick] (s2) -- node[right,font=\footnotesize]{$\langle H\rangle=v/\sqrt2$} (s3);
\foreach \x/\t in {0.4/(1),2.1/(2),3.8/(3)} {\node[lab] at (\x,-0.35) {\t};}
\draw[->,>=stealth,black!70] (0.7,-0.35) -- node[below=1pt,font=\tiny]{$\langle\Sigma_i\rangle,\Delta_{L,R}$} (1.8,-0.35);
\draw[->,>=stealth,black!70] (2.4,-0.35) -- node[below=1pt,font=\tiny]{$\langle H\rangle$} (3.5,-0.35);
\end{scope}
\begin{scope}[shift={(7.6,0)}]
\node[font=\footnotesize\bfseries] at (2.1,5.85) {charge $2/3$};
\draw[lvl,red!45!black] (0,1.9) -- ++(0.8,0); \node[font=\footnotesize,above] at (0.4,1.9) {$m_Q$};
\draw[lvl,red!45!black] (0,2.6) -- ++(0.8,0); \node[font=\footnotesize,above] at (0.4,2.6) {$m_S$};
\draw[lvl,red!45!black] (1.7,1.9) -- ++(0.8,0); \node[font=\footnotesize,below] at (2.1,1.92) {$X_{2/3}$};
\draw[lvl,blue!55!black] (1.7,2.45) -- ++(0.8,0); \node[font=\footnotesize,above] at (2.1,2.43) {$T$};
\draw[lvl,blue!55!black] (1.7,3.4) -- ++(0.8,0); \node[font=\footnotesize,above] at (2.1,3.4) {$\widetilde T$};
\draw[lvl,black!70] (1.7,0.03) -- ++(0.8,0); \node[font=\footnotesize,above] at (2.1,0.03) {$t$};
\draw[con] (0.8,1.9) -- (1.7,1.9);
\draw[con] (0.8,1.9) -- (1.7,2.45);
\draw[con] (0.8,2.6) -- (1.7,3.4);
\foreach \ya in {0.03,1.9,2.45,3.4} {\foreach \yb in {0.2,1.7,2.35,3.6} {\draw[black!30] (2.5,\ya) -- (3.4,\yb);}}
\draw[lvl,red!45!black] (3.4,1.7) -- ++(0.8,0) node[right,font=\footnotesize,black]{$T_{\rm light}$};
\draw[lvl,red!45!black] (3.4,2.35) -- ++(0.8,0) node[right,font=\footnotesize,black]{$T_2$};
\draw[lvl,red!45!black] (3.4,3.6) -- ++(0.8,0) node[right,font=\footnotesize,black]{$T_3$};
\draw[lvl,black!70] (3.4,0.2) -- ++(0.8,0) node[right,font=\footnotesize,black]{$t$\ ($m_t$)};
\node[lab,align=center] at (2.95,4.3) {mixing by $Y_*v/\sqrt2$};
\foreach \x/\t in {0.4/(1),2.1/(2),3.8/(3)} {\node[lab] at (\x,-0.35) {\t};}
\draw[->,>=stealth,black!70] (0.7,-0.35) -- node[below=1pt,font=\tiny]{$\langle\Sigma_i\rangle,\Delta_{L,R}$} (1.8,-0.35);
\draw[->,>=stealth,black!70] (2.4,-0.35) -- node[below=1pt,font=\tiny]{$\langle H\rangle$} (3.5,-0.35);
\end{scope}
\begin{scope}[shift={(14.6,0)}]
\node[font=\footnotesize\bfseries] at (2.1,5.85) {charges $5/3$ and $-1/3$};
\draw[lvl,red!45!black] (0,1.9) -- ++(0.8,0); \node[font=\footnotesize,above] at (0.4,1.9) {$m_Q$};
\draw[lvl,red!45!black] (1.7,1.9) -- ++(0.8,0); \node[font=\footnotesize,below] at (2.1,1.92) {$X_{5/3}$};
\draw[lvl,blue!55!black] (1.7,2.45) -- ++(0.8,0); \node[font=\footnotesize,above] at (2.1,2.43) {$B$};
\draw[lvl,black!70] (1.7,0.03) -- ++(0.8,0); \node[font=\footnotesize,above] at (2.1,0.03) {$b$};
\draw[con] (0.8,1.9) -- (1.7,1.9);
\draw[con] (0.8,1.9) -- (1.7,2.45);
\draw[con] (2.5,1.9) -- (3.4,1.9); \draw[con] (2.5,2.45) -- (3.4,2.45); \draw[con] (2.5,0.03) -- (3.4,0.03);
\draw[lvl,red!45!black] (3.4,1.9) -- ++(0.8,0) node[right,font=\footnotesize,black]{$X_{5/3}$ ($m_Q$)};
\draw[lvl,blue!55!black] (3.4,2.45) -- ++(0.8,0) node[right,font=\footnotesize,black]{$B$ ($\simeq m_Q/c_L$)};
\draw[lvl,black!70] (3.4,0.03) -- ++(0.8,0) node[right,font=\footnotesize,black]{$b$};
\node[lab,align=center] at (2.95,3.4) {no mixing\\(up to $y_b$)};
\foreach \x/\t in {0.4/(1),2.1/(2),3.8/(3)} {\node[lab] at (\x,-0.35) {\t};}
\draw[->,>=stealth,black!70] (0.7,-0.35) -- node[below=1pt,font=\tiny]{$\langle\Sigma_i\rangle,\Delta_{L,R}$} (1.8,-0.35);
\draw[->,>=stealth,black!70] (2.4,-0.35) -- node[below=1pt,font=\tiny]{$\langle H\rangle$} (3.5,-0.35);
\end{scope}
\end{scope}
\end{tikzpicture}}
\caption{(a) Multiplets of the composite sector, eqs.~\eqref{eq:higgsmatrix}--\eqref{eq:vectormultiplets}, drawn as weight diagrams: each state is placed at its values of $(T^3_L,T^3_R)$, and its electric charge is $T^3_L+T^3_R+X$. Left: the top partners, the bidoublet $\Psi=(Q,X)$, with the doublet $Q$ that mixes with $q_L$ and the custodians $X$, and the singlet $\widetilde T$ that mixes with $t_R$. Centre: the Higgs bidoublet $\Sigma_H$, whose neutral weights are $\phi^0$ and $\phi^{0*}$, with $\phi^0=(v+h+i\pi^0)/\sqrt2$; $\pi^\pm$ and $\pi^0$ become the longitudinal $W$ and $Z$, and the open point is the singlet $\eta$ of \HVTPeta{}. Right: the spin-one resonances; the grey points $\rho_R^\pm$ are not included in \HVTP{}, and $B'$ is the hypercharge combination of $\rho_R^3$ and $\rho_X$. (b) Schematic path from the Lagrangian masses (1) to the masses after the link vacua $\langle\Sigma_i\rangle$ and the elementary--composite mixings $\Delta_{L,R}$, before EWSB (2), and to the physical masses after the Higgs vacuum (3); the left block shows the symmetry at each step, with $\mathcal G_{0,1}$ the gauge groups of the two sites and $SO(4)\times U(1)_X$ the global symmetry of the composite sector. The mixings raise the partners of $q_L$ and $t_R$ (blue); EWSB mixes all charge-$2/3$ states (grey lines), while $X_{5/3}$ and $B$ keep their masses. The levels are not to scale and their ordering depends on the parameters.}\label{fig:multiplets}
\end{figure}

\paragraph{Global symmetry and multiplets.} The composite fields are organised by the global symmetry of the strong sector, $SO(5)\times U(1)_X$ together with colour, spontaneously broken at the scale $f$ to $SO(4)\times U(1)_X\simeq SU(2)_L\times SU(2)_R\times U(1)_X$. Hypercharge and electric charge are
\begin{equation}
 Y=T^3_R+X,\qquad Q_{\rm em}=T^3_L+T^3_R+X .
\end{equation}
A multiplet $(\mathbf2,\mathbf2)$ of $SU(2)_L\times SU(2)_R$ can be written as a $2\times2$ matrix on which $SU(2)_L$ acts from the left and $SU(2)_R$ from the right, $\mathcal M\to U_L\mathcal M U_R^\dagger$; its rows have $T^3_L=+\frac12,-\frac12$ and its columns $T^3_R=-\frac12,+\frac12$. The Higgs doublet $H=(\phi^+,\phi^0)^T$ then reads
\begin{equation}
 \Sigma_H=\big(\widetilde H,\ H\big)=\begin{pmatrix}\phi^{0*}&\phi^+\\-\phi^-&\phi^0\end{pmatrix}\sim(\mathbf2,\mathbf2)_0 .
 \label{eq:higgsmatrix}
\end{equation}
The top partners form a $\mathbf5_{2/3}$ of $SO(5)$, which decomposes under $SO(4)$ as $\mathbf5=\mathbf4\oplus\mathbf1$:
\begin{equation}
 \Psi_{\mathbf5}=\begin{pmatrix}\Psi_{\mathbf4}\\ \widetilde T\end{pmatrix},\qquad
 \Psi_{\mathbf4}=\frac1{\sqrt2}\begin{pmatrix}iB-iX_{5/3}\\ B+X_{5/3}\\ iT+iX_{2/3}\\ -T+X_{2/3}\end{pmatrix},
 \label{eq:fiveplet}
\end{equation}
where the fourplet is written in the $SO(4)$ basis of ref.~\cite{hunter}. In the $SU(2)_L\times SU(2)_R$ notation the fourplet is the bidoublet and the fifth component the singlet,
\begin{equation}
 \Psi=\big(Q,\ X\big)=\begin{pmatrix}T&X_{5/3}\\B&X_{2/3}\end{pmatrix}\sim(\mathbf2,\mathbf2)_{2/3},\qquad
 S=\widetilde T\sim(\mathbf1,\mathbf1)_{2/3}.
 \label{eq:multiplets}
\end{equation}
The vector resonances are
\begin{equation}
 \rho_L=\frac12\begin{pmatrix}\rho_L^3&\sqrt2\,\rho_L^+\\\sqrt2\,\rho_L^-&-\rho_L^3\end{pmatrix}\sim(\mathbf3,\mathbf1)_0,\qquad
 \rho_c=\rho_c^A\frac{\lambda^A}{2}\sim\mathbf8 ,
 \label{eq:vectormultiplets}
\end{equation}
together with $\rho_Y$, a singlet of $SU(2)_L$ coupled to hypercharge.
Each entry is a state of definite $(T^3_L,T^3_R)$, which gives the weight diagrams of figure~\ref{fig:multiplets}(a):
\begin{itemize}
\item $\Sigma_H$ contains the four real pNGBs of $SO(5)/SO(4)$: the Higgs boson $h$ and the Goldstones $\pi^\pm,\pi^0$ eaten by the $W$ and the $Z$. Its vacuum, $\langle\Sigma_H\rangle=(v/\sqrt2)\,\mathbb 1$, preserves the diagonal (custodial) $SU(2)_V$. In the next-to-minimal coset $SO(6)/SO(5)$ the pNGBs form a $\mathbf5$ of $SO(5)$, which adds to $\Sigma_H$ the singlet $\eta$ of \HVTPeta{} (section~\ref{sec:etamodel});
\item the two columns of $\Psi$ are $SU(2)_L$ doublets: $Q=(T,B)$, with $Y=1/6$, has the quantum numbers of $q_L$, and the custodians $X=(X_{5/3},X_{2/3})$ have $Y=7/6$. The singlet $\widetilde T$ has the quantum numbers of $t_R$. Since only $SO(4)$ is linearly realised in the effective theory, the fourplet and the singlet of eq.~\eqref{eq:fiveplet} have independent Dirac masses $m_Q$ and $m_S$~\cite{hunter};
\item the vectors of a complete strong sector fill the adjoint of $SU(2)_L\times SU(2)_R\times U(1)_X$, a triplet $\rho_L$, a triplet $\rho_R$ and a singlet $\rho_X$, together with the colour octet $\rho_c$. \HVTP{} retains the vectors with the quantum numbers of the SM gauge bosons, which mix with them and are produced in Drell--Yan: $\rho_L$, whose components give $W'^\pm$ and $Z'$; $\rho_Y$, the combination of $\rho_R^3$ and $\rho_X$ that couples to hypercharge, which gives $B'$; and $\rho_c$, which gives $G'$. Since $\rho_Y$ contains $\rho_R^3$ but not $\rho_R^\pm$, this truncation breaks the custodial symmetry of the vector sector (section~\ref{sec:bprime}).
\end{itemize}
The elementary doublet $q_L=(t_L,b_L)$ mixes with $Q$, and $t_R$ with $\widetilde T$, through the mass parameters $\Delta_L$ and $\Delta_R$ of section~\ref{sec:fermions}; $b_R$ and the light fermions do not mix.

\paragraph{From Lagrangian to physical masses.} The masses of the physical states arise in two steps, shown schematically in figure~\ref{fig:multiplets}(b) and listed in table~\ref{tab:spectrum}. Before EWSB the heavy vectors have masses $M_i$, the elementary--composite mixings of section~\ref{sec:composite} push the partners of the top components upwards, $Q$ to $m_Q/c_L$ and $\widetilde T$ to $m_S/c_R$, and leave the custodians at $m_Q$. EWSB then mixes all states of equal electric charge. The heavy vectors mix with the $W$ and the $Z$, which changes their masses by relative amounts of order $v^2/M_i^2$. The composite Yukawa coupling mixes the four charge-$2/3$ states $t$, $T$, $X_{2/3}$ and $\widetilde T$ into the top quark and three heavy eigenstates, which we call $T_{\rm light}$, $T_2$ and $T_3$ in order of increasing mass; their masses and compositions can differ substantially from the Lagrangian values when the partners are close in mass. $X_{5/3}$, which has no partner of equal charge, stays at $m_Q$, and $B$, which mixes with $b$ only through the small bottom Yukawa coupling, stays at $m_Q/c_L$. The package accepts the physical masses as inputs and performs the inversion to the Lagrangian parameters (section~\ref{sec:usage}); the names of the states in the UFO are given in table~\ref{tab:names}.

\begin{table}[htbp]\centering\small
\begin{tabular}{@{}lllll@{}}\toprule
Field & $(SU(3)_c,SU(2)_L)_Y$ & Components & Mass before EWSB & Mass eigenstates\\\midrule
$Q$ & $(\mathbf3,\mathbf2)_{1/6}$ & $T,\ B$ & $m_Q/c_L$ & $T\to T_{\rm light},T_2,T_3$;\ $B$\\
$X$ & $(\mathbf3,\mathbf2)_{7/6}$ & $X_{5/3},\ X_{2/3}$ & $m_Q$ & $X_{5/3}$;\ $X_{2/3}\to T_{\rm light},T_2,T_3$\\
$S$ & $(\mathbf3,\mathbf1)_{2/3}$ & $\widetilde T$ & $m_S/c_R$ & $\widetilde T\to T_{\rm light},T_2,T_3$\\
$\rho_L$ & $(\mathbf1,\mathbf3)_0$ & $\rho_L^\pm,\ \rho_L^3$ & $M_2$ & $W'^\pm,\ Z'$\\
$\rho_Y$ & $(\mathbf1,\mathbf1)_0$ & $\rho_Y$ & $M_1$ & $B'$\\
$\rho_c$ & $(\mathbf8,\mathbf1)_0$ & $\rho_c^A$ & $M_3$ & $G'$ ($m_{G'}=M_3$)\\\bottomrule
\end{tabular}
\caption{Heavy fields of \HVTP{} under the diagonal SM gauge group, their masses before EWSB and the mass eigenstates to which they contribute. The three heavy charge-$2/3$ eigenstates are superpositions of $T$, $X_{2/3}$, $\widetilde T$ and of the elementary top, labelled by increasing mass.}\label{tab:spectrum}
\end{table}

\paragraph{States not included.} The right-handed charged vectors $\rho_R^\pm$, the component of $\rho_X$ orthogonal to hypercharge, the axial resonances in the coset $SO(5)/SO(4)$, which the Weinberg sum rules typically place above the vectors~\cite{Redi:2012ha,Marzocca:2012zn}, and the fermionic multiplets that would contribute to the bottom mass~\cite{wp} are decoupled; the bottom quark receives its mass from an elementary Yukawa coupling. These vectors have no elementary counterpart: they couple to the light quarks only through electroweak mixing, of relative order $v^2/M^2$, and are produced in vector-boson fusion or in pairs rather than in Drell--Yan. Their main effect on the present analysis, when they are lighter than the retained vectors, would be to open further decay channels of $W'$, $Z'$ and of the partners, making the vectors broader but possibly reducing the branching fractions into the partner channels of section~\ref{sec:broad}. They also contribute to the electroweak precision observables; in particular, a complete $SU(2)_R$ triplet degenerate with $B'$ would remove the tree-level contribution of $B'$ to $\hat T$ (section~\ref{sec:bprime}). Finally, the model keeps the leading interactions after EWSB: effects of higher order in $v/f$ depend on the nonlinear completion and are not included. In particular, the $\mathcal O(\xi)$ modifications of the Higgs couplings, which give the bound $\xi\lesssim0.1$ of section~\ref{sec:intro}, are absent, and $f$ does not appear as an independent parameter: the bound on $\xi$ enters our discussion only through the naturalness argument for light partners. Throughout, $M_i$ denote vector masses before EWSB and $m_Q,m_S$ the composite Dirac masses; all thresholds in the phenomenological analysis use physical masses.

The tree-level Lagrangian is
\begin{align}
\mathcal L_{\rm \HVTP{}}={}&\mathcal L_{\rm gauge}+\mathcal L_H+
 \sum_{f=q_L,u_R,d_R,\ell_L,e_R}\sum_{n=1}^3
 \bar f_n i\!\not\!D_0 f_n+\sum_{\Psi=Q,X,S}\bar\Psi i\!\not\!D_1\Psi
 +\mathcal L_{m,Y}\nonumber\\
 &-\sum_{\ell=\mu,\tau}(y_\ell\bar L_\ell H\ell_R+\mathrm{h.c.}) ,
\end{align}
where the elementary fermions, including the top and bottom quarks, live on site 0 and the composite fermions on site 1. We now describe the three non-trivial terms $\mathcal L_{\rm gauge}$, $\mathcal L_H$ and $\mathcal L_{m,Y}$.

\subsection{Gauge sector}\label{sec:gauge}
For each gauge factor $i$, let $A_{0i}$ and $A_{1i}$ be the canonically normalised site fields, with covariant derivatives $D_r=\partial-ig_{ri}A_{ri}^at^a$ and field strengths $F_{ri}^a$. In the link-unitary gauge
\begin{equation}
 {\cal L}_{\rm gauge}=-\tfrac14\sum_{r,i}(F_{ri}^a)^2+
 \sum_i\frac{M_i^2}{2(g_{0i}^2+g_{1i}^2)}
 (g_{1i} A_{1i}^a-g_{0i} A_{0i}^a)^2 .
\end{equation}
With $k_i=\cot\theta_i=g_{1i}/g_{0i}$ and $g_i=g_{0i}\cos\theta_i=g_{1i}\sin\theta_i$, the massless field $A_i$ and the heavy vector $\rho_i$ are defined by $A_{0i}=c_iA_i-s_i\rho_i$ and $A_{1i}=s_iA_i+c_i\rho_i$, with $c_i=\cos\theta_i$ and $s_i=\sin\theta_i$, which is the inverse of eq.~\eqref{eq:gaugesuperposition}. An elementary field therefore couples to $g_i(A_i-\rho_i/k_i)$ and a composite field to $g_i(A_i+k_i\rho_i)$: the heavy vectors couple weakly, as $1/k_i$, to the elementary fermions and strongly, as $k_i$, to the composite fermions and to the Higgs. For the Higgs current and the light elementary fermions the weak triplet reproduces the heavy-vector-triplet (HVT) parameterisation~\cite{hvt}, in its strongly coupled version (model B), with $g_H\equiv g_Vc_H=-g_2k_2$ and $g_F\equiv g_2^2c_F/g_V=g_2/k_2$, at leading order in the electroweak mixing, after the field redefinition $\rho_L\to-\rho_L$ that matches the sign convention of the experimental combinations. The site kinetic terms also fix all interactions with two or more vectors. The composite couplings $g_ik_i$, like the composite Yukawa coupling $Y_*$ introduced below, must stay below the strong-coupling limit $4\pi$; this holds throughout the parameter space considered in this paper.\footnote{In our scans $k_i\le5.6$, so that $g_2k_2\le3.7$, $g_1k_1\le2.0$ and $g_3k_3\lesssim6$ at TeV scales, corresponding to loop factors $g_*^2/(16\pi^2)\le0.08$, $0.03$ and $0.22$, and below the naive-dimensional-analysis estimate $4\pi/\sqrt N\simeq7$ for a strong sector with $N=3$ colours; the top mass fixes $Y_*\simeq2.9$ for $s_L=0.55$, $s_R=0.6$, so that $N_cY_*^2/(16\pi^2)\simeq0.16$. A stronger practical limitation comes from the widths: with the light partner spectrum of section~\ref{sec:broad}, a 4 TeV triplet has $\Gamma/m\simeq0.3$ at $k_2=5$, where its description as a single tree-level resonance becomes less accurate.}

\subsection{Higgs and fermion sectors}\label{sec:fermions}
The Higgs lives on the composite site, with $Y_H=1/2$, $\langle H\rangle=(0,v/\sqrt2)^T$ and
\begin{equation}
 {\cal L}_H=|D_1H|^2-\lambda_H(H^\dagger H-v^2/2)^2,\qquad m_h^2=2\lambda_H v^2 .
\end{equation}
With the multiplets of eq.~\eqref{eq:multiplets}, $\widetilde H=i\sigma^2H^*$ and the hypercharges $1/6$, $7/6$ and $2/3$ of $Q$, $X$ and $S$, the mass and Yukawa terms are
\begin{align}\label{eq:LmY}
 -{\cal L}_{m,Y}={}&m_Q(\bar Q_LQ_R+\bar X_LX_R)+m_S\bar S_LS_R
 +\Delta_L\bar q_LQ_R+\Delta_R\bar S_Lt_R+y_b^{\rm el}\bar q_L H b_R\nonumber\\
 &+Y_*\big[(\bar Q_L\widetilde H+\bar X_LH)S_R+
 \bar S_L(\widetilde H^\dagger Q_R+H^\dagger X_R)\big]+\mathrm{h.c.}
\end{align}
The first line contains the composite Dirac masses and the mixings $\Delta_{L,R}$ of $q_L$ with $Q$ and of $t_R$ with $S$, which fix the degrees of compositeness of eq.~\eqref{eq:superposition} through $\Delta_L=m_Qs_L/c_L$ and $\Delta_R=m_Ss_R/c_R$. The second line is the composite Yukawa interaction; in the matrix notation of eqs.~\eqref{eq:higgsmatrix} and~\eqref{eq:multiplets} it reads $Y_*[{\rm Tr}(\bar\Psi_L\Sigma_H)S_R+\bar S_L{\rm Tr}(\Sigma_H^\dagger\Psi_R)]+{\rm h.c.}$ and is therefore invariant under $SU(2)_L\times SU(2)_R$; we take equal couplings for the two chiralities, as implied by a left--right parity~\cite{protect,custodians}, and the relative sign of the $X$ term fixes the phase of the custodians. The bottom and lepton Yukawa couplings are elementary, and light-quark masses and CKM mixing are neglected.

After EWSB the mass matrices of the states of charge $2/3$ and $-1/3$, in the bases $(t,T,X_{2/3},S)$ and $(b,B)$, are
\begin{equation}
 {\cal M}_u=\begin{pmatrix}0&\Delta_L&0&0\\0&m_Q&0&a\\0&0&m_Q&a\\
 \Delta_R&a&a&m_S\end{pmatrix},\qquad
 {\cal M}_d=\begin{pmatrix}y_b^{\rm el}v/\sqrt2&\Delta_L\\0&m_Q\end{pmatrix},
 \qquad a=\frac{Y_*v}{\sqrt2},
\end{equation}
while $X_{5/3}$ does not mix and has mass $m_Q$. The matrices are diagonalised by biunitary transformations, $U_L^\dagger{\cal M}U_R={\rm diag}(m_i)$ with positive masses, and all couplings are rotated accordingly: a gauge current $J_\chi$ of chirality $\chi$ becomes $U_\chi^\dagger J_\chi U_\chi$, charged currents involve the rotations of both charge sectors, and the Higgs couplings are $U_L^\dagger(\partial{\cal M}/\partial v)U_R$. The leading-order relation $m_t=Y_*s_Ls_Rv/\sqrt2$ receives corrections of order $v^2/m_Q^2$; we therefore fix $Y_*$ by requiring that the lightest eigenvalue of ${\cal M}_u$ equals the physical top mass. This exact treatment retains the electroweak mass splittings and all transitions between partners, which become important when the partners are close in mass.

\subsection{Transition couplings}\label{sec:lambda}
The interactions of the partners with the third-generation quarks and the longitudinal gauge bosons are most transparent in terms of the Goldstone bosons. Collecting the Higgs doublet and the custodial bidoublet into $2\times2$ matrices,
\begin{equation}
 {\cal H}=(\widetilde H,H)=\begin{pmatrix}\phi^{0*}&\phi^+\\-\phi^-&\phi^0\end{pmatrix},\qquad
 {\cal Q}=(Q,X)=\begin{pmatrix}T&X_{5/3}\\B&X_{2/3}\end{pmatrix},
\end{equation}
with $\phi^0=(v+h+i\pi^0)/\sqrt2$ and $\phi^\pm=\pi^\pm$, the composite Yukawa interaction of eq.~\eqref{eq:LmY} takes the manifestly $SU(2)_L\times SU(2)_R$ invariant form
\begin{equation}
 -{\cal L}_{Y}=Y_*\big[{\rm Tr}(\bar{\cal Q}_L{\cal H})\,S_R+\bar S_L\,{\rm Tr}({\cal H}^\dagger{\cal Q}_R)\big]+\mathrm{h.c.}
\end{equation}
Here $S\equiv\widetilde T$ is the composite singlet of eq.~\eqref{eq:LmY}, and only composite fields appear. The elementary quarks enter through the mixings $\Delta_{L,R}$, which make the light quarks superpositions of elementary and composite fields, eq.~\eqref{eq:superposition}: $(t_L,b_L)$ has a component $s_L$ along $(T,B)$, and $t_R$ a component $s_R$ along $S$. Rotating to the light and heavy states of the elementary--composite system, and denoting the heavy states by the same symbols as the composite fields, the Yukawa interaction contains, before EWSB mixing, the transitions
\begin{align}
 -{\cal L}_{Y}\supset{}&Y_*s_Lc_R\,(\bar t_L\phi^{0*}-\bar b_L\phi^-)\,\widetilde T_R\nonumber\\
 &+Y_*s_R\big[c_L(\bar T_L\phi^{0*}-\bar B_L\phi^-)+\bar X_{5/3L}\,\phi^++\bar X_{2/3L}\,\phi^0\big]t_R+\mathrm{h.c.},
 \label{eq:goldstone}
\end{align}
with $c_{L,R}=\sqrt{1-s_{L,R}^2}$~\cite{Vignaroli:2012nf}. Each term couples a partner to a third-generation quark and to a Goldstone boson or to the Higgs. We define the transition coupling $\lambda$ of each partner as the coefficient of its charged-Goldstone term,
\begin{equation}
 \lambda_{\widetilde T}=Y_*s_Lc_R,\qquad \lambda_B=Y_*c_Ls_R,\qquad \lambda_{X_{5/3}}=Y_*s_R .
 \label{eq:lambdaLO}
\end{equation}
The singlet couples to the left-handed doublet $(t_L,b_L)$ through the left mixing, the doublet partners to $t_R$ through the right mixing. The neutral transitions, with $\pi^0$ or $h$, have couplings $\lambda/\sqrt2$, because $\phi^0=(v+h+i\pi^0)/\sqrt2$.

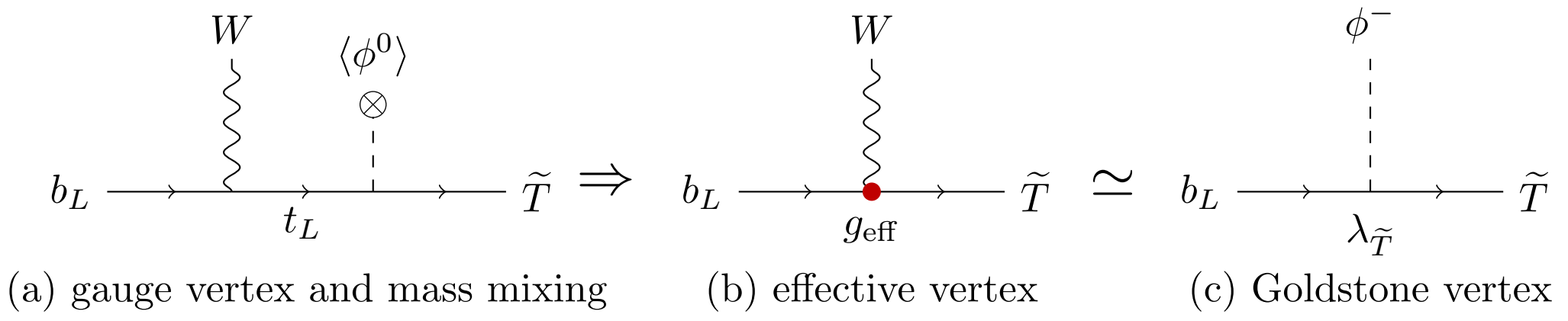
\begin{figure}[htbp]\centering
\tikzset{ferm/.style={postaction={decorate},decoration={markings,mark=at position .55 with {\arrow{>}}}},
weak/.style={decorate,decoration={snake,amplitude=.7mm,segment length=2.6mm}}}
\begin{tikzpicture}[x=.72cm,y=.72cm,font=\small]
\begin{scope}
\draw[ferm] (0,0) node[left]{$b_L$} -- (1.5,0);
\draw[ferm] (1.5,0) -- (3.2,0);\node[below] at (2.35,0){$t_L$};
\draw[ferm] (3.2,0) -- (4.8,0) node[right]{$\widetilde T$};
\draw[weak] (1.5,0) -- (1.5,1.6) node[above]{$W$};
\draw[dashed] (3.2,0) -- (3.2,.9);
\node at (3.2,1.05){$\otimes$};\node[above] at (3.2,1.2){$\langle\phi^0\rangle$};
\node at (2.4,-1.2){(a) gauge vertex and mass mixing};
\end{scope}
\node at (6.0,.1){\Large$\Rightarrow$};
\begin{scope}[shift={(7.6,0)}]
\draw[ferm] (0,0) node[left]{$b_L$} -- (1.6,0);
\draw[ferm] (1.6,0) -- (3.2,0) node[right]{$\widetilde T$};
\draw[weak] (1.6,0) -- (1.6,1.6) node[above]{$W$};
\fill[red!75!black] (1.6,0) circle (.11);
\node[below] at (1.6,-.12){$g_{\rm eff}$};
\node at (1.6,-1.2){(b) effective vertex};
\end{scope}
\node at (12.1,.1){\Large$\simeq$};
\begin{scope}[shift={(13.6,0)}]
\draw[ferm] (0,0) node[left]{$b_L$} -- (1.6,0);
\draw[ferm] (1.6,0) -- (3.2,0) node[right]{$\widetilde T$};
\draw[dashed] (1.6,0) -- (1.6,1.6) node[above]{$\phi^-$};
\node[below] at (1.6,-.08){$\lambda_{\widetilde T}$};
\node at (1.6,-1.2){(c) Goldstone vertex};
\end{scope}
\end{tikzpicture}
\caption{Origin of the $W b\widetilde T$ coupling. (a) The $W$ couples $b_L$ to $t_L$, and the Higgs vacuum mixes $t_L$ with the singlet through the first term of eq.~\eqref{eq:goldstone}. (b) The resulting effective vertex is $g_{\rm eff}=(g/\sqrt2)\sin\theta$, with $\sin\theta\simeq\lambda_{\widetilde T}v/(\sqrt2m_{\widetilde T})$. (c) For a longitudinal $W$ the effective vertex is enhanced by $m_{\widetilde T}/m_W$ and reproduces the Goldstone coupling $\lambda_{\widetilde T}$, as required by the equivalence theorem.}\label{fig:goldstone}
\end{figure}

Figure~\ref{fig:goldstone} shows how these Goldstone couplings become couplings to the physical gauge bosons. The $W$ couples $b_L$ to $t_L$ with the SM strength, and the vacuum value of $\phi^0$ in eq.~\eqref{eq:goldstone} mixes $t_L$ with $\widetilde T$. For $\lambda_{\widetilde T}v\ll m_{\widetilde T}$ the mixing angle is $\sin\theta\simeq\lambda_{\widetilde T}v/(\sqrt2\,m_{\widetilde T})$, and the $Wb\widetilde T$ vertex is $g_{\rm eff}=(g/\sqrt2)\sin\theta$. The emission of a longitudinal $W$ carries an extra factor $m_{\widetilde T}/m_W$, and with $m_W=gv/2$ the amplitude becomes $g_{\rm eff}\,m_{\widetilde T}/m_W=\lambda_{\widetilde T}$, the Goldstone coupling. The small mixing angle is thus compensated by the longitudinal enhancement, and the decays and the single production of a heavy partner are controlled by $\lambda$, with no suppression by $v/m_{\widetilde T}$.

The equivalence theorem also gives the branching fractions directly. For a partner of mass $M$ much heavier than the top and the gauge bosons, a charged transition gives $\Gamma\simeq\lambda^2M/(32\pi)$, and each neutral transition, into $Zt$ or $ht$, half of it. Equation~\eqref{eq:goldstone} then yields the familiar pattern: the singlet decays as $\Gamma(Wb):\Gamma(Zt):\Gamma(ht)\simeq2:1:1$, $T$ and $X_{2/3}$ into $Zt$ and $ht$ with equal rates, and $B$ and $X_{5/3}$ into $Wt$~\cite{aguilar09,aguilar13}.

Equation~\eqref{eq:goldstone} holds at leading order in $v/m_Q$ and for isolated partners. After EWSB the partners also mix with each other, the mixing angles are no longer small when two partners are close in mass, and the couplings must be taken from the mass eigenstates. In \HVTP{} we define them from the gauge current between a partner $Q$ of physical mass $M$, a light electroweak boson $V$ and a third-generation quark $q$,
\begin{equation}
 {\cal L}_{VQq}=V_\mu\bar Q\gamma^\mu(g_LP_L+g_RP_R)q+\mathrm{h.c.},\qquad
 \lambda^a_{Q,Vq}=\frac{Mg_a-m_qg_{\bar a}}{m_V}
 \quad(a=L,R),
 \label{eq:partnerlambda}
\end{equation}
where $g_{L,R}$ include the fermion and gauge rotations and $\bar a$ denotes the opposite chirality; here and in the following, $Q$ denotes a generic partner. A common rephasing of the fermion fields does not change the observables; the overall phase of $\lambda^a$ is therefore conventional, while the relative phase of the two chiralities, which enters the widths through ${\rm Re}(g_Lg_R^*)$, is physical. The contours and tables of this paper use $|\lambda^a|$. This is the construction of figure~\ref{fig:goldstone} with exact rotations: $\lambda^a$ is the coupling of the longitudinal $V$, it reproduces eq.~\eqref{eq:lambdaLO} at leading order and includes all mixing effects in general. Table~\ref{tab:partnerlambda} lists the couplings relevant for each partner together with their leading expressions.

\begin{table}[htbp]\centering
\begin{tabular}{@{}lllll@{}}\toprule
Partner & Transition & \HVTP{} coupling & Leading magnitude & UFO vertex\\\midrule
$\widetilde T$ & $W^+ b_L$ & $\lambda_{\widetilde T}\equiv|\lambda^L_{\widetilde T,Wb}|$ & $|Y_*s_Lc_R|$ & \verb|tp3~ b W+|\\
$B$ & $W^- t_R$ & $\lambda_B\equiv|\lambda^R_{B,Wt}|$ & $|Y_*c_Ls_R|$ & \verb|Bh~ t W-|\\
$X_{5/3}$ & $W^+ t_R$ & $\lambda_{X_{5/3}}\equiv|\lambda^R_{X_{5/3},Wt}|$ & $|Y_*s_R|$ & \verb|x53~ t W+|\\
$T$ & $Zt_R$ & $\lambda_{T,Z}\equiv|\lambda^R_{T,Zt}|$ & $|Y_*c_Ls_R|/\sqrt{2}$ & \verb|tp2~ t Z|\\
$X_{2/3}$ & $Zt_R$ & $\lambda_{X_{2/3},Z}\equiv|\lambda^R_{X_{2/3},Zt}|$ & $|Y_*s_R|/\sqrt{2}$ & \verb|tp1~ t Z|\\\bottomrule
\end{tabular}
\caption{Transition couplings of the partners. The fourth column gives the leading Goldstone coupling of eq.~\eqref{eq:goldstone} for an isolated state of the indicated multiplet. The last column gives the corresponding UFO vertex at the default point of the package, where \texttt{tp1}, \texttt{tp2} and \texttt{tp3} are mostly $X_{2/3}$, $T$ and $\widetilde T$; at other points the composition is given in the param card (section~\ref{sec:usage}).}\label{tab:partnerlambda}
\end{table}

Electroweak mixing redistributes these couplings among the charge-$2/3$ mass eigenstates, especially when two of them are close in mass, so that for a realistic spectrum we always use the rotated couplings of eq.~\eqref{eq:partnerlambda}. Appendix~\ref{app:matching} relates them to the conventions of the experimental searches.

\subsection{Vector mass matrices and gauge vertices}\label{sec:vectormass}
In the charged basis $(W,\rho_L)$ and the neutral basis $(W^3,B,\rho_L^3,\rho_Y)$, with $g=g_2$ and $g'=g_1$, the vector mass matrices are
\begin{equation}
 {\cal M}_{C,N}^2=D_{C,N}+\frac{v^2}{4}u_{C,N}u_{C,N}^T,\qquad
 u_C=(g,g k_2)^T,\quad u_N=(g,-g',g k_2,-g' k_1)^T,
\end{equation}
with $D_C=\mathrm{diag}(0,M_2^2)$ and $D_N=\mathrm{diag}(0,0,M_2^2,M_1^2)$. The neutral matrix has exactly one vanishing eigenvalue, the photon. If $O$ diagonalises ${\cal M}^2$, the coupling of the Higgs to two vectors is $v(O^Tu)_i(O^Tu)_j/2$ (times $ig_{\mu\nu}$), and the $hhVV$ coupling is the same expression divided by $v$. The trilinear and quartic gauge couplings are obtained by rotating each leg of the Yang--Mills terms of the two sites; for instance, with the SM tensor structure stripped off,
\begin{equation}
 g_{N_j W_i^- W_l^+}=\sum_{r=0,1}g_{r2}\,
 O^{(3)}_{rj}O^{(+)}_{ri}O^{(+)}_{rl},
\end{equation}
where $O^{(3)}$ and $O^{(+)}$ project the mass eigenstates onto the neutral and charged weak site fields. These couplings control the decays $W'\to WZ$ and $Z'\to WW$, whose widths grow as $k_2^2$, and they guarantee the cancellations required by gauge invariance in vector-boson scattering.

\subsection{The composite pseudoscalar of \HVTPeta{}}\label{sec:etamodel}
Non-minimal cosets contain additional pNGBs. The simplest example, $SO(6)/SO(5)$, completes the Higgs $\mathbf4$ into a $\mathbf5$ of $SO(5)$ by adding a singlet $\eta$ (figure~\ref{fig:multiplets}(a)). This coset is locally isomorphic to $SU(4)/Sp(4)$, and it is the minimal one that admits an ultraviolet completion in terms of a four-dimensional gauge theory with only fermionic matter: the minimal coset $SO(5)/SO(4)$ cannot be obtained in this way, whereas $SU(4)/Sp(4)$ arises from an $SU(2)$ gauge theory with two Dirac fermions in the fundamental representation~\cite{Cacciapaglia:2014uja,Ferretti:2013kya}. The singlet $\eta$ thus appears in the simplest composite Higgs models with an underlying fermionic description, and its couplings to the top partners can change their decay pattern substantially~\cite{Serra:2015xfa}. Exotic decays of this kind are generic in composite models with an underlying fermionic description~\cite{Bizot:2018tds,Xie:2019gya}, they can open gaps in the coverage of the standard searches~\cite{Cacciapaglia:2019zmj}, and bounds for arbitrary combinations of standard and exotic decays were derived in ref.~\cite{Chala:2017xgc}. Such composite pseudoscalars, with couplings to gluons and photons generated by the composite fermions and by topological (Wess--Zumino--Witten) interactions, and their collider signatures were studied in refs.~\cite{Molinaro:2015cwg,Molinaro:2016oix,Molinaro:2017mwb}. We implement this possibility in \HVTPeta{}, in which a CP-odd singlet on the composite site couples to the composite multiplets,
\begin{align}
 \mathcal L_\eta={}&\tfrac12(\partial\eta)^2-\tfrac12m_\eta^2\eta^2
 -i\eta\big[\kappa_Q(\bar Q\gamma_5Q+\bar X\gamma_5X)+\kappa_S\bar S\gamma_5S\big]\nonumber\\
 &+\frac{\eta}{8\pi f_\eta}\Big[\alpha_sK_G\,G^a_{\mu\nu}\widetilde G^{a\mu\nu}
 +\alpha_2K_W\,W^a_{\mu\nu}\widetilde W^{a\mu\nu}
 +\alpha_1K_B\,B_{\mu\nu}\widetilde B^{\mu\nu}\Big],
 \label{eq:eta}
\end{align}
The couplings to the mass eigenstates follow from the same rotations as the gauge and Higgs currents, $K=U_L^T\,{\rm diag}(0,\kappa_Q,\kappa_Q,\kappa_S)\,U_R$ for the charge-$2/3$ states and analogously for the bottom sector. A singlet-like partner therefore couples to $t\eta$ with a strength of order $\kappa_Ss_R$ and a doublet-like one with $\kappa_Qs_L$, in close analogy with the transitions of table~\ref{tab:partnerlambda}. The last term of eq.~\eqref{eq:eta}, with $\alpha_{2,1}=g^2/4\pi,\,g'^2/4\pi$, contains the local ultraviolet contributions to the couplings of $\eta$ to the gauge bosons, written as operators of the unbroken gauge group with dimensionless coefficients $K_{G,W,B}$ and a reference scale $f_\eta$. They include the Wess--Zumino--Witten interactions fixed by the anomalies of the underlying theory~\cite{Molinaro:2017mwb,Cacciapaglia:2019bqz} and, more generally, the matching contributions of states not included in the model. The explicit fermions add loop contributions, so that the physical amplitudes are described by
\begin{equation}
 \mathcal L_{\rm eff}\supset\frac{\alpha_sc_G}{8\pi}\,\eta\,G^a_{\mu\nu}\widetilde G^{a\mu\nu}+\frac{\alpha\,c_A}{8\pi}\,\eta\,F_{\mu\nu}\widetilde F^{\mu\nu}+C_{ZZ}\,\eta Z_{\mu\nu}\widetilde Z^{\mu\nu}+C_{Z\gamma}\,\eta Z_{\mu\nu}\widetilde F^{\mu\nu}+C_{WW}\,\eta W^+_{\mu\nu}\widetilde W^{-\mu\nu}.
 \label{eq:etaeff}
\end{equation}
Each coefficient is the sum of two parts: the loops of the fermions of the model, and the ultraviolet terms of eq.~\eqref{eq:eta}. For gluons the ultraviolet part is $K_G/f_\eta$, and for photons it is $(K_W+K_B)/f_\eta$; for the $Z$ and $W$ bosons it is obtained by rotating $K_W$ and $K_B$ from the gauge basis to the mass basis (appendix~\ref{app:widths}). For $\kappa_Q=\kappa_S=0$ and vanishing ultraviolet terms ($K_G=K_W=K_B=0$ in the param card, section~\ref{sec:usage}) $\eta$ decouples and the model reduces exactly to \HVTP{}. The phenomenology of $\eta$ and its effect on the partners are discussed in section~\ref{sec:eta}.

The non-Abelian part of $W^a_{\mu\nu}\widetilde W^{a\mu\nu}$ also gives the contact vertices $\eta W^+W^-\gamma$ and $\eta W^+W^-Z$, which are required by gauge invariance in processes with $W$ bosons, such as vector-boson fusion. The loop contributions to $C_{ZZ}$, $C_{Z\gamma}$ and $C_{WW}$ are the exact one-loop contributions of all fermion mass eigenstates, the top and bottom quarks and the partners, including the off-diagonal couplings of $\eta$ and of the $Z$ and $W$ bosons between different eigenstates (appendix~\ref{app:widths}). In the limit of heavy partners each vector-like multiplet with coupling $K$ to $\eta$ and mass $m$ contributes a local term $2N_c\,{\rm Tr}(T^2)\,K/m$ to the coefficients of $\alpha_2W^a_{\mu\nu}\widetilde W^{a\mu\nu}/8\pi$ and $\alpha_1B_{\mu\nu}\widetilde B^{\mu\nu}/8\pi$, with ${\rm Tr}(T^2)=1/2$ for an $SU(2)_L$ doublet and the sum of $Y^2$ over the components for hypercharge~\cite{Molinaro:2016oix,Molinaro:2017mwb}; the top quark, which is chiral and couples to $\eta$ only after EWSB, interferes destructively with the partners, and the mixing of the charge-$2/3$ states modifies their couplings to the $Z$, so that the exact loops reduce the width into $W^+W^-$ by about a factor of two with respect to this limit. In the range considered in this paper the loops give $\mathcal B(\eta\to WW,ZZ)$ from about 1\% at $m_\eta=200$ GeV to about 10\% near the $t\bar t$ threshold (appendix~\ref{app:etaprod}).

These terms are the part of the gauged Wess--Zumino--Witten action that involves a single pNGB. The complete action, whose coefficients depend on the embedding of the electroweak group in the coset, on the vacuum misalignment angle and on the dimension of the hypercolour representation, also contains interactions of $\eta$ with the Higgs boson and the gauge bosons, which are not included in \HVTPeta{} and were studied in detail in refs.~\cite{Molinaro:2017mwb,Vignaroli:2017eps}. These topological interactions dominate the production of $\eta$ in vector-boson fusion, in association with a $W$, a $Z$ or a Higgs boson, and therefore give direct access to the underlying dynamics: at a 100 TeV collider with 3 ab$^{-1}$ they can be tested down to hypercolour representations of dimension $d(R)\simeq7$ for $m_\eta=1$ TeV and a misalignment angle $\sin\theta\simeq0.2$~\cite{Molinaro:2017mwb,Vignaroli:2017eps}. At the LHC, however, the cross sections of these processes are small~\cite{Vignaroli:2017eps}, and a test with sufficient significance is beyond the reach of the HL-LHC.

\section{The public models}\label{sec:usage}
\HVTP{} and \HVTPeta{} are distributed as a small Python package that writes a UFO model~\cite{Degrande:2011ua,Darme:2023jdn} for MadGraph5\_aMC@NLO~\cite{Alwall:2014hca} for a chosen parameter point. The package, together with the inputs and scripts that reproduce the results of this paper, is available at \url{https://doi.org/10.5281/zenodo.24345218}~\cite{hvtpzenodo}. This DOI always points to the latest version of the package; the version used in this paper, 1.0.0, has DOI \href{https://doi.org/10.5281/zenodo.24345219}{10.5281/zenodo.24345219}. This section summarises what the package does, its inputs, a minimal example and the reference points; appendix~\ref{app:usage} gives the particle names, the options of the scripts, practical recommendations for event generation and the details of the validation. The package requires Python~3 with NumPy and SciPy; the computation of the widths requires MadGraph5\_aMC@NLO, with which it has been tested in version 3.8.1.

\subsection{How the package works}
All masses, mixing angles and couplings of \HVTP{} follow from the diagonalisation of the gauge and fermion mass matrices described in section~\ref{sec:model}. These relations are nonlinear, and the package solves them numerically before writing the model. As a consequence, the masses of all particles and all couplings are stored in the UFO as fixed numbers. The only parameters read from the param card are the strong coupling $\alpha_s$, whose running MadGraph applies to all vertices of QCD order, and the total widths in the \texttt{DECAY} block. In particular, \emph{editing the masses in the \texttt{MASS} block of a param card has no effect}: MadGraph lists them only for the interface to parton showers. The package provides a ready-to-use model at a default point; any other point is obtained with a single command (section~\ref{sec:example}), which writes the corresponding model and param card, and scans are automated in the same way.

\subsection{Inputs}
Table~\ref{tab:inputs} lists the inputs. The Lagrangian inputs are the parameters of the two-site Lagrangian of section~\ref{sec:model} before EWSB: the vector masses $M_i$ and coupling ratios $k_i$, the Dirac masses $m_Q$ of the bidoublet and $m_S$ of the singlet and the degrees of compositeness $s_L,s_R$ of $q_L$ and $t_R$, together with the SM inputs. The composite Yukawa coupling $Y_*$ is not an input: it is fixed by requiring that the lightest charge-$2/3$ eigenvalue equals $m_t$. After EWSB the physical masses differ from the inputs: those of $W'$, $Z'$ and $B'$ through the mixing with the SM gauge bosons, and those of the three heavy charge-$2/3$ states through the Yukawa mixing, by large amounts when the partners are close in mass; only $m_{G'}=M_3$ and $m_{X_{5/3}}=m_Q$ are exact. The package therefore also accepts the physical masses in the lower part of the table, each of which replaces the Lagrangian input that it determines (appendix~\ref{app:usage}).

\begin{table}[htbp]\centering\footnotesize
\begin{tabular}{@{}ll>{\raggedright\arraybackslash}p{6.0cm}l@{}}\toprule
Key & Symbol & Meaning & Default\\\midrule
\multicolumn{4}{@{}l}{\textit{Lagrangian inputs}}\\
\verb|M1|, \verb|M2|, \verb|M3| & $M_1,M_2,M_3$ & masses of $\rho_Y$, $\rho_L$, $\rho_c$ before EWSB [GeV] & 4400, 4400, 5000\\
\verb|k1|, \verb|k2|, \verb|k3| & $k_i=\cot\theta_i$ & composite-to-elementary coupling ratios, $k_i>0$ & 3, 3, 3\\
\verb|mQ|, \verb|mS| & $m_Q,m_S$ & Dirac masses of the bidoublet and the singlet [GeV] & 1900, 2400\\
\verb|sL|, \verb|sR| & $s_L,s_R$ & degrees of compositeness of $q_L$ and $t_R$, $0<s<1$ & 0.55, 0.60\\
\verb|alpha_inverse|, \verb|alpha_s| & $1/\alpha$, $\alpha_s$ & electromagnetic and strong couplings at $M_Z$ & 127.9, 0.1184\\
\verb|v|, \verb|MZ| & $v$, $m_Z$ & Higgs vev and $Z$ mass, which fix $g$ and $g'$ [GeV] & 246.22, 91.1876\\
\verb|mt|, \verb|mb|, \verb|mh| & $m_t,m_b,m_h$ & physical masses [GeV]; $m_t$ fixes $Y_*$ & 172.5, 4.7, 125\\
\verb|Ymax| & $Y_{\max}$ & upper end of the search interval for $Y_*$ & 6\\\midrule
\multicolumn{4}{@{}l}{\textit{Physical inputs (optional; each replaces the input in brackets)}}\\
\verb|mWp|, \verb|mBp| & $m_{W'}$, $m_{B'}$ & physical $W'$ and $B'$ masses [GeV] (\verb|M2|, \verb|M1|) & --\\
\verb|mGp|, \verb|mX53| & $m_{G'}$, $m_{X_{5/3}}$ & $G'$ and $X_{5/3}$ masses [GeV] (\verb|M3|, \verb|mQ|) & --\\
\verb|mtp1| & $m_{T_{\rm light}}$ & mass of the lightest heavy charge-$2/3$ state [GeV] (\verb|mS|) & --\\
\verb|Ystar| & $Y_*$ & composite Yukawa coupling (\verb|sL|) & --\\\midrule
\multicolumn{4}{@{}l}{\textit{Additional inputs of \HVTPeta{}}}\\
\verb|m_eta| & $m_\eta$ & mass of the pseudoscalar [GeV] & 500\\
\verb|kQ|, \verb|kS| & $\kappa_Q,\kappa_S$ & couplings of $\eta$ to the bidoublet and the singlet, $\ge0$ & 1, 1\\
\verb|KG|, \verb|KW|, \verb|KB| & $K_G,K_W,K_B$ & optional UV terms $K_{G,W,B}/f_\eta$ in $c_G,c_W,c_B$; $c_A$ gets $(K_W+K_B)/f_\eta$ & 0, 0, 0\\
\verb|f_eta| & $f_\eta$ & normalisation of the UV terms [GeV] & 1000\\
\verb|cancel_heavy| & --- & if 1, sets $K_{G,W,B}$ to cancel the partner loops (heavy limit) & 0\\\bottomrule
\end{tabular}
\caption{Inputs of the point generators, with the keys used in the JSON files and on the command line. The SM inputs fix $g$ and $g'$ through $\alpha$, $v$ and the light neutral eigenvalue $m_Z$; the $W$ mass is then predicted.}\label{tab:inputs}
\end{table}

\subsection{A minimal example}\label{sec:example}
From the directory of the package, the command
\begin{verbatim}
python3 prepare_point.py my_point --set mWp=4000 --set k2=3 \
        --set mtp1=1800 --mg5 /path/to/MG5_aMC/bin/mg5_aMC
\end{verbatim}
creates the directory \verb|my_point| for the default point modified by the requested inputs. It resolves the physical inputs, writes the UFO \verb|HVTP_UFO|, computes all two-body widths with MadWidth~\cite{Alwall:2014bza} and writes the complete \verb|param_card.dat| of the point, with all branching fractions; \verb|prepare_point_eta.py| does the same for \HVTPeta{}. The model is then used as any other model. For example, the MadGraph commands
\begin{verbatim}
import model ./my_point/HVTP_UFO
generate p p > x53 tp1~
output x53_tp1
launch x53_tp1
\end{verbatim}
import the model of the point, generate the production of $X_{5/3}$ together with the light partner, $pp\to X_{5/3}\bar T_{\rm light}$, which at this point proceeds mainly through the $W'^+$ resonance and gives the same-sign dilepton signature discussed in section~\ref{sec:broad}, and create and run the corresponding process directory; the undecayed partners can then be decayed by MadSpin or by the parton shower.
Three points deserve attention in every application. First, the param card that MadGraph creates in the process directory has no branching fractions. The complete param card of the point, \verb|my_point/param_card.dat|, written by \verb|prepare_point.py|, must be used whenever the new states are decayed by the parton shower: it can be copied into the \verb|Cards| directory of the process, or its path can be entered at the card-editing prompt of \verb|launch|. Second, when a vector is broad the full $2\to n$ process, as in the example, should be generated rather than a decay chain in the narrow-width approximation (section~\ref{sec:fw}). Third, each coupling carries, besides its \verb|QCD| and \verb|QED| orders, an order \verb|HV| equal to the number of heavy-vector legs, so that QCD production, vector exchange and the top partners alone (\verb|HV=0|, or vectors decoupled at large $M_i$) can be selected explicitly. Appendix~\ref{app:usage} describes these and further aspects: the particle names, the scans, the flavour scheme, the running of $\alpha_s$, the interface to parton showers and detector simulation, and the accuracy of the models.

\subsection{Reference points}\label{sec:points}
Table~\ref{tab:points} lists the two points used most often in this paper. The default point of the package has the light partner spectrum of section~\ref{sec:broad}, with $W'$ and $B'$ at 4.4 TeV and $k_1=k_2=3$ and the octet at 5 TeV with $k_3=3$, which illustrates the opening of the partner channels. It is not excluded by the limits derived in sections~\ref{sec:vlqlimits}, \ref{sec:vectorsearches} and~\ref{sec:broad} when the narrow-width interpretation of the searches is applied; since its vectors are broad, this conclusion is indicative and should be confirmed with signal predictions that include the widths and the efficiencies of the partner final states. It gives $m_{T_{\rm light}}=1730$ GeV, with $(\mathcal B_{ht},\mathcal B_{Zt})=(0.76,0.21)$, $m_{X_{5/3}}=1900$ GeV with $\mathcal B(Wt)=0.84$, and a $W'$ with $\Gamma/m=15\%$ that decays into partners in 87\% of the cases, 59\% into $X_{5/3}\bar T_{\rm light}$. The test point P0, used to illustrate the mixing effects in sections~\ref{sec:currents}, \ref{sec:vlqlimits} and~\ref{sec:nonstandard}, is excluded by current searches and is provided as \verb|examples/point_P0.json|. Its large splitting of the charge-$2/3$ states from the input masses illustrates why the physical masses, not the inputs, must be used for thresholds and comparisons with searches.

\begin{table}[htbp]\centering\small
\begin{tabular}{@{}lcccccccccc@{}}\toprule
Point & $M_1$ & $M_2$ & $M_3$ & $k_1,k_2$ & $k_3$ & $m_Q$ & $m_S$ & $m_{W'}$ & $m_{T_{\rm light}}$ & $Y_*$\\\midrule
Default & 4400 & 4400 & 5000 & 3 & 3 & 1900 & 2400 & 4407 & 1730 & 2.89\\
P0 & 5000 & 2700 & 2700 & 3 & 2.27 & 1800 & 1440 & 2711 & 1295 & 2.80\\\bottomrule
\end{tabular}
\caption{Reference points (masses in GeV; $s_L=0.55$, $s_R=0.60$ and the SM inputs of table~\ref{tab:inputs} in both). At P0 the heavy charge-$2/3$ masses are 1295, 1977 and 2541 GeV, $m_B=2155$ GeV and $m_{X_{5/3}}=1800$ GeV; at the default point they are 1730, 2169 and 3327 GeV, $m_B=2275$ GeV and $m_{X_{5/3}}=1900$ GeV.}\label{tab:points}
\end{table}

\subsection{Validation}\label{sec:validation}
The two-body widths computed by MadWidth agree with the analytic formulas of appendix~\ref{app:widths}, evaluated with an independent implementation of the mass matrices, for all channels of the partners and of the vectors; for $Z'$ and $B'$ the agreement is within $4\times10^{-5}$ in every channel. Dedicated scripts verify on the UFO the conservation of charge and colour at every vertex, the universality of the photon and gluon couplings, the hermiticity of the fermion couplings, the Ward identities of amplitudes with external photons and gluons, including the $\eta ggg$ contact term of \HVTPeta{}, and the cancellation of the terms that grow with energy in $W^+W^-\to W^+W^-$ and $f_i\bar f_j\to W^+W^-$, as required by the unitarity sum rules. Events generated with MadGraph5\_aMC@NLO 3.8.1 reproduce the expected masses and scale dependence, and have been processed with Pythia~8~\cite{Bierlich:2022pfr} and Delphes~\cite{deFavereau:2013fsa} without errors. Appendix~\ref{app:usage} gives the details; the scripts that reproduce the figures and tables of this paper are distributed with the models.

\section{Decays of the vector resonances}\label{sec:currents}
With the model and its implementation in place, we now turn to its phenomenology, starting from the vector resonances. This section shows how the partial compositeness of the top sector shapes their couplings, which decay channels open as the partners become accessible and how their widths grow, and relates \HVTP{} to the simplified models used in the experimental searches. The phenomenology of the partners themselves is discussed in section~\ref{sec:vlq}.

\subsection{Currents and widths}
The couplings of a heavy vector to the quarks follow from rotating its current, which has opposite-sign components on the elementary and composite fields, to the mass basis. For a chiral pair with degree of compositeness $s$ and $c=\sqrt{1-s^2}$ (eq.~\eqref{eq:superposition}), in units of the corresponding gauge coupling, one finds
\begin{equation}
 C(k,s)=
 \begin{pmatrix}
 k s^2-c^2/k &(k+1/k)sc\\
 (k+1/k)sc &k c^2-s^2/k
 \end{pmatrix},
\end{equation}
where the rows and columns refer to the light and heavy states, and the weak charged current carries an additional factor $1/\sqrt2$. Three regimes are apparent. A mostly elementary quark ($s\to0$) couples with $-1/k$, which suppresses Drell--Yan production at large $k$. The heavy--light transition, proportional to $(k+1/k)sc$, is large for the top quark, whose components are sizeably composite, and mediates decays into one partner and one SM quark. The partners themselves couple with nearly the full composite strength $k$; in particular the custodians $X_{5/3}$ and $X_{2/3}$, which do not mix with the elementary quarks, couple to $W'$ with strength $gk_2/\sqrt2$.

The corresponding widths, collected in appendix~\ref{app:widths}, scale accordingly. The widths into light fermions decrease as $1/k_2^2$, those into $WZ$ and $Wh$ grow as $g^2k_2^2m_{W'}/(192\pi)$, and the width into a pair of degenerate custodians of mass $M_X$,
\begin{equation}
 \Gamma(W'^+\to X_{5/3}\bar X_{2/3})=
 \frac{g_2^2 k_2^2 m_{W'}}{8\pi}(1+2r)\sqrt{1-4r},\qquad r=\frac{M_X^2}{m_{W'}^2},
\end{equation}
is larger by a factor of about 24 than each of the diboson widths away from threshold. Once the partner channels open, a strongly coupled vector therefore becomes broad and decays mainly into partners. This is the central feature of the phenomenology discussed in this paper.

\begin{figure}[htbp]\centering
\includegraphics[width=.99\linewidth]{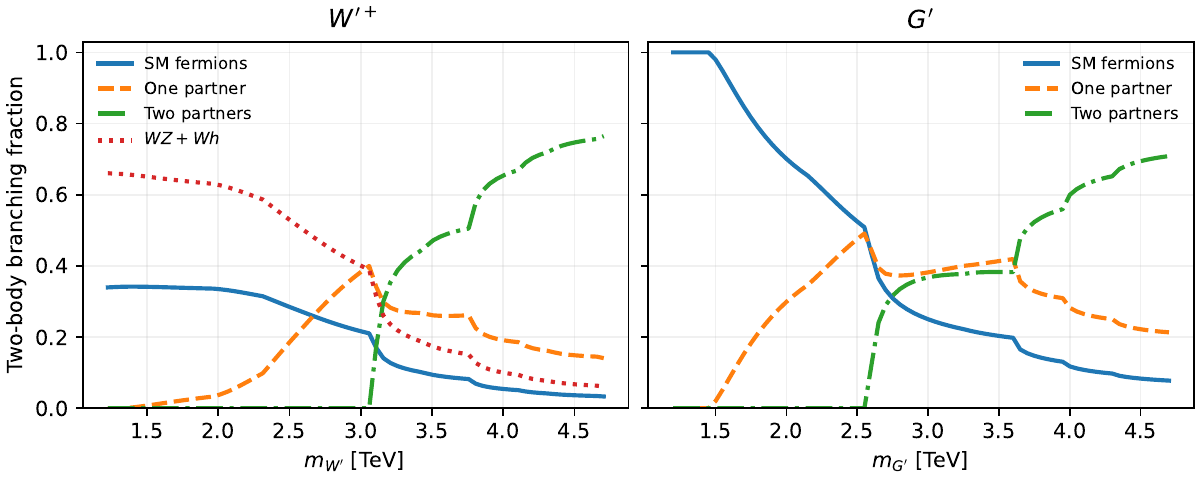}
\caption{Tree-level branching fractions of the vector resonances as a function of their mass, for $M_2=M_3$ varied between 1.2 and 4.7 TeV and the other inputs of the test point P0 (table~\ref{tab:points}). Fermion channels are grouped by the number of heavy daughters; the bosonic curve of the charged vector sums $WZ$ and $Wh$.}\label{fig:br}
\end{figure}

Figure~\ref{fig:br} shows how the decay pattern evolves with the vector mass at a fixed partner spectrum. Below the lightest partner threshold the octet decays only into SM quarks, mostly light-quark pairs and $t\bar t$. Above the threshold for one partner and one top, at about 1.5 TeV for the 1.3 TeV lightest partner of this example, the heavy--light channels take over, and above the threshold for two partners the heavy pairs dominate. The charged vector shows the same pattern with different combinations: electric charge selects $T\bar b$, $t\bar B$ and, above about 3.1 TeV, the pair of the lightest charge-$2/3$ partner and $X_{5/3}$.

\subsection{Reconstructing both resonances}
The heavy--light window $M+m_q<m_{V'}<2M$, with $M$ the mass of the lightest partner, is particularly interesting, as originally shown in refs.~\cite{gluon,Barcelo:2011wu}, because the partner can be reconstructed inside the decay of the vector. For example, lowering $M_3$ to 2.4 TeV at the test point P0 gives $\mathcal B(G'\to T_{\rm light}\bar t+\bar T_{\rm light}t)=0.38$ and $\mathcal B(G'\to t\bar t)=0.17$. Since $\mathcal B(T_{\rm light}\to ht)=0.68$, the dominant heavy--light final state is $t\bar th$, with a fraction of 0.26, while $Wtb$ receives only 0.07. The invariant mass $m(ht)$ then reconstructs the partner and $m(ht\bar t)$ the parent, and the relative weight of the $Wtb$, $Zt\bar t$ and $ht\bar t$ topologies measures the decay pattern of the partner, which mixing moves away from the singlet and doublet limits.

The charged vector also decays into a pair of custodians, $W'^+\to X_{5/3}\bar X_{2/3}$, even when the $T\bar B$ channel is closed~\cite{wp}. The decay $X_{5/3}\to W^+t\to W^+W^+b$ then gives two same-sign leptons, a signature with a much smaller background than the opposite-sign dileptons from top pairs, while the $Z$ or Higgs boson from the other partner tests its neutral decays. The production through an $s$-channel resonance also gives the partners harder spectra and different angular distributions than QCD pair production at the same mass. Additional bottom-sector multiplets would give further signatures, such as $ZZb\bar b$ and $hhb\bar b$ from octet decays~\cite{Vignaroli:2015ama}, and the chirality of the induced $Wtb$ current affects flavour observables such as $b\to s\gamma$~\cite{Vignaroli:2012si}.

Vector-boson fusion offers a complementary production mode. Its rate grows with the coupling to longitudinal gauge bosons, $(g_2k_2)^2$, while Drell--Yan production from light quarks scales as $(g_2/k_2)^2$, so that fusion becomes increasingly important at large $k_2$, with the forward tagging jets as an additional handle~\cite{Mohan:2015doa}; for multi-TeV vectors at the LHC, however, its rate remains small (section~\ref{sec:vectorsearches}).

\subsection{Relation to existing models}\label{sec:existing}
\HVTP{} unifies models that have been used separately. The two-site description supplies the elementary--composite rotations~\cite{two}. The heavy-gluon study of ref.~\cite{gluon} considered the decays $G'\to Q\bar q+\bar Qq$, which give access to partners below the pair threshold; its original implementation also contained a bottom singlet, which is decoupled here. The charged-vector study of ref.~\cite{wp} introduced the channels $W'\to T\bar b,t\bar B$ and the custodian pairs. In \HVTP{} the two production mechanisms share the same partners, so that they can be tested against the same reconstructed final states.

Most experimental searches for $W'$ and $Z'$ bosons are interpreted in the sequential SM (SSM)~\cite{Altarelli:1989ff}, in which the new vectors couple to the SM fermions as the $W$ and the $Z$. Their couplings to the SM gauge bosons are either absent or, in the extended gauge model of ref.~\cite{Altarelli:1989ff}, arise from mixing and are suppressed by $m_W^2/m_{V'}^2$. The emission of longitudinal bosons compensates this suppression~\cite{hvt}, but since the couplings to the fermions are not suppressed the diboson and $Vh$ branching fractions remain at the per-cent level. This is the opposite of the pattern expected for the resonances of a strong sector, which couple weakly to the light fermions, with strength of order $g/k$, and strongly, with strength of order $gk$, to the longitudinal $W$ and $Z$ and to the Higgs, where $k\gg1$ is the ratio of the composite to the elementary coupling (section~\ref{sec:composite}), i.e.\ to the Goldstone bosons of the strong sector, so that their dominant SM decays are into $WZ$, $WW$ and $Vh$. The heavy vector triplet (HVT) framework~\cite{hvt} was introduced to describe both regimes with a few parameters. With the partners and the other vectors decoupled, the couplings of the weak triplet of \HVTP{} to the Higgs current and to the light fermions match the heavy vector triplet of ref.~\cite{hvt}, with $g_H=-g_2k_2$ for the longitudinal bosons and $g_F=g_2/k_2$ for the elementary fermions, at leading order in the electroweak mixing (section~\ref{sec:model}).\footnote{The third generation is different: with finite compositeness the left-handed coupling of the triplet to $tb$ is proportional to $k_2s_L^2-c_L^2/k_2$ rather than to $-1/k_2$, e.g.\ 0.68 instead of $-0.33$ for $k_2=3$ and $s_L=0.55$, and this non-universality does not disappear when the partners are made heavy at fixed $s_L$.} With $g_V=g_2k_2$ these values of $g_H$ and $g_F$ correspond to $c_H=-1$ and $c_F=1$, i.e.\ to the strongly coupled benchmark model B of ref.~\cite{hvt}, which describes composite resonances, rather than to model A, an extended gauge group in which $c_F\simeq1$ and the coupling to the Higgs current is suppressed, $c_H\simeq-g^2/g_V^2=-1/k_2^2$, as in the SSM. \HVTP{} extends this simplified description in the direction required by a natural composite Higgs, in which the top partners are light, in three respects: the partner channels open new decays and make the vectors broad, the vectors $B'$ and $G'$ add independent production amplitudes, and the non-universal third-generation couplings follow from the top mixings rather than from free parameters.

Simplified models of top partners organise their interactions with the SM and their direct production~\cite{hunter}. \HVTP{} adds the vector propagators and therefore the resonant production of partners. The decays of electroweak vectors into partners and their competition with the conventional searches were studied in refs.~\cite{hunt,liu,model}, which showed that light partners can substantially weaken the bounds on the vectors. \HVTP{} builds on these studies in four respects. First, electroweak and colour vectors are produced and decay with the same partners, so that $W'$, $Z'$, $B'$ and $G'$ signals can be tested on the same reconstructed final states. Second, the couplings are not expanded in $v/m_Q$ or in the mixings: the three heavy charge-$2/3$ states, the custodians and the transitions between partners are obtained from the exact diagonalisation, which matters when partners are close in mass (section~\ref{sec:nonstandard}). Third, the model is distributed as a validated public UFO with physical masses as inputs, which can be used directly in experimental analyses, also for top partners alone. Fourth, we confront it with the most recent searches at 13 and 13.6 TeV, including the width-dependent interpretations and the finite-width effects in resonant partner production.

\section{Phenomenology of the top partners}\label{sec:vlq}
\begin{figure}[htbp]\centering
\tikzset{ferm/.style={postaction={decorate},decoration={markings,mark=at position .52 with {\arrow{>}}}},
weak/.style={decorate,decoration={snake,amplitude=.75mm,segment length=3mm}},
gluon/.style={decorate,decoration={coil,amplitude=.9mm,segment length=1.5mm,aspect=.55}}}
\begin{tikzpicture}[x=.82cm,y=.82cm,font=\small]
\begin{scope}
\draw[ferm] (0,1) node[left]{$q$} -- (1.1,0);
\draw[ferm] (1.1,0) -- (0,-1) node[left]{$\bar q$};
\draw[gluon] (1.1,0) -- (3,0);\node at (2,.4){$g$};
\draw[ferm] (3,0) -- (4,1) node[right]{$Q$};
\draw[ferm] (4,-1) node[right]{$\bar Q$} -- (3,0);
\node at (2,-1.55){QCD pair production};
\end{scope}
\begin{scope}[xshift=6.9cm]
\draw[ferm] (0,1) node[left]{$u$} -- (1,1);
\draw[ferm] (1,1) -- (3,1.25) node[right]{$d$};
\draw[weak] (1,1) -- (1,-.55);\node at (.48,.21){$W^+$};
\draw[ferm] (0,-1) node[left]{$b$} -- (1,-.55);
\draw[ferm] (1,-.55) -- (2.9,-.55);\node at (2,-.18){$T$};
\draw[ferm] (2.9,-.55) -- (4,-.05) node[right]{$t$};
\draw[dashed] (2.9,-.55) -- (4,-1) node[right]{$h$};
\node at (2,-1.55){Single production};
\end{scope}
\begin{scope}[yshift=-3.4cm]
\draw[ferm] (0,1) node[left]{$q$} -- (1,0);
\draw[ferm] (1,0) -- (0,-1) node[left]{$\bar q$};
\draw[gluon] (1,0) -- (2.4,0);\node at (1.7,.4){$G'$};
\draw[ferm] (2.4,0) -- (3.55,.6);\node at (2.95,.68){$B$};
\draw[ferm] (3.55,.6) -- (4.65,1.05) node[right]{$t$};
\draw[weak] (3.55,.6) -- (4.65,.15) node[right]{$W^-$};
\draw[ferm] (3.7,-1) node[right]{$\bar b$} -- (2.4,0);
\node at (2,-1.55){Heavy--light decay};
\end{scope}
\begin{scope}[xshift=6.9cm,yshift=-3.4cm]
\draw[ferm] (0,1) node[left]{$u$} -- (1,0);
\draw[ferm] (1,0) -- (0,-1) node[left]{$\bar d$};
\draw[weak] (1,0) -- (2.4,0);\node at (1.7,.4){$W'^+$};
\draw[ferm] (2.4,0) -- (3.6,.65);\node at (2.95,.8){$X_{5/3}$};
\draw[ferm] (3.6,.65) -- (4.7,1.1) node[right]{$t$};
\draw[weak] (3.6,.65) -- (4.7,.15) node[right]{$W^+$};
\draw[ferm] (3.7,-1) node[right]{$\bar X_{2/3}$} -- (2.4,0);
\node at (2,-1.55){Resonant custodian pair};
\end{scope}
\end{tikzpicture}
\caption{Representative partonic diagrams. The continuum also includes gluon fusion. In single production the recoiling light quark gives a forward jet and the $ht$ system reconstructs the partner. In the octet cascade $m(Wt)$ and $m(Wt\bar b)$ reconstruct two masses. In the charged-vector cascade the top on the exotic leg supplies a second $W^+$, allowing a same-sign dilepton tag. Charge conjugates are implicit.}\label{fig:diagrams}
\end{figure}
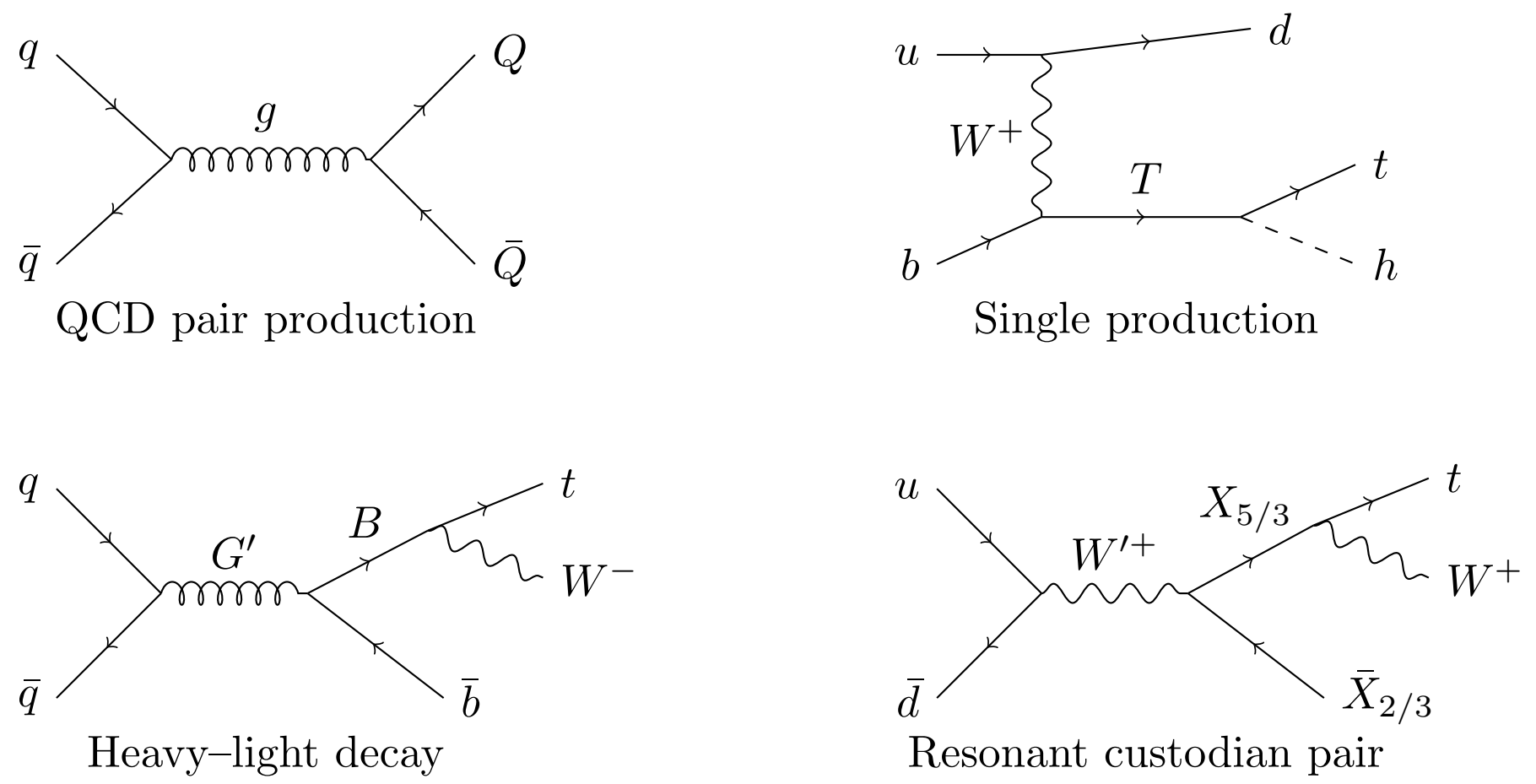

We first summarise the decays of the top partners, which define the final states of all searches, and then their production. The partners can be produced in pairs through QCD, singly through their electroweak transitions, or through the decays of the vector resonances. Figure~\ref{fig:diagrams} shows representative diagrams for the three mechanisms, which probe different combinations of masses and couplings.

\subsection{Decays}
For isolated partners much heavier than the top, the branching fractions derived in section~\ref{sec:lambda} from the equivalence theorem, such as $\Gamma(Wb):\Gamma(Zt):\Gamma(ht)\simeq2:1:1$ for the singlet, reproduce the benchmarks used in the experimental searches~\cite{aguilar09,aguilar13}. In \HVTP{} the three heavy charge-$2/3$ states mix with each other as well as with the top. A single mixing angle cannot then describe all vertices, and the branching fractions follow from the full diagonalisation of section~\ref{sec:model}, which generalises the mass-basis construction of ref.~\cite{aguilar13} to several multiplets. When a partner is heavier than another by more than a $W$ mass, decays such as $X_{5/3}\to W^+T_{\rm light}$ also open, as discussed in section~\ref{sec:nonstandard}.

\subsection{Pair production through QCD}
All partners are colour triplets, and their pair production through $gg,q\bar q\to Q\bar Q$ depends only on their mass. Since the gluon couples to the identity in flavour space, its coupling remains diagonal in the mass basis: QCD produces pairs of the same species even when two charge-$2/3$ states are strongly mixed. Pair production therefore provides a model-independent mass reach, but the signal seen by a search is the QCD rate times a sum of products of branching fractions and channel-dependent efficiencies. Limits derived for a degenerate doublet count the production of both members and must be applied to the actual spectrum of \HVTP{}.

The colour octet adds an $s$-channel amplitude in $q\bar q$ annihilation, which can also produce different partners through its off-diagonal couplings and modifies the kinematics of the pairs. In MadGraph, \texttt{generate p p > x53 x53\string~ QCD=2 QED=0 HV=0} gives the QCD continuum, whereas the default \texttt{generate p p > x53 x53\string~} also includes the octet exchange (section~\ref{sec:usage}). Same-sign leptons from $X_{5/3}\to W^+t\to W^+W^+b$ tag a particularly clean subset of pair events~\cite{wp}.

\subsection{Electroweak single production}
Single production gives access to both the mass of a partner and its mixing with the top sector~\cite{Vignaroli:2012nf,Vignaroli:2012sf}. It proceeds through the transition couplings $\lambda$ of section~\ref{sec:lambda}, to which the rate is proportional: at fixed mass, the rate reconstructed in a decay mode $f$ scales as $\lambda^2\mathcal B(T\to f)$. Measuring several single-production modes at the same mass thus constrains both the branching fractions and the strength of the heavy--light current. For a singlet the latter is $Y_*s_Lc_R$, while the top mass depends on $Y_*s_Ls_R$, so that the two measurements together probe the right-handed mixing. In $qb\to q'T$, through $t$-channel $W$ exchange, the partonic threshold is close to $m_T$ rather than $2m_T$, so that single production probes larger parton luminosities and overtakes pair production at high mass, despite its electroweak coupling, at a crossover that depends on $\lambda$.

In $qb\to q'T$ the $W$ is exchanged in the $t$ channel, so that the light quark that emits it recoils with a transverse momentum of order $m_W$ and is produced at large rapidity. This forward jet identifies the production mechanism and, since it is rare in the dominant $t\bar t$ and $W/Z+$jets backgrounds, it is also a powerful handle to separate the signal from the background, as emphasised in refs.~\cite{Vignaroli:2012nf,Vignaroli:2012sf}. In the $T\to ht$ channel the Higgs candidate gives a peak in $m_{b\bar b}$ and the top--Higgs system a second peak in $m_{ht}$, which allows a Higgs-tagged search with a direct measurement of the partner mass~\cite{Vignaroli:2012nf}. When the partner is produced in the decay of a heavy vector, as in $W'\to T\bar b$, the invariant mass of the partner and the accompanying quark supplies a second resonance, so that forward-jet and heavy-parent categories separate the light and heavy charged currents.

\HVTP{} can also be used as a model for the top partners alone, with the vectors decoupled or removed with \verb|HV=0| (section~\ref{sec:usage}). This is particularly useful for single production. In most interpretations the single-production coupling and the branching fractions are treated as independent parameters, whereas in \HVTP{} both follow from the same diagonalisation of the mass matrices, without expansions in $v/m_Q$ or in the mixings. The rate in each final state, proportional to $\lambda^2\mathcal B(T\to f)$, is then predicted consistently for every partner, including the mixing among the three charge-$2/3$ states, the custodians and the decays between partners when they are open. Single-production searches can therefore be interpreted directly in terms of the masses and degrees of compositeness of a concrete model, and limits from different production modes and decay channels can be combined at the same parameter point.

\section{Current limits on the top partners}\label{sec:vlqlimits}
In this section we consider the limits from QCD pair production and from single production through the electroweak transitions of the partners, which decay into SM final states. They apply directly when the vector resonances are heavy enough for their contribution to the partner rate to be negligible; the resonant production of partners by lighter vectors is discussed in section~\ref{sec:broad}. The comparison with the experimental results then requires only the masses and branching fractions of the \HVTP{} mass eigenstates. These generally differ from the singlet and doublet benchmarks used by the experiments, because the three heavy charge-$2/3$ states mix. As an illustration, consider a spectrum in which the singlet is isolated: with $(M_1,M_2,M_3)=(10,9,8.5)$ TeV, $m_Q=5$ TeV and the other inputs of the test point P0 (table~\ref{tab:points}), the lightest partner has a mass of 1.711 TeV and is mostly a singlet, while the next partner lies at 5 TeV. Its width is 54 GeV and its branching fractions,
\begin{equation}
 \big(\mathcal B(Wb),\mathcal B(Zt),\mathcal B(ht)\big)=(0.482,0.232,0.286),
\end{equation}
are close to the singlet pattern $(1/2,1/4,1/4)$. The excess of $ht$ over $Zt$ comes from the residual mixing with the doublets, which corrects the Higgs coupling of the singlet at relative order $Y_*v/(\sqrt2m_Q)\simeq0.1$ and therefore decreases only as $1/m_Q$; the Goldstone-equivalence pattern is recovered for $m_Q\to\infty$. In less isolated configurations the deviations are much larger, as shown in section~\ref{sec:nonstandard}.

\subsection{Pair production}
Table~\ref{tab:pair} lists the strongest observed limits from QCD pair production for the decay hypotheses relevant to the custodial multiplet.
\begin{table}[htbp]\centering\small
\begin{tabular}{@{}llrl@{}}
\toprule Partner & Decay hypothesis & $m_{95}$ [TeV] & Source\\\midrule
$T, X_{2/3},\widetilde T$ & Any mixture of $Wb,Zt,ht$ & 1.48 & CMS~\cite{exp220907327}\\
 & Singlet, $(Wb,Zt,ht)=(\tfrac12,\tfrac14,\tfrac14)$ & 1.40$^*$ & ATLAS~\cite{exp260528538}\\
 & Doublet, $\mathcal B(Zt)=\mathcal B(ht)=\tfrac12$ & 1.56 & ATLAS~\cite{exp260528538}\\
 & $\mathcal B(ht)=1$ & 1.66 & ATLAS~\cite{exp260528538}\\
 & $\mathcal B(Zt)=1$ & 1.60 & ATLAS~\cite{exp221015413}\\
 & $\mathcal B(Wb)=1$ & 1.70 & ATLAS~\cite{exp240117165}\\
$B$ & $\mathcal B(Wt)=1$ & 1.56 & CMS~\cite{exp220907327}\\
$X_{5/3}$ & $\mathcal B(Wt)=1$ & 1.63 & CMS~\cite{exp261114335}\\
$(T,B)$ degenerate & Doublet & 1.59 & ATLAS~\cite{exp221205263}\\
\bottomrule
\end{tabular}
\caption{Observed 95\% CL lower mass limits from QCD pair production at 13 TeV. The charge-$2/3$ labels denote the heavy eigenstates identified by their multiplet content. The first row is the weakest limit over the full $(Wb,Zt,ht)$ triangle and therefore applies to any charge-$2/3$ eigenstate whose two-body decays are exhausted by these modes. The degenerate-doublet row counts the production rates of both members. $^*$Result of the dedicated ATLAS analysis; the CMS limit of the first row, 1.48 TeV, also applies to the singlet and is stronger.}\label{tab:pair}
\end{table}
The most robust constraint for \HVTP{} is the first row, which holds for any charge-$2/3$ partner whose decays are exhausted by $Wb$, $Zt$ and $ht$, independently of the branching fractions: partners lighter than about 1.5 TeV are excluded. The exclusive rows show how strongly the reach of each analysis depends on the decay hypothesis, from 1.40 TeV in the ATLAS search for a singlet, which is weaker than the CMS limit of the first row, to 1.70 TeV for a partner that decays only into $Wb$; only a quarter of the pairs of a singlet decay into the $WbWb$ topology targeted by that search~\cite{exp240117165}. Two effects must be kept in mind for a realistic spectrum. Nearly degenerate partners contribute separate QCD rates, so that the limit on a degenerate doublet counts the yields of both members~\cite{exp221205263}. When transitions between partners are open, as $X_{5/3}\to W^+T_{\rm light}$, the exclusive limits overestimate the signal and the cascade must be included explicitly (section~\ref{sec:nonstandard}).

\subsection{Single production}
\begin{figure}[htbp]\centering
\includegraphics[width=.72\linewidth]{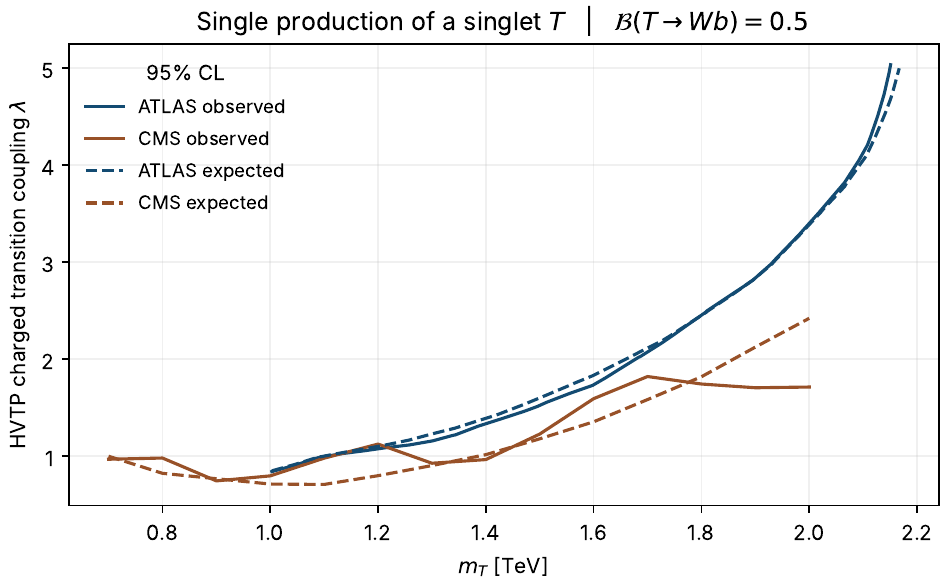}
\caption{Reinterpretation: observed (solid) and median expected (dashed) 95\% CL upper limits on the \HVTP{} transition coupling $\lambda_{\widetilde T}$ for an isolated singlet-like partner. The ATLAS $ht,Zt$ combination uses 139 fb$^{-1}$ and the CMS $Wb$ analysis uses 138 fb$^{-1}$ at 13 TeV~\cite{exp240808789,cmsvlq}. Each curve follows the published coupling limit as a function of mass. The CMS curves stop at 2 TeV, the stated upper mass for the narrow-width interpretation of the singlet. ATLAS retains its finite-width domain $\Gamma_T/m_T<0.5$.}\label{fig:single}
\end{figure}

Single production probes the transition couplings of table~\ref{tab:partnerlambda} as a function of the mass. For a singlet-like partner, $Wb$ fusion probes $\lambda_{\widetilde T}$ even when the partner is reconstructed in $Zt$ or $ht$. Figure~\ref{fig:single} translates the ATLAS combination of the $Zt$ and $ht$ channels and the CMS $Wb$ search into the \HVTP{} coupling, assuming the singlet branching fractions; appendix~\ref{app:matching} gives the conversion between the conventions.
The two searches are complementary. The CMS charged-current search is the stronger one between 1.2 and 2.0 TeV and limits $\lambda_{\widetilde T}\lesssim1$ near 1.4 TeV~\cite{cmsvlq}, while above 2 TeV only the ATLAS neutral-current combination provides a result, which remains valid for widths up to half of the mass and excludes $m_T<2.1$ TeV for a coupling $\kappa=0.5$ in the experimental convention of appendix~\ref{app:matching}~\cite{exp240808789}. The ATLAS search in $W(\ell\nu)b$, which includes the interference with SM single-top production, constrains the singlet coupling in the range 1.15--2.30 TeV~\cite{exp250615515}.

\begin{figure}[htbp]\centering
\includegraphics[width=\linewidth]{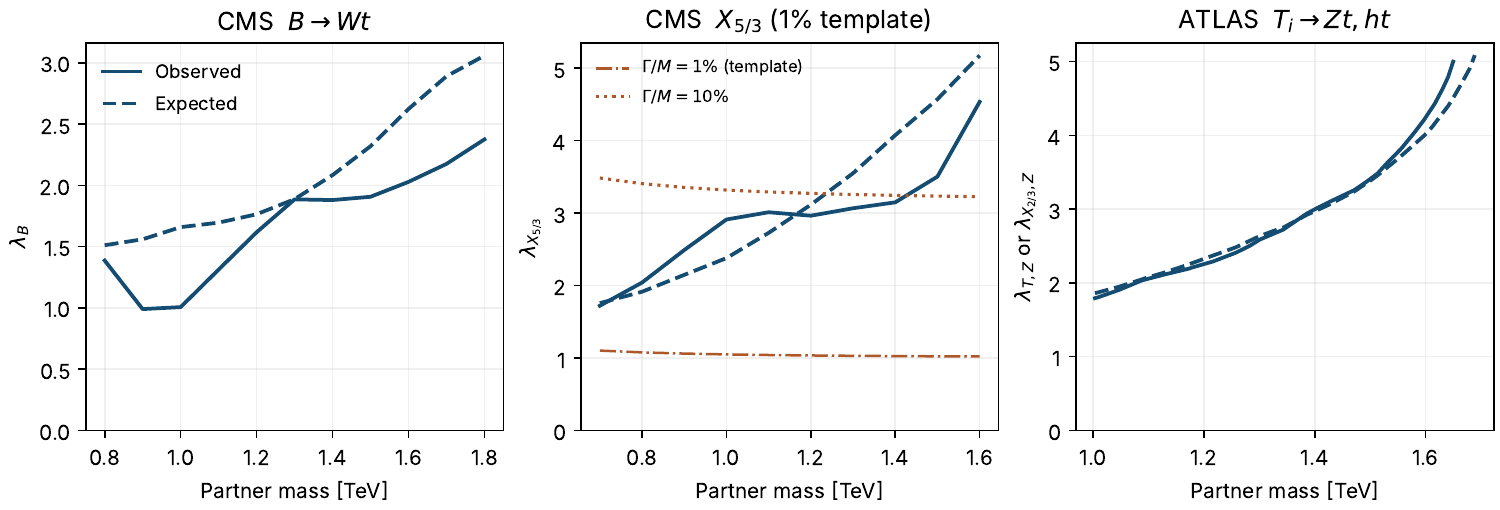}
\caption{Reinterpretation: single-production sensitivity at 13 TeV in the \HVTP{} conventions of table~\ref{tab:partnerlambda}. Solid and dashed lines denote observed and median expected 95\% CL results. Left: CMS doublet $B$, with $\mathcal B(Wt)=1$~\cite{expB2G24013}. Centre: rate interpretation of the CMS right-handed $X_{5/3}$ signal, with $\mathcal B(Wt)=1$ and a fixed 1\% width template~\cite{exp180908597,partnerRates}. The dash-dotted and dotted lines mark physical widths of 1\%, the width of the template, and 10\%. Right: ATLAS neutral-current combination for one isolated doublet-like eigenstate, with equal $Zt$ and $ht$ fractions~\cite{exp240808789}.}\label{fig:partners}
\end{figure}

The custodial doublets couple mainly to $t_R$, and their charged currents produce $B\bar tj$ or $X_{5/3}\bar tj$. The charge of the $W$ from the partner decay distinguishes the two signals: $X_{5/3}\to W^+t$ gives two same-sign $W$ bosons on the resonance side, $B\to W^-t$ opposite-sign ones~\cite{expB2G24013,exp180908597}. Figure~\ref{fig:partners} shows the corresponding limits together with the neutral-current production of a doublet-like partner.
At 1.4 TeV the CMS doublet search gives $\lambda_B<1.88$. For $X_{5/3}$ the observed rate at the same mass corresponds to $\lambda_{X_{5/3}}\simeq3.15$, but such a coupling implies a width of about 9\% of the mass, well beyond the 1\% template used in the search; a reliable limit in this region requires a signal simulated at the physical width. This illustrates a general point that recurs for the vector resonances: in composite models large couplings imply large widths, and the experimental results must be interpreted with signal shapes that account for them. Neutral-current production, $pp\to T\bar tj$, remains sensitive even when the $Wb$ coupling vanishes; the ATLAS combination reaches 1.4 TeV for $\Gamma/m=0.2$ and 1.6 TeV for $\Gamma/m=0.5$ in the doublet benchmark~\cite{exp240808789}, and applies to either $T$ or $X_{2/3}$ when one of them dominates the signal. At small transition couplings, pair production of table~\ref{tab:pair} provides the complementary reach.

\section{Current limits on the vector resonances}\label{sec:vectorsearches}
We now turn to the vector resonances. In this section we translate the most recent LHC searches at 13 and 13.6 TeV into the \HVTP{} parameter space, first for decoupled partners, where the vectors decay only into SM particles, and then for a finite partner spectrum, where part of their width goes into partners. The searches used and their inputs are listed in appendix~\ref{sec:inventory}. We concentrate on masses above 2 TeV: for the couplings and widths considered here lighter vectors are excluded by the same searches, which extend down to a few hundred GeV, and in minimal models they are also disfavoured by the $S$ parameter (section~\ref{sec:intro}).

In this and the following sections we distinguish three kinds of results, and the captions of the figures state which one is shown. Reinterpretations are published experimental limits translated into the \HVTP{} parameter space under stated assumptions, such as the transfer of the efficiencies of a reference model or the use of narrow-width templates; estimates rely on simplified efficiencies or on approximations beyond those of the experimental analyses; projections extrapolate to the HL-LHC. In all figures solid and dashed lines denote observed and median expected results, unless stated otherwise.

\subsection{The electroweak triplet}
The weak triplet is produced from light quarks with a coupling $g_2/k_2$, which also controls its decays into leptons, whereas its decays into $WZ$, $WW$ and $Wh$ are controlled by the coupling to the Higgs doublet, $g_2k_2$. The ratio of the bosonic to the leptonic widths therefore grows as $k_2^4$: leptonic searches dominate at weak coupling and diboson searches at strong coupling, and the two classes of final states cover complementary regions of the mass--coupling plane even before any partner channel opens.

With the partners decoupled, the couplings of the weak triplet of \HVTP{} to the Higgs current and to the light fermions are those of a heavy vector triplet with $g_H=-g_2k_2$ and $g_F=g_2/k_2$, up to the non-universal third-generation couplings (sections~\ref{sec:model} and~\ref{sec:existing}), so that \HVTP{} describes the line $g_Hg_F=-g_2^2$ in the HVT parameter plane; the HVT benchmark B lies on it at $k_2\simeq4.5$. The most recent combination, CMS B2G-25-003 with 138 fb$^{-1}$ at 13 TeV~\cite{cms}, publishes separate limits for the lepton, boson and quark categories, which we rescale to the \HVTP{} production rate and branching fractions at each point (appendix~\ref{app:vectorrates}). We add the CMS $\ell\nu$ and $\ell^+\ell^-$ searches at 13.6 TeV with 62 and 283 fb$^{-1}$~\cite{expEXO24021,expEXO25021}, and the separate ATLAS $WZ,WW$ and $Wh,Zh$ searches at 13 TeV with 139 fb$^{-1}$~\cite{atlasVV,atlasVH}.

\begin{figure}[htbp]\centering
\includegraphics[width=\linewidth]{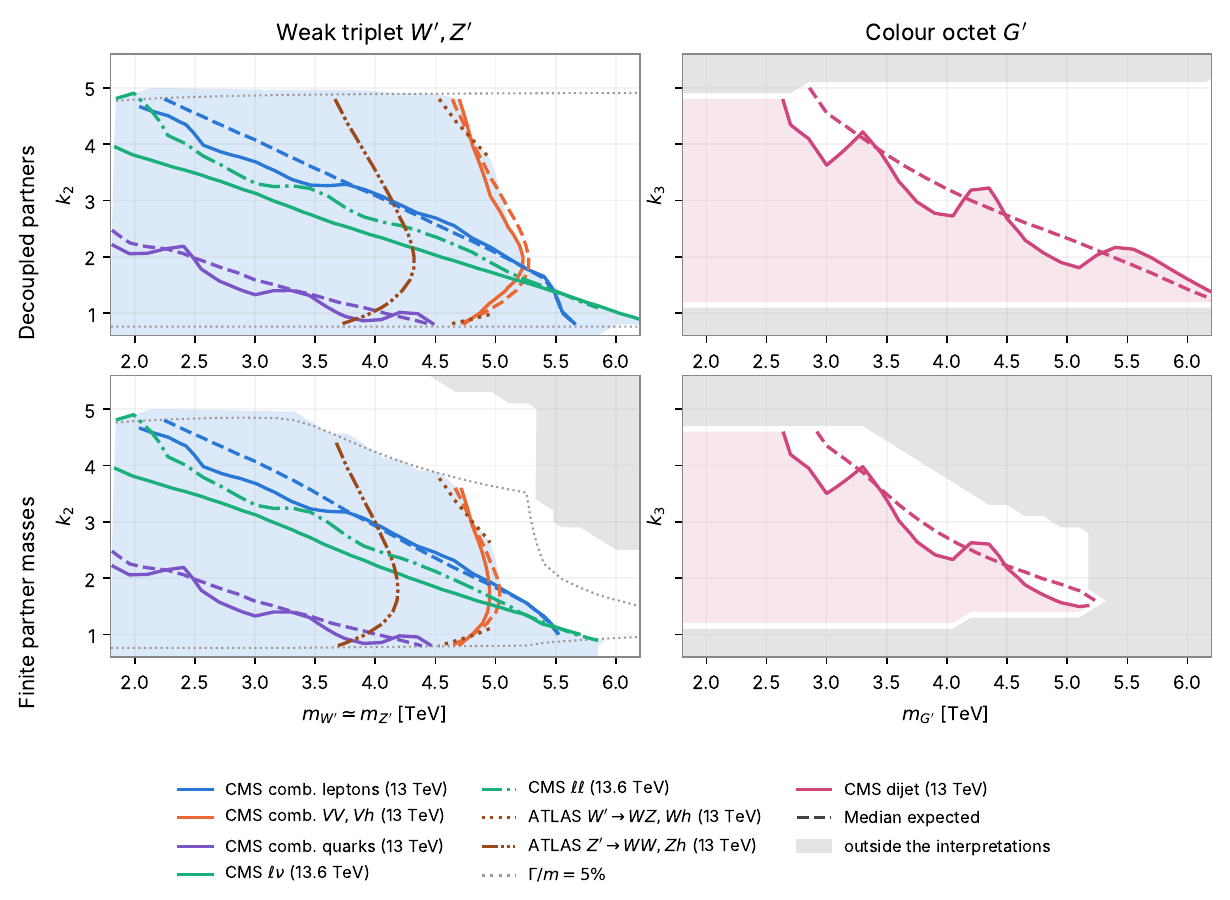}
\caption{Reinterpretation: 95\% CL exclusions on the weak triplet (left) and the colour octet (right) in the \HVTP{} mass--coupling plane, for decoupled partners (upper row) and for the finite spectrum $(m_Q,m_S)=(2.70,3.40)$ TeV (lower row). Solid lines are observed limits and the shaded regions are excluded by at least one of them; dashed lines give the median expected limits of the CMS combination categories and of the dijet search. Triplet: CMS combination categories at 13 TeV~\cite{cms} and CMS $\ell\nu$, $\ell^+\ell^-$ at 13.6 TeV~\cite{expEXO24021,expEXO25021}. Octet: CMS dijet search at 13 TeV with the $q\bar q$ template~\cite{exp191103947}. Dotted grey lines mark $\Gamma/m=5\%$: narrow-template interpretations are restricted to $\Gamma/m\le5\%$, while the width-dependent CMS dilepton limits are used up to 10\%. Grey regions ($\Gamma/m>10\%$ for the triplet, $>5\%$ for the octet) are outside all the interpretations shown; see section~\ref{sec:broad}. Brown lines: ATLAS $VV,Vh$ searches at 13 TeV~\cite{atlasVV,atlasVH}, using for each point the channel with the best expected sensitivity.}\label{fig:vectorlimits}
\end{figure}

Figure~\ref{fig:vectorlimits} shows the result. With decoupled partners our reinterpretation excludes the triplet below about 6.0, 5.1 and 5.0 TeV for $k_2=1$, 2 and 3. These values rely on the rescaling of appendix~\ref{app:vectorrates}: at the highest masses, where the $\ell\nu$ search dominates, the signal efficiency depends on the off-shell component of the resonance, which our rescaling does not model, so that the bounds above about 5 TeV should be regarded as estimates. As expected, the leptonic searches dominate at small $k_2$, with the 13.6 TeV $\ell\nu$ search giving the strongest bound, while the diboson searches dominate at large $k_2$ and keep the bound at about 4.7 TeV up to $k_2\simeq4.6$. The ATLAS diboson searches give a comparable bound for the charged vector and a weaker one, 3.9--4.3 TeV, for the neutral vector, and the dijet channel is weaker throughout. When the partners are lighter, here $(m_Q,m_S)=(2.7,3.4)$ TeV, part of the width goes into heavy--light and heavy-pair channels and the conventional signals are diluted: the reach decreases to 5.7, 4.8 and 4.7 TeV for $k_2=1$, 2 and 3, and the bosonic bound to 3.5 TeV at $k_2=4.6$. In both cases a triplet lighter than about 3.5 TeV is excluded for $k_2\lesssim4.6$. The exclusion becomes increasingly uncertain where the resonance is broad, as indicated by the grey regions of the figure; we return to this point in section~\ref{sec:broad}.

Electroweak vectors can also be produced by vector-boson fusion, $pp\to V'jj$, through their coupling to longitudinal $W$ and $Z$ bosons. This rate grows as $(g_2k_2)^2$, whereas Drell--Yan production falls as $(g_2/k_2)^2$, so that vector-boson fusion probes precisely the strongly coupled region in which the vectors are broad and the Drell--Yan channels are suppressed, and the forward tagging jets provide an additional handle~\cite{Mohan:2015doa}. ATLAS and CMS have searched for this production mode in the $WZ$, $WW$ and $Vh$ final states~\cite{ATLAS:2022jsi,atlasVV,atlas,cms}, reaching about 2 TeV for a vector produced only by fusion~\cite{cms}. To assess its role at the masses relevant here, we have computed both production modes with MadGraph at LO for $m_{W'}=4$ TeV and the default partner spectrum, requiring $m_{jj}>500$ GeV for the fusion process. The ratio $\sigma_{\rm VBF}/\sigma_{\rm DY}$ grows from $3\times10^{-4}$ at $k_2=2$ to $1\times10^{-3}$ at $k_2=4$ and $5\times10^{-3}$ at $k_2=6$, where the fusion cross section is about 2 ab at 13 TeV. These numbers refer to the production of the vector itself, computed with an on-shell final state; at the largest couplings the vector is very broad ($\Gamma/m\simeq0.4$ at $k_2=6$), and a quantitative comparison would require the full processes with decays and the selections of the searches. They nevertheless indicate that at the LHC vector-boson fusion is not competitive for multi-TeV vectors, even at strong coupling; it becomes a powerful complementary probe of large $k_2$ at higher-energy colliders, where the luminosities of the electroweak bosons in the proton grow and the Drell--Yan rate of strongly coupled vectors is small~\cite{Mohan:2015doa}.

\subsection{The colour octet}
The octet couples to elementary quarks with $-g_s/k_3$ and to the top quark with a strength that grows with its compositeness. At weak coupling it therefore decays mostly into light-quark pairs and is constrained by dijet searches, while at strong coupling the $t\bar t$ channel and, when open, the partner channels take over: at $m_{G'}=3$ TeV and with the finite spectrum, $(\mathcal B(jj),\mathcal B(t\bar t))$ changes from $(0.97,0.02)$ at $k_3=1.6$ to $(0.49,0.48)$ at $k_3=3$. The dijet search excludes $1.8<m_{G'}<4.65$ TeV at $k_3=2.2$ and $1.8<m_{G'}<4.35$ TeV at $k_3=3$ for decoupled partners, and up to 4.35 and 3.6 TeV with the finite spectrum. At larger $k_3$ the $t\bar t$ searches become competitive, but their interpretation requires the octet line shape and its interference with QCD~\cite{exp251217856,expB2G23010}.

\subsection{The hypercharge vector}\label{sec:bprime}
\begin{figure}[htbp]\centering
\includegraphics[width=\linewidth]{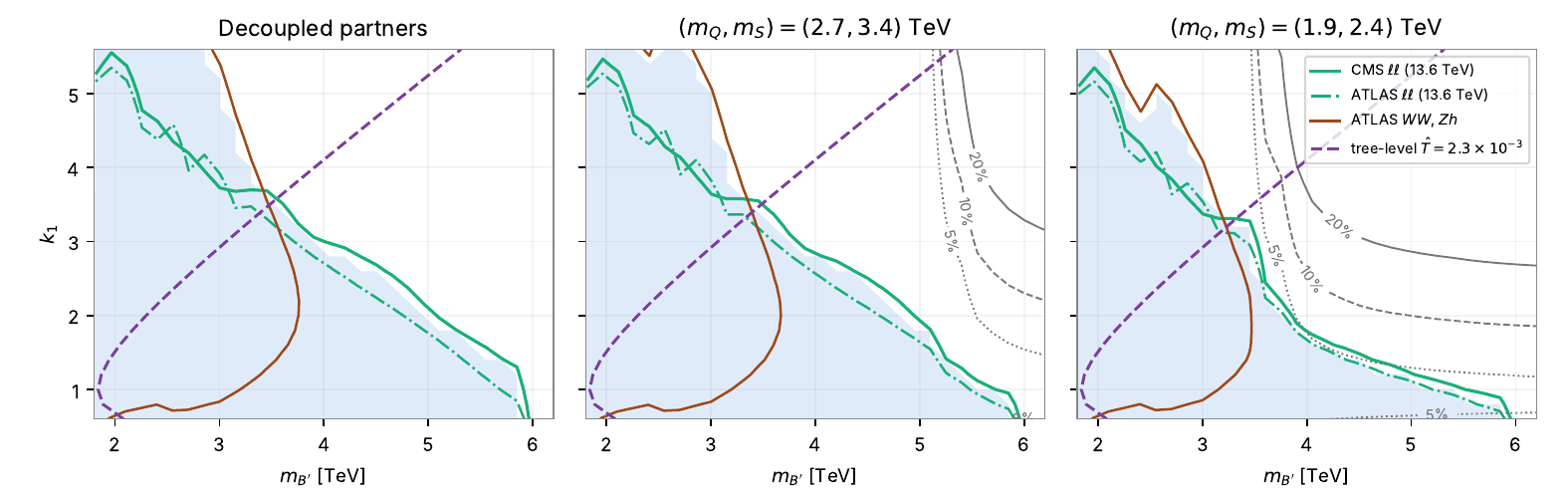}
\caption{Reinterpretation: 95\% CL exclusions on the hypercharge vector $B'$ for decoupled partners, for $(m_Q,m_S)=(2.7,3.4)$ TeV and for the light spectrum $(1.9,2.4)$ TeV, with $W'$, $Z'$ and $G'$ decoupled. Green: CMS (solid) and ATLAS (dash-dotted) dilepton searches at 13.6 TeV; brown: ATLAS $WW$ and $Zh$ searches. The shaded region is excluded. Grey lines are contours of $\Gamma_{B'}/m_{B'}$. Dashed purple: tree-level $\hat T=2.3\times10^{-3}$, excluded to the left (indicative, see text).}\label{fig:bprime}
\end{figure}

The neutral vector $B'$ associated with $U(1)_Y$ couples to the elementary fermions with strength of order $g'Y/k_1$ and to the composite fields with strength $g'Yk_1$. Since $g'$ is smaller than $g$ and $B'$ has no charged partner, it is produced less abundantly than the triplet and is constrained mainly by dilepton searches. We scan $(m_{B'},k_1)$ with the other vectors decoupled and apply the CMS and ATLAS dilepton searches at 13.6 TeV~\cite{expEXO25021,exp260728334} with the width-dependent templates as for the $Z'$ (appendix~\ref{app:vectorrates}), together with the ATLAS $WW$ and $Zh$ searches~\cite{atlasVV,atlasVH}. Figure~\ref{fig:bprime} shows that the dilepton bound dominates: for decoupled partners $B'$ is excluded below about 5.9, 5.0, 3.9 and 2.7 TeV at $k_1=1$, 2, 3 and 4, and the diboson searches exclude masses up to 3.0--3.75 TeV, depending on $k_1$. Because of its weaker coupling to light quarks, the $B'$ bound lies below the triplet one for $k_1=k_2\gtrsim2$. Electroweak precision data give a complementary constraint. Since the composite site gauges only the hypercharge, $B'$ does not belong to a complete multiplet of the custodial symmetry, and its coupling $g'k_1$ to the Higgs current generates a tree-level contribution to $\hat T$, which approaches $g'^2k_1^2v^2/(4M_1^2)$ at large $k_1$; the exact tree-level value, computed from the mass eigenstates, is $\hat T\simeq1.1\times10^{-3}$ at the default point of section~\ref{sec:points}. Requiring $\hat T<2.3\times10^{-3}$ ($T\lesssim0.3$) gives the dashed purple line of figure~\ref{fig:bprime}, which is stronger than the dilepton bound for $k_1\gtrsim3.5$. This line is indicative, since the tree-level contribution is a consequence of retaining only the hypercharge component of the composite vectors: a complete $SU(2)_R$ triplet degenerate with $B'$ would cancel it (section~\ref{sec:model}).

The partners contribute to $\hat T$ at one loop. We have computed the fermionic contribution with the general expressions of ref.~\cite{Lavoura:1992np}, using the SM $SU(2)_L\times U(1)_Y$ currents rotated with the exact fermion mixing matrices of section~\ref{sec:model}; the mixing of $W$ and $Z$ with the heavy vectors is neglected in this loop. The calculation reproduces the closed form for a single singlet partner. Relative to the SM top--bottom loop the contribution is negative, $\Delta\hat T\simeq-1.7\times10^{-3}$ both at the default point and at P0, and decreases roughly as $1/m_Q^2$, to $-0.6\times10^{-3}$ when all partner masses are doubled; for larger left mixing, $s_L=0.85$, it reaches $-2.5\times10^{-3}$. It is comparable in size to the tree-level contribution of $B'$ and of opposite sign, so that the sums are about $-0.6\times10^{-3}$ at the default point and $-0.9\times10^{-3}$ at P0, where the tree-level term is $0.85\times10^{-3}$. These numbers show that the two effects can partly compensate; a statement on the compatibility with the data requires a global fit, including the $S$ parameter with its correlation with $T$, the non-universal effects discussed below and the contact interactions of section~\ref{sec:dilepton}.

The $Zb_L\bar b_L$ coupling deserves a separate comment. The left--right parity of section~\ref{sec:composite} ensures that the mixing of $b_L$ with $B$, which has the same $T^3_L=T^3_R$, does not change its coupling to the SM currents. In \HVTP{}, however, the physical $Z$ contains a small admixture of the heavy vectors, which couple with different strengths to the elementary and composite components of $b_L$, and the parity protection would require the complete custodial vector sector, which the model truncates (section~\ref{sec:model}). The result is a non-universal tree-level shift of the $b_L$ coupling with respect to that of $d_L$, which has the same quantum numbers but is elementary: $\delta g_{b_L}\simeq+5.5\times10^{-4}$ in units of $g/c_W$ at the default point and $+1.4\times10^{-3}$ at P0, where $W'$ is at 2.7 TeV; it decreases as $1/M^2$, to $1.5\times10^{-4}$ for vectors at 8 TeV. The one-loop corrections from the partners are suppressed by $y_t^2/(16\pi^2)$ and by $v^2/m_Q^2$, up to a logarithm of the partner mass, and are of order a few $\times10^{-4}$ for the partner masses considered here~\cite{Grojean:2013qca}. Both effects are of the order of the precision of the LEP and SLD measurements, about $10^{-3}$: the $Zb\bar b$ vertex is therefore a relevant constraint for light vectors combined with a sizeable $s_L$, and should be included in the global fit mentioned above.

\subsection{Searches for a vector decaying into a partner}
A heavy--light decay produces a second resonance inside the vector decay, which is the most direct test of the vector--partner coupling. The most recent search of this type is the CMS search for $W'\to Tb,tB$ in the all-hadronic $thb$ and $tZb$ final states with 138 fb$^{-1}$ at 13 TeV~\cite{expB2G20002}. It provides upper limits on $\sigma(pp\to W')\,\mathcal B$ for three partner masses, $m_{\rm VLQ}=m_{W'}/2$, $2m_{W'}/3$ and $3m_{W'}/4$. We compute the corresponding \HVTP{} rate along the same three slices, choosing $m_Q$ at each point so that the doublet-like charge-$2/3$ eigenstate has the required mass, with $m_S=1.3\,m_Q$. In \HVTP{} the bidoublet $B$ decays almost exclusively to $Wt$ or to a lighter partner, so the signal comes from $W'\to T\bar b$ with $T\to ht,Zt$. Partner-to-partner decays of the doublet $T$ reduce its fraction into the target final states to $\mathcal B(T\to ht)+\mathcal B(T\to Zt)\simeq0.43$, and the total signal fraction of the $W'$ is about 0.10 for $k_2=3$ and $m_T=2m_{W'}/3$, compared with up to 0.5 in the benchmark adopted by CMS. 

\begin{figure}[htbp]\centering
\includegraphics[width=\linewidth]{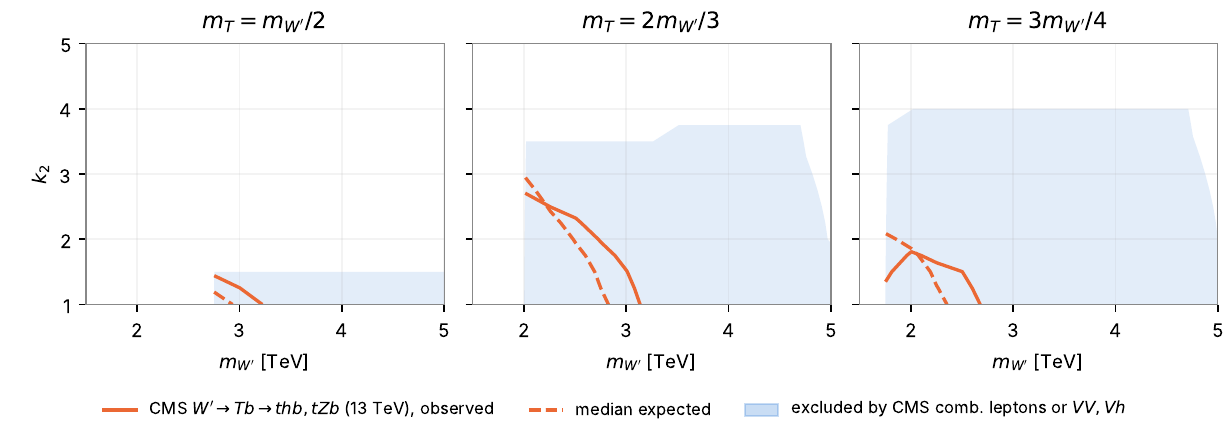}
\caption{Reinterpretation of the CMS $W'\to Tb\to thb,tZb$ search at 13 TeV~\cite{expB2G20002} for the three partner-mass hypotheses. The region to the left of the orange curves is excluded. The blue shading shows the region excluded at the same points, with the same partner spectrum, by the leptonic and bosonic categories of the CMS combination (figure~\ref{fig:vectorlimits}), applied for $\Gamma/m\le5\%$. Points with partners below about 1 TeV, where the top mass cannot be reproduced with $m_S=1.3\,m_Q$, are not shown; they are in any case excluded by pair production (table~\ref{tab:pair}).}\label{fig:cascade}
\end{figure}

Figure~\ref{fig:cascade} shows that the cascade search excludes $m_{W'}\lesssim3.1$ TeV at $k_2=1$ and $m_{W'}\lesssim2.7$ TeV at $k_2=2$ for $m_T=2m_{W'}/3$, and no region for $k_2\gtrsim3$. At every point excluded by the cascade search, the leptonic and bosonic categories of the CMS combination, evaluated for the same spectrum and widths, also exclude the vector. The reason is structural: in \HVTP{} the same elementary coupling $g_2/k_2$ controls the Drell--Yan rate and the leptonic width, while the heavy--light and diboson widths both grow with $k_2$, so that whenever the cascade is large the diboson signal is also large. For the partner-mass slices covered by the search, the cascade channel therefore does not provide the leading constraint with current data, although it remains a direct test of the heavy--light current once a resonance is observed. The situation changes when the partners are light enough for the vectors to decay into partner pairs, as we discuss in section~\ref{sec:broad}. The colour-octet cascade $G'\to T\bar t$ has been searched for only with 35.9 fb$^{-1}$ and excluded $1.5<m_{G'}<2.4$ TeV for the partner masses tested~\cite{exp181206489}; above 1.8 TeV this interval is already excluded in \HVTP{} by the dijet search for $1.2\lesssim k_3\lesssim4.8$.

\subsection{Indirect constraints from dilepton tails}\label{sec:dilepton}
Vectors heavier than the reach of the resonance searches still affect the high-mass tails of the Drell--Yan spectra through their virtual exchange. At energies well below their mass, the exchange of the weak triplet and of $B'$ between elementary currents generates the universal operators parameterised by $\mathcal W$ and $\mathcal Y$, with $\mathcal W=m_W^2/(k_2M_2)^2$ and $\mathcal Y=m_W^2/(k_1M_1)^2$ at leading order in $1/M_i^2$ and neglecting electroweak mixing. The CMS analysis of the dilepton tails~\cite{expEXO25001} gives $\mathcal W<1.6\times10^{-4}$ and $\mathcal Y<3.5\times10^{-4}$ when one operator is fitted at a time, which translates into
\begin{equation}
 k_2 M_2 \gtrsim 6.3~{\rm TeV}\quad(\mathcal Y=0),\qquad
 k_1 M_1 \gtrsim 4.3~{\rm TeV}\quad(\mathcal W=0).
\end{equation}
These bounds constrain the product of mass and coupling rather than the mass alone. At $k_2\simeq1$ the triplet bound, $M_2\gtrsim6.3$ TeV, is comparable to the direct limit, whereas at larger $k_i$, where the coupling to the light fermions is suppressed, the tails are much less constraining than the resonance searches. They rely on the validity of the effective description, i.e.\ on partonic energies well below the vector mass, and on the absence of significant non-universal contributions from the third generation. When both vectors are at finite mass, as in \HVTP{}, the two one-dimensional bounds should be replaced by the two-dimensional contour in the $(\mathcal W,\mathcal Y)$ plane.

\section{Broad resonances and partner channels}\label{sec:broad}
The limits of section~\ref{sec:vectorsearches} rest on the assumption that the vector resonances are narrow and decay mainly into SM particles. In composite models this assumption fails precisely where the theory is most interesting: when the top partners are light, as naturalness suggests, the vectors decay into them and become broad. In this section we study this region, ask what the current data still constrain there, and identify the signatures that carry the signal.

\subsection{When are the vectors broad?}
A limit obtained with a narrow signal template can be applied to a broader resonance as long as its width stays below the experimental mass resolution, typically a few per cent for dileptons and 5--10\% for dijets and boosted bosons. Beyond this point the reconstructed peak is diluted, part of the signal moves to lower masses, where it is enhanced by the parton luminosities, and the interference with SM production becomes relevant. Several searches now publish limits for explicit width hypotheses, for instance CMS dileptons up to $\Gamma/m=10\%$~\cite{expEXO25021}, ATLAS dileptons~\cite{exp260728334}, CMS $W'\to tb$ up to 30\%~\cite{exp231019893} and ATLAS $t\bar t$~\cite{exp251217856}. In figure~\ref{fig:vectorlimits} we used the width-dependent dilepton limits up to 10\% and the narrow templates only up to 5\%; the grey regions mark where the published results cannot be applied without a dedicated signal prediction, not where the data lose sensitivity.

With decoupled partners \HVTP{} produces broad vectors only at the edges of the parameter space, at very large $k_2$ through the diboson widths, or at small and very large $k_3$ for the octet. The situation changes when the vectors can decay into two partners. Since the partner widths grow as $k^2$ and are larger than the diboson widths by an order of magnitude (section~\ref{sec:currents}), above the pair threshold the partner channels dominate and the width grows rapidly with the coupling. For the spectrum of figure~\ref{fig:vectorlimits}, however, the pair thresholds lie above 5 TeV, where the production rate is small. The effect is most relevant when the partners are as light as the current bounds allow.

\subsection{A light partner spectrum}
We therefore consider $(m_Q,m_S)=(1.9,2.4)$ TeV with the other inputs of the test point P0 (table~\ref{tab:points}). The lightest partner, mostly $X_{2/3}$, has a mass of 1.73 TeV and decays into $ht$, $Zt$ and $Wb$ with fractions 0.76, 0.21 and 0.04; the $X_{5/3}$ lies at 1.90 TeV, the other charge-$2/3$ states at 2.17 and 3.33 TeV and the $B$ at 2.28 TeV. All partners are just above the pair-production limits of table~\ref{tab:pair}, and their transition couplings are below the single-production limits of section~\ref{sec:vlqlimits}.

\begin{figure}[htbp]\centering
\includegraphics[width=\linewidth]{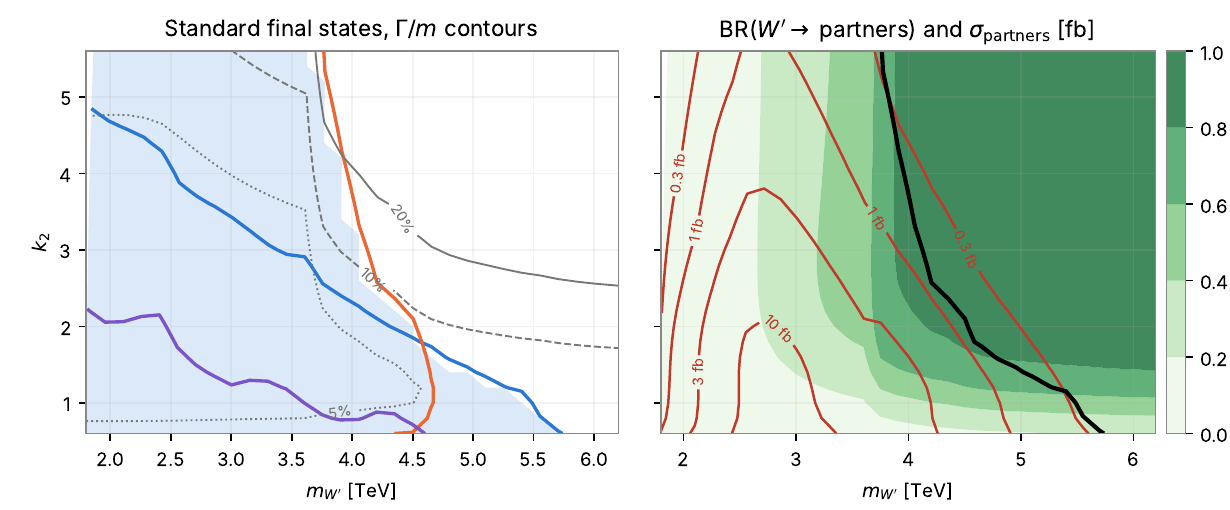}
\caption{Weak triplet with the light partner spectrum $(m_Q,m_S)=(1.9,2.4)$ TeV. Left (reinterpretation): observed exclusions from the lepton (blue), boson (orange) and quark (purple) categories of the CMS combination~\cite{cms}, obtained by extrapolating the narrow templates without a width restriction; grey lines are contours of $\Gamma_{W'}/m_{W'}$. Right (model prediction): fraction of $W'$ decays into final states with at least one partner (shading) and the 13 TeV rate $\sigma_{\rm partners}$ of resonant $W'$-mediated partner production, including finite-width effects (red contours), with the boundary of the left panel's exclusion in black.}\label{fig:lightpartners}
\end{figure}

Figure~\ref{fig:lightpartners} shows the consequences for the weak triplet. Below about 3.6 TeV the vector decays mostly into SM particles and the exclusion is close to the decoupled case. At 3.6 TeV the channel $W'^+\to X_{5/3}\bar T_{\rm light}$ opens, and the fraction of decays into partners rises quickly to 80--90\%, while the width grows with $k_2$. At $m_{W'}=4.2$ TeV and $k_2=3$, for example, $\Gamma/m=13\%$ and the $W'^+$ decays into $X_{5/3}\bar T_{\rm light}$ in 60\% of the cases and into $WZ$ or $Wh$ in only 9\%. The predicted rates in the leptonic and bosonic categories are then 0.12 and 0.75 of the observed limits if the narrow templates are applied, so that the point is not excluded by this interpretation; given its width, a firm conclusion requires a signal prediction for a broad resonance. As a result, the reach of the standard searches at $k_2=3$ decreases from about 5.0 TeV for decoupled partners to 4.1 TeV, and to 3.6 TeV if the narrow templates are trusted only up to $\Gamma/m=5\%$.

In this region the partner channels carry most of the signal. The resonant production of partners amounts to $\sigma_{\rm partners}\simeq0.3$--1.7 fb at 13 TeV for $3.6\lesssim m_{W'}\lesssim4.2$ TeV and $k_2\simeq3$--4, i.e.\ to some 40--240 events in the Run~2 data set, and more than twenty times as many at the HL-LHC. These events have distinctive features. The dominant pair $X_{5/3}\bar T_{\rm light}$ has mixed electric charges and is not produced by QCD; the decay $X_{5/3}\to W^+t$ yields same-sign dileptons, while $T_{\rm light}\to ht$ provides a reconstructable Higgs and top; and the invariant mass of the two partners reconstructs the parent vector, even if it is broad. These are the signatures proposed for the charged vector in ref.~\cite{wp} and, for heavy--light final states of the octet, in refs.~\cite{gluon,Vignaroli:2015ama}. Note that the existing cascade search in the $thb$ and $tZb$ final states~\cite{expB2G20002} captures only a small part of this signal, because the $B$ decays into $Wt$ and $WT_{\rm light}$ rather than into $hb$ or $Zb$.

\subsection{Resonant partner production}\label{sec:fw}
Since the dominant partner channel opens only about one width below the pole, the narrow-width factorisation $\sigma(pp\to W')\,\mathcal B$ is not reliable for it. We have therefore computed the full process $pp\to X_{5/3}\bar T_{\rm light}+{\rm c.c.}$ at leading order with MadGraph5\_aMC@NLO~\cite{Alwall:2014hca} and the \HVTP{} UFO at the benchmark points of table~\ref{tab:dedicated}. The full cross section is smaller than the narrow-width estimate by a factor between 0.73 and 0.46, the reduction growing with the width from 5 to 16\%. Two effects contribute: the part of the Breit--Wigner distribution below the threshold does not contribute, and near the threshold the partial width is suppressed by phase space; the rise of the parton luminosities at lower masses does not compensate. A semi-analytic convolution of the Breit--Wigner with the off-shell partial widths reproduces these factors to within 10\%, and we use it, calibrated to MadGraph, for all resonant rates in this section.

The neutral vectors partially compensate this loss. The $Z'$, degenerate with the $W'$, decays into $T_{\rm light}\bar T_{\rm light}$ and $X_{5/3}\bar X_{5/3}$ in 30--40\% of the cases and adds 0.2--0.75 fb of partner pairs at the benchmarks. A hypercharge vector $B'$ degenerate with the triplet ($M_1=M_2$, $k_1=k_2$) is not excluded by the dilepton searches at the three benchmarks with $k\ge3$ (section~\ref{sec:bprime}) and adds a similar amount, mainly in the same channels, which raises the estimates below by 5--25\%. We quote the results for a decoupled $B'$ and give the degenerate case in parentheses.

\subsection{Sensitivity of partner searches}
\begin{table}[htbp]\centering\small
\begin{tabular}{@{}rrrrrrrrrrr@{}}\toprule
$m_{W'}$ & $k_2$ & $\Gamma/m$ & $\mathcal B_{\rm partners}$ & $\sigma_{\rm partners}$ & $R_{\rm QCD}$ & $r_{\rm pair}$ & \multicolumn{2}{c}{Run 2 standard} & $N_{\rm ss}$ & $Z_{\rm HL}$\\
{[TeV]} & & & & [fb] & & & $r_{\ell}$ & $r_{VV,Vh}$ & Run 2 & cons./opt.\\\midrule
3.8 & 2 & 0.046 & 0.73 & 2.8 & 3.3 & 2.4 & 3.6 & 5.6 & 1.2 & 3.9 / 9.9\\
3.8 & 3 & 0.090 & 0.79 & 1.13 & 1.3 & 1.3 & 0.36 & 2.8 & 0.5 & 1.7 / 4.7\\
4.0 & 3 & 0.111 & 0.83 & 0.85 & 0.98 & 1.1 & 0.22 & 1.6 & 0.4 & 1.4 / 4.0\\
3.8 & 4 & 0.158 & 0.79 & 0.52 & 0.61 & 0.91 & 0.06 & 1.6 & 0.2 & 0.8 / 2.3\\\bottomrule
\end{tabular}
\caption{Benchmark points with $(m_Q,m_S)=(1.9,2.4)$ TeV. $\mathcal B_{\rm partners}$ is the fraction of $W'$ decays with at least one partner and $\sigma_{\rm partners}$ the corresponding 13 TeV rate of resonant production including finite-width effects (section~\ref{sec:fw}), normalised to the CMS HVT $W'$ cross section as in appendix~\ref{app:vectorrates}; the dilepton estimates use LO rates; higher-order corrections to the resonant signal are not included. $R_{\rm QCD}$ is its ratio to the NNLO QCD pair-production rate of all partners below 2.4 TeV. The ratios $r$ compare the predicted signal with the observed 95\% CL upper limit of a search: $r>1$ means that the benchmark would be excluded by that search, $r<1$ that it is not. $r_{\rm pair}$ refers to the ATLAS $T\to ht$ pair search~\cite{exp260528538}, with the partner-pair signal given by QCD plus $W'$- and $Z'$-mediated production (estimate, appendix~\ref{app:vectorrates}); for QCD production alone it is 0.55. $r_\ell$ and $r_{VV,Vh}$ refer to the lepton and boson categories of the CMS combination of $W'$ searches~\cite{cms}, applied without a width restriction. $N_{\rm ss}$ is the expected number of selected same-sign dilepton signal events with 140 fb$^{-1}$ at 13 TeV, and $Z_{\rm HL}$ the expected discovery significance with 3 ab$^{-1}$ at 14 TeV, for the conservative background and for the optimistic one (both $W'$ charges, which increase the background by a factor 1.5, and a background reduced by a factor of ten, i.e.\ $0.15$ times the conservative one). $r_{\rm pair}$ and $Z_{\rm HL}$ are estimates; see the text for their uncertainties.}\label{tab:dedicated}
\end{table}

Table~\ref{tab:dedicated} summarises the benchmark points and highlights three properties of the partner signatures.

First, the vectors are a copious source of partners. Resonant $W'$ production alone gives between 0.6 and 3.3 times the QCD pair-production rate ($R_{\rm QCD}$), in mixed-charge combinations that QCD does not produce and with the parent mass imprinted on the pair, and the neutral vectors add further pairs. The existing pair-production searches are therefore sensitive to this region, provided that the resonant contribution is included in the signal. To estimate their response without a dedicated simulation, we take the efficiency of the ATLAS $T\to ht$ search to be proportional to the probability that at least one partner decays into $ht$ (appendix~\ref{app:vectorrates}). The column $r_{\rm pair}$ gives the ratio of the predicted signal to the observed limit of this search, so that values above one indicate that the benchmark would be excluded. QCD production alone gives 0.55, i.e.\ the benchmarks are not excluded by the pair search as it is currently interpreted, whereas including the resonant pairs gives 0.9--2.4 (1.0--2.8 with a degenerate $B'$): three of the four benchmarks would be excluded once the vector-mediated pairs are included in the signal. A cross-check with the ATLAS search with one lepton and missing momentum~\cite{exp221205263} gives 0.37 for QCD alone and 0.6--1.6 (0.8--2.0) with the resonant pairs. The efficiency rule reproduces the published singlet and $ht$-only benchmarks within 10 and 20\%, but it has not been validated for mixed pairs, which dominate the resonant signal; at each benchmark the two searches differ by a factor of about 1.5, which we take as indicative of the uncertainty of this estimate, while the ranges quoted reflect the variation across the benchmarks. A reinterpretation of the Run~2 pair searches that includes resonant production would nevertheless be sensitive to most of the benchmark points, which the standard vector searches constrain only through the extrapolation of narrow templates. A definite statement requires the efficiencies for mixed pairs, which the collaborations can provide directly.

Second, the reconstruction of narrow partner peaks reduces the reliance on a narrow template for the parent resonance, although the signal rate and acceptance remain sensitive to its line shape (section~\ref{sec:fw}). This matters because at the points with $k_2\ge3$ the Run~2 boson-category limits are exceeded, i.e.\ $r_{VV,Vh}>1$, only by a factor 1.6--2.8, and only if narrow templates are applied to resonances with $\Gamma/m=9$--16\%. The partner channels thus provide a test that relies much less on this extrapolation.

Third, the cleanest channel, same-sign dileptons from $X_{5/3}\to W^+t$, is also the rarest. Rescaling the analysis of ref.~\cite{wp} to the present spectrum (appendix~\ref{app:vectorrates}), it retains about 0.6\% of the $X_{5/3}\bar T_{\rm light}$ rate, i.e.\ about 0.3\% of the total partner rate, a fraction confirmed by a detector-level simulation (appendix~\ref{app:vectorrates}), and needs the HL-LHC: at the three points with $k_2\ge3$ it reaches 2.3--4.7 standard deviations with an optimistic background and 0.8--1.7 with a conservative one ($Z_{\rm HL}$), without counting the same-sign pairs from $Z',B'\to X_{5/3}\bar X_{5/3}$. Final states with one lepton or none, with boosted top and Higgs tagging, would select an order of magnitude more of the partner rate. Since this analysis is a rescaled parton-level study, the significances $Z_{\rm HL}$ should be read as order-of-magnitude estimates.

\subsection{Projections for the HL-LHC}\label{sec:hl}
\begin{figure}[htbp]\centering
\includegraphics[width=\linewidth]{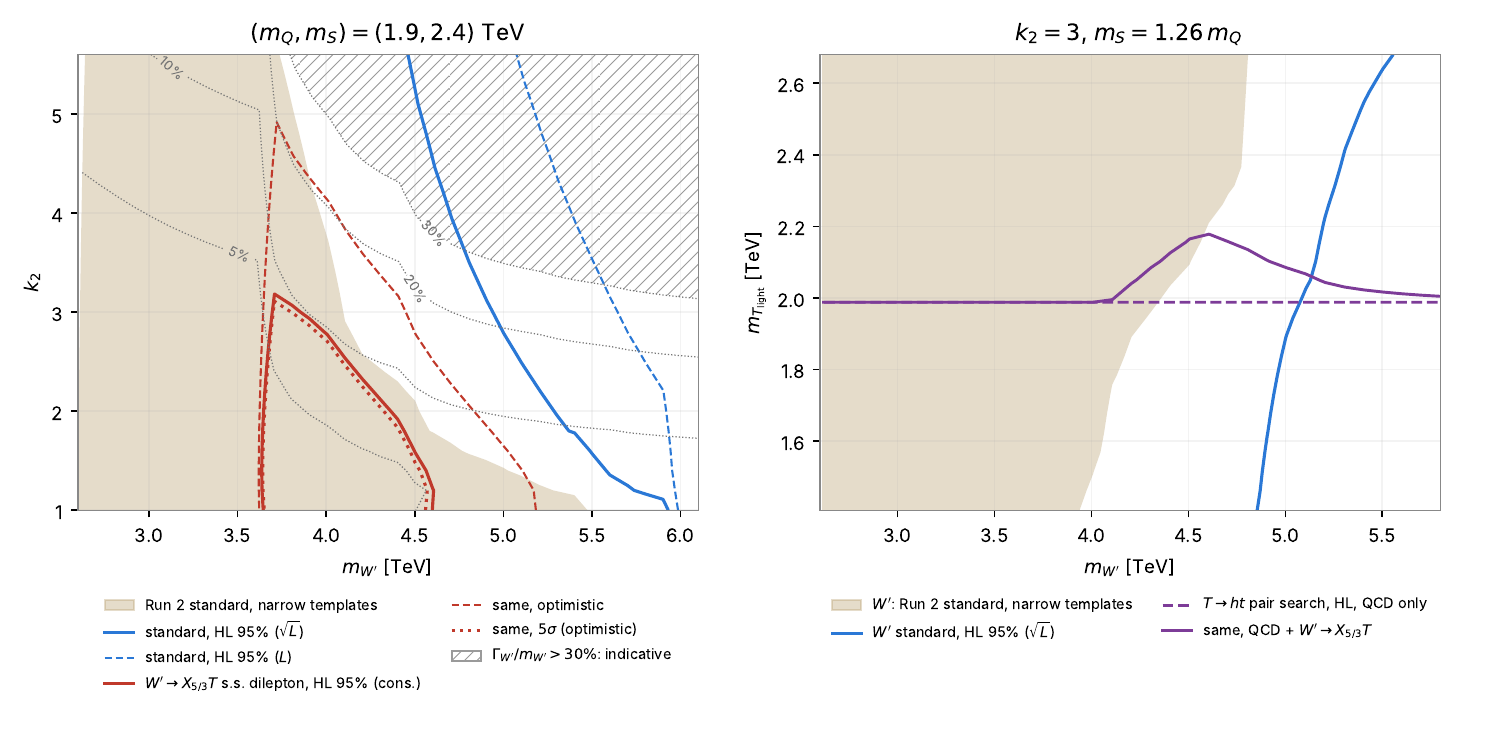}
\caption{Projections for the HL-LHC (14 TeV, 3 ab$^{-1}$). Left: light partner spectrum in the $(m_{W'},k_2)$ plane. Light brown: excluded by the Run~2 standard searches when narrow templates are applied, also to broad resonances. Blue: projected 95\% CL reach of the standard channels with $\sqrt L$ (solid) and $L$ (dashed) scaling of the expected limits. Red: same-sign dilepton channel from $W'\to X_{5/3}\bar T_{\rm light}$, 95\% CL exclusion with conservative (solid) and optimistic (dashed) background, and $5\sigma$ discovery with optimistic background (dotted). Dotted grey lines: $\Gamma_{W'}/m_{W'}$; hatched: $\Gamma_{W'}/m_{W'}>30\%$, where the projections are only indicative. Right: $k_2=3$ in the $(m_{W'},m_{T_{\rm light}})$ plane, with $m_S=1.26\,m_Q$. Purple: projected 95\% CL reach of the ATLAS $T\to ht$ pair search for QCD production only (dashed) and including resonant $W'\to X_{5/3}\bar T_{\rm light}$ production (solid); $Z'$ and $B'$ contributions are not included.}\label{fig:hl}
\end{figure}

To compare the two strategies at the HL-LHC we rescale the present expected limits to 14 TeV and 3 ab$^{-1}$, assuming either a background-dominated ($\sqrt L$) or a background-free ($L$) scaling (appendix~\ref{app:vectorrates}). The results are shown in figure~\ref{fig:hl}.

For the light spectrum the projected reach of the standard channels is 5.2, 4.9, 4.6 and 4.5 TeV for $k_2=2$, 3, 4 and 5 with $\sqrt L$ scaling, about 0.8 TeV beyond the Run~2 exclusion. At these masses, however, the $W'$ width is 11, 22, 33 and 48\%: for $k_2\gtrsim3$ the whole gain lies in the broad region, and the projection holds only if the searches are extended to large widths. Above $\Gamma/m\simeq30\%$, hatched in figure~\ref{fig:hl}, the $s$ dependence of the width and the interference with SM production are no longer small corrections, and the projections for $k_2=4$ and 5 are only indicative. The same-sign dilepton channel reaches 4.3 and 3.8 TeV for $k_2=2$ and 3 with the conservative background, and 4.8, 4.4 and 4.0 TeV for $k_2=2$, 3 and 4 with the optimistic one, inside the standard reach. This does not reflect a small signal. At the benchmarks the rate of resonant partner production is 20--200 times the $W'\to e\nu,\mu\nu$ rate and three to five times the $W'\to WZ,Wh$ rate, but the same-sign selection, which requires the two $W$ bosons from $X_{5/3}\to W^+t\to W^+W^+b$ to decay leptonically, retains only about 0.6\% of the $X_{5/3}\bar T_{\rm light}$ rate. An analysis that selects ten times more signal events with the same signal-to-background ratio, as could be obtained by adding final states with one lepton or none and boosted top and Higgs tagging, would extend the exclusion to 4.85, 4.46 and 4.06 TeV for $k_2=2$, 3 and 4, close to the reach of the standard channels. The partner channels thus become competitive as soon as they capture a larger fraction of the partner rate, with the advantage of not relying on a narrow-width template for the vector.

The right panel shows the role of resonant production for the partners themselves. QCD production alone brings the reach of the $T\to ht$ pair search to about $m_{T_{\rm light}}\simeq2.0$ TeV ($\sqrt L$) or 2.2 TeV ($L$), where above 2 TeV, the end of the published mass range, the expected limit is kept constant, so the light spectrum considered here will be probed at the HL-LHC independently of the vectors. Resonant $X_{5/3}\bar T_{\rm light}$ production extends the reach by up to 200 GeV, before including the neutral vectors. In this region the $W'$ is also within the projected standard reach, so that a discovery of partners would come together with a resonance in the partner-pair mass.

\subsection{Implications for the HL-LHC programme}
For a heavy partner sector, the searches for narrow resonances in SM final states are the most sensitive probes of the vectors. When the partners are lighter than the vectors and close to their present bounds, as in natural scenarios, the picture changes qualitatively: the vectors become broad and decay mainly into partners, the standard final states lose a factor of five or more in rate, and their constraints rely on narrow-width templates. The partner channels then carry most of the signal, enhance partner production to a level comparable with or above QCD, and reconstruct both the partners and the parent. A programme that covers this scenario at the HL-LHC should include:
\begin{enumerate}
\item interpretations of the partner pair-production searches that add vector-mediated production to the QCD rate, with the mixed-charge pairs such as $X_{5/3}\bar T_{\rm light}$ and $T\bar B$ and the branching fractions computed consistently, as \HVTP{} does;
\item searches for a resonance in the invariant mass of two reconstructed partners, or of a partner and a third-generation quark, in which the narrow partner peaks provide the background rejection while the parent may be broad;
\item cascade final states beyond same-sign dileptons, in particular single-lepton and fully hadronic channels with boosted top, $W/Z$ and Higgs tagging, which retain an order of magnitude more of the signal;
\item limits on vector resonances for widths up to 20--30\%, including the interference with SM production, so that the standard channels can be applied in the same region.
\end{enumerate}
None of these requires new detector capabilities; they require that the signal hypotheses of the vector and partner searches be defined together. The standard and cascade channels are complementary: the former reach the broad region only if extended to large widths, the latter only with acceptances well beyond the same-sign dilepton channel, and only the latter measure the partner spectrum.

\section{Nonstandard partner decays and extended scalar sectors}\label{sec:nonstandard}
The searches for top partners assume that they decay into a third-generation quark and a $W$, $Z$ or Higgs boson. In composite models this assumption can fail in two ways: a partner can decay into a lighter partner, and it can decay into new scalars of the strong sector. Both effects are included in \HVTP{} and \HVTPeta{}.

\subsection{Cascades between partners}
Electroweak mixing splits the partners, and when the splitting exceeds the $W$ mass the heavier ones can decay into the lighter ones. These transitions are not suppressed by any mixing angle: within a composite multiplet the $W$ current has full strength, and the decay into the longitudinal $W$ is further enhanced by the Goldstone equivalence theorem. After the diagonalisation the currents between mass eigenstates carry composition factors, so that the size of each transition must be computed; in the cases discussed below they dominate once the splitting exceeds $m_W$ by a few hundred GeV. The typical example is provided by the custodians. Before EWSB the two members of $X$ are degenerate; the Yukawa interaction then mixes $X_{2/3}$ with the singlet and lowers the lightest charge-$2/3$ state, while $X_{5/3}$, which cannot mix, remains at $m_Q$. The decay $X_{5/3}\to W^+T_{\rm light}$ thus opens as soon as the splitting exceeds $m_W$.

At the test point P0 the lightest partner, at 1.30 TeV, decays into $Wb$, $Zt$ and $ht$ with fractions $(0.19,0.13,0.68)$, far from the singlet pattern because of the mixing among the three heavy states, and the $X_{5/3}$, at 1.80 TeV, decays into $W^+T_{\rm light}$ in 72\% of the cases and into $W^+t$ only in 28\%. Only 8\% of its pairs then decay into the $WtWt$ final state targeted by the searches, and the $B$ behaves in the same way, with $\mathcal B(W^-T_{\rm light})=0.76$. The exclusive limits of table~\ref{tab:pair} cannot be applied to such a spectrum: the cascades populate final states with additional $W$ bosons, which the searches select with different efficiencies. For the light spectrum of section~\ref{sec:broad} the splitting is only 170 GeV and $\mathcal B(X_{5/3}\to W^+T_{\rm light})=0.17$.

\subsection{Decays into a composite pseudoscalar}\label{sec:eta}
\HVTPeta{} (section~\ref{sec:etamodel}) allows a direct assessment of how a composite pseudoscalar modifies the phenomenology of the top partners, since the partner decays into $\eta$ and the production of $\eta$ are fixed by the same couplings. Two questions are relevant: how much $\eta$ can change the decays of the partners and the searches for them, and which couplings are allowed by the direct searches for $\eta$.

\begin{table}[htbp]\centering\small\setlength{\tabcolsep}{5pt}
\begin{tabular}{@{}lrrrrr@{}}\toprule
Spectrum & $m_\eta$ [GeV] & $\kappa$ & $\mathcal B(T_{\rm light}\to t\eta)$ & $\mathcal B(T_{\rm light}\to ht)$ & $\mathcal B(B\to b\eta)$\\\midrule
P0 ($m_{T_{\rm light}}=1.30$ TeV) & 200 & 1 & 0.13 & 0.59 & 0.03\\
 & 200 & 2 & 0.37 & 0.43 & 0.10\\
 & 200 & 3 & 0.57 & 0.29 & 0.20\\
Light partners ($m_{T_{\rm light}}=1.73$ TeV) & 200 & 1 & 0.06 & 0.71 & 0.07\\
 & 200 & 2 & 0.20 & 0.60 & 0.23\\
 & 200 & 3 & 0.36 & 0.48 & 0.41\\
 & 500 & 3 & 0.33 & 0.50 & 0.39\\\bottomrule
\end{tabular}
\caption{Branching fractions into the composite pseudoscalar for $\kappa_Q=\kappa_S=\kappa$ and no ultraviolet term, for the test point P0 and for the light partner spectrum $(m_Q,m_S)=(1.9,2.4)$ TeV of section~\ref{sec:broad}. For $m_\eta=200$ GeV and the light spectrum $\eta$ decays into $gg$ (88\%), $b\bar b$ (7\%), $\gamma\gamma$ (3.5\%) and $W^+W^-$ (1\%); above the $t\bar t$ threshold almost exclusively into $t\bar t$. Without an ultraviolet term the rows with $m_\eta=200$ GeV are excluded by the diphoton search (figure~\ref{fig:eta}); they illustrate the effect on the partners in models in which ultraviolet terms suppress the couplings of $\eta$ to gluons and photons (see text).}\label{tab:eta}
\end{table}

Table~\ref{tab:eta} shows that for couplings $\kappa\sim2$--3, comparable to the composite Yukawa coupling $Y_*\simeq2.8$--2.9 of table~\ref{tab:points}, the decay $T_{\rm light}\to t\eta$ competes with $ht$, which otherwise dominates. The searches based on $ht$ lose sensitivity in proportion to the reduced $ht$ fraction, but the effect on the partner bounds is moderate: the limit from the $T\to ht$ pair search on the light spectrum, 1.60 TeV without $\eta$, decreases to 1.57, 1.54 and 1.49 TeV for $\kappa=2$, 3 and 4. The $X_{5/3}$ cannot decay directly into $\eta$ and a SM quark, and its same-sign signatures are affected only through the cascade $X_{5/3}\to W^+T_{\rm light}$, $T_{\rm light}\to t\eta$, which is small for the light spectrum, where $\mathcal B(X_{5/3}\to W^+T_{\rm light})=0.17$. The final states change more significantly. Below the $t\bar t$ threshold $\eta$ decays mostly into gluons, so that pair production gives $t\bar t+4j$ and, more rarely, $t\bar t+jj\gamma\gamma$; the searches for $T\to t\phi$ with $\phi\to b\bar b$ or $\gamma\gamma$~\cite{exp251025874,exp260728381,expB2G24016,expB2G22001} assume unit decay into the reconstructed mode and must be rescaled, and the dominant $gg$ mode is not targeted by any current search. Above the threshold $\eta\to t\bar t$ turns pair production into final states with four or more top quarks.

\begin{figure}[htbp]\centering
\includegraphics[width=.75\linewidth]{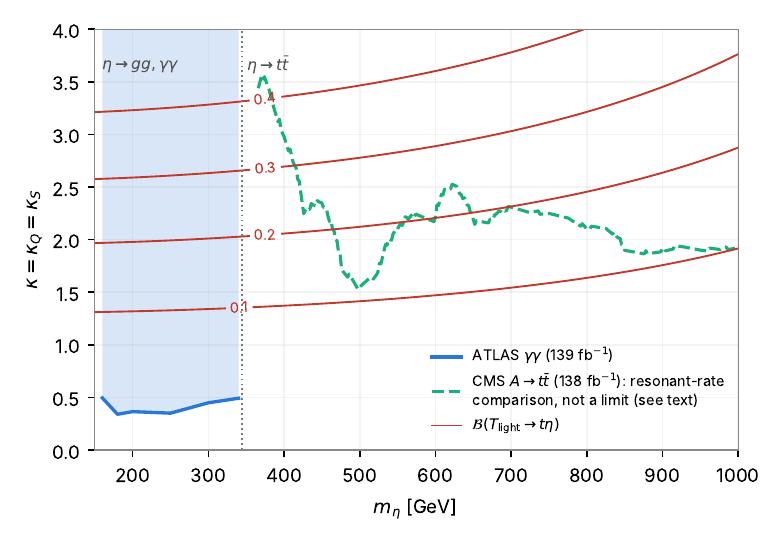}
\caption{Composite pseudoscalar in \HVTPeta{} with $\kappa_Q=\kappa_S=\kappa$, the light partner spectrum $(m_Q,m_S)=(1.9,2.4)$ TeV and no ultraviolet term. Blue (reinterpretation): region excluded by the ATLAS diphoton search~\cite{ATLAS:2021uiz}, with $\sigma(pp\to\eta)$ from MadGraph at LO. Red: contours of $\mathcal B(T_{\rm light}\to t\eta)$. To the right of the dotted line $\eta$ decays into $t\bar t$; green dashed: value of $\kappa$ at which the product of the production and decay couplings of $\eta$ equals that of a pseudoscalar at the observed CMS $A\to t\bar t$ limit for $\Gamma/m=1\%$, i.e.\ equal resonant rates at a common width with $\eta_t$ in the background~\cite{CMS:2025kzt}; this comparison ignores the different phase of the production amplitude and is not a limit (see text and appendix~\ref{app:etaprod}).}\label{fig:eta}
\end{figure}

The couplings that make these decays sizeable also produce $\eta$ in gluon fusion, through the loops of the top quark and of the partners (appendix~\ref{app:etaprod}). Below the $t\bar t$ threshold $\eta$ decays mostly into gluons, but its diphoton branching fraction of a few per cent is enough for the ATLAS diphoton search~\cite{ATLAS:2021uiz} to exclude $\kappa\gtrsim0.35$--0.5 for $160<m_\eta<2m_t$ (figure~\ref{fig:eta}), where $\mathcal B(T_{\rm light}\to t\eta)\lesssim0.02$. Sizeable partner decays into a light $\eta$ therefore require its couplings to gluons and photons to be suppressed. This happens naturally in models with an underlying fermionic description in which $\eta$ is made of constituents without colour and has no anomaly with electromagnetism, such as the singlet of the electroweak coset $SU(4)/Sp(4)$: its Wess--Zumino--Witten couplings to gluons and photons vanish~\cite{Arbey:2015exa}, and anomaly matching~\cite{tHooft:1979rat} requires the low-energy theory to contain local terms that cancel the loop contributions of the heavy partners, so that only those of the light states, in particular of the top quark, remain. The precise matching conditions depend on the embedding of the partners in the underlying theory. In \HVTPeta{} the local terms are the independent coefficients $K_G$, $K_W$ and $K_B$ of eq.~\eqref{eq:eta}; in the limit of heavy partners the cancellation corresponds to $K_G/f_\eta=-\sum K_{ii}/m_i$ and $(K_W+K_B)/f_\eta=-2\sum N_cQ_i^2K_{ii}/m_i$, with the sums over the partners, and to the analogous conditions for $K_W$ and $K_B$ separately, which the point generator can impose (table~\ref{tab:inputs}). Setting $K_B=-K_W$ would only remove the local diphoton term and leave the partner loops. The remaining couplings to $WW$ and $ZZ$ may then control the decays of a light $\eta$ (section~\ref{sec:etamodel}). Above the threshold $\eta$ decays into $t\bar t$ and the diphoton constraint disappears. The searches for $A\to t\bar t$~\cite{ATLAS:2024vxm,CMS:2025kzt} then become the relevant probe, but their limits cannot be applied directly to $\eta$: the partner loops change both the size and the phase of its production amplitude, and with them the interference with QCD $t\bar t$ production on which these searches rely (appendix~\ref{app:etaprod}). Near the threshold the $\eta$ amplitude also interferes with toponium formation, since both are in the colour-singlet ${}^1S_0$ channel. For a pseudoscalar produced through the top loop and decaying into $t\bar t$, unitarity fixes the imaginary part of the Fano parameter of the line shape, and the interference strongly distorts the resonance and reduces its rate by 70--90\% for masses up to about 380~GeV~\cite{Djouadi:2026toponium}. In \HVTPeta{} the partner loops add to the dispersive part of the production amplitude, so that a quantitative comparison in this region requires a dedicated treatment of the threshold. A comparison of the couplings at a common width, shown in figure~\ref{fig:eta} for orientation, indicates that these searches become sensitive for $\kappa\simeq1.6$--3.

The implications for the top partners can be summarised as follows. A composite pseudoscalar with couplings to the partners comparable to $Y_*$ can reduce $\mathcal B(T_{\rm light}\to ht)$ by up to a factor of two, and replace it with final states, $t\bar t+4j$ below the $t\bar t$ threshold and, above it, four top quarks when one partner decays into $t\eta$ and six when both do, that are not targeted by current searches. Below the threshold this requires the couplings of $\eta$ to gluons and photons to be suppressed, as in models in which $\eta$ has no anomaly with QCD and electromagnetism; above it the searches for $A\to t\bar t$ test the same couplings that control the partner decays. Since \HVTPeta{} fixes the partner decays and the production of $\eta$ with the same parameters, it allows these complementary searches to be interpreted consistently in a single model.

\subsection{Partners decaying into a lighter vector}\label{sec:inverted}
If the vectors were lighter than the partners, the decays $Q\to qG'$ or $X_{5/3}\to tW'^+$ would open; such inverted cascades were examined in ref.~\cite{liu}. For moderate couplings the current limits of section~\ref{sec:vectorsearches}, $m_{W'},m_{Z'}\gtrsim5$ TeV and $m_{G'}\gtrsim4.5$ TeV, would place the partners largely beyond the reach of the LHC in this ordering; since these limits depend on the couplings and weaken for broad resonances, the ordering is not excluded in general. \HVTP{} describes it consistently, but we do not pursue it further.

\subsection{Further extensions of the scalar sector (not implemented)}\label{sec:scalars}
\HVTP{} describes the Higgs as a linear doublet, with SM-like self-interactions, and \HVTPeta{} adds a single pNGB coupled to the partners. Two further extensions are natural in composite models, in particular in those with an underlying fermionic description, where light scalars are generic~\cite{Cacciapaglia:2019bqz}. They are not part of the released models, but could be implemented within the same framework.

\paragraph{A nonlinear Higgs.} In the minimal $SO(5)/SO(4)$ model the Higgs is the angular coordinate $\varphi$ of the vacuum, and its potential can be written in terms of $s_\varphi=\sin(\varphi/f)$. Retaining two harmonics, $V=\alpha s_\varphi^2+\beta s_\varphi^4$, the minimum sets $\xi=v^2/f^2=-\alpha/(2\beta)$ and
\begin{equation}
 m_h^2=\frac{8\beta\xi(1-\xi)}{f^2},\qquad
 \frac{\lambda_{hhh}}{\lambda_{hhh}^{\rm SM}}=
 \frac{1-2\xi}{\sqrt{1-\xi}} ,
\end{equation}
so that, at fixed $m_h$ and $v$, the scale $f$ determines both coefficients and the trilinear coupling. The expansion of the Goldstone kinetic term also modifies the $hVV$ and $hhVV$ couplings by $\sqrt{1-\xi}$ and $1-2\xi$, and the embedding of the fermions fixes the corresponding Yukawa vertices. These effects modify single-Higgs, double-Higgs and longitudinal vector-boson scattering processes in a correlated way~\cite{mchm,holo}. Their implementation requires a choice of $f$ and of the embedding, while a prediction of the potential from the partner spectrum would require further ultraviolet information.

\paragraph{CP-even singlets and dark matter.} A real CP-even singlet $s$ can couple to the singlet partner through $-y_s\,s\,\bar S_Lt_R+{\rm h.c.}$, which after the rotation to the mass basis induces the decay $\widetilde T\to ts$ with
\begin{equation}
 \Gamma(\widetilde T\to ts)\simeq\frac{|y_{\widetilde Tt}|^2m_{\widetilde T}}{32\pi}\Big(1-\frac{m_s^2}{m_{\widetilde T}^2}\Big)^2
\end{equation}
for negligible $m_t$, where $y_{\widetilde Tt}\simeq y_sc_R$ is the coupling to the mass eigenstates, of a single chirality. Depending on the mixing of $s$ with the Higgs or on its loop-induced couplings, it decays into $b\bar b$, gauge bosons or photons, as constrained by the searches of table~\ref{tab:searches}~\cite{scalar}. If the singlet is instead stable under a $Z_2$ symmetry, it is a dark-matter candidate~\cite{singlet}, and the decays of the partners give $t+E_T^{\rm miss}$ final states. A consistent realisation requires the $Z_2$ to be respected by the elementary--composite mixings, for instance with an odd partner that does not take part in the generation of the top mass.

\section{Conclusions}\label{sec:conclusions}
In composite Higgs models the vector resonances and the top partners belong to the same strongly coupled sector, and partial compositeness ties their interactions to the mixings that generate the top mass. Experimental searches, however, usually treat them separately, with simplified models in which the vectors decay only into SM particles and the partners are produced only through QCD or through their electroweak couplings. \HVTP{} provides a common description of both: an electroweak triplet, a hypercharge vector and a colour octet coupled to the custodial top-partner multiplet, the $\mathbf5_{2/3}$ of $SO(5)$, with all couplings obtained from the exact diagonalisation of the mass matrices at each parameter point. The vector--partner transitions, the partner decays and the modified couplings of the SM particles are therefore mutually consistent, also for large mixings and nearly degenerate partners. The custodians, among the lightest partners for the large left mixing required by the top mass, open the decays of the vectors into a partner and a top quark and into pairs of partners, which drive the phenomenology discussed here.

To make this description directly usable in experimental analyses, we release \HVTP{} and its extension \HVTPeta{}, which adds a composite pseudoscalar, as public UFO models for MadGraph. They extend the heavy-vector-triplet and vector-like-quark models used by ATLAS and CMS, which are recovered when the partners or the vectors are decoupled, and can be used at any parameter point specified by the physical masses of the resonances. A single model thus describes the signals of the vectors, those of the partners alone and their interplay.

The current LHC data exclude top partners up to 1.4--1.7 TeV, depending on their decays, and, for a heavy partner sector, the electroweak triplet up to 5--6 TeV and the colour octet up to about 4.5 TeV for moderate couplings, under the assumption of narrow resonances decaying into SM particles. When the partners are lighter than the vectors, as expected in natural scenarios, this assumption fails: the vectors decay mainly into partners and become broad, with widths of 5--16\% at our benchmark points for partners close to their present limits, and their limits in the standard final states rely on narrow-width templates. At the same time the vectors become a copious source of partners. Resonant production, including finite-width effects and the neutral vectors, is comparable to or larger than QCD pair production, and our rate estimates indicate that the existing pair searches, once resonant production is included, are sensitive to most of our benchmarks, which they do not exclude when only QCD production is considered; establishing an exclusion requires the efficiencies and templates for the mixed pairs produced through the vectors. Indirect tests, from electroweak precision, flavour and Higgs measurements, are complementary to these direct searches; \HVTP{} is designed for the latter, and a quantitative assessment of the former requires a dedicated global analysis.

These results have direct implications for the experimental programme. At the HL-LHC the standard resonance searches can reach about 0.8 TeV beyond the Run~2 exclusion, but for $k_2\gtrsim3$ this gain lies entirely in the broad region and holds only if the searches are extended to large widths. The partner channels offer a complementary route. The rate of resonant partner production is one to two orders of magnitude larger than the leptonic rate of the vectors, but the same-sign dilepton selections, originally designed for $X_{5/3}$ pairs produced through QCD and requiring two leptonic $W$ decays, retain only about 0.6\% of the $X_{5/3}\bar T_{\rm light}$ pairs produced in $W'$ decays. Analyses that capture a larger fraction, with one lepton or none and boosted top and Higgs tagging, would reach a sensitivity comparable to that of the standard channels without relying on narrow templates, and would also measure the partner spectrum, since a discovery of partners would then be accompanied by a resonant structure in the partner-pair mass. We therefore suggest four additions to the experimental programme: including vector-mediated production in the signal hypotheses of the partner searches, searching for resonances in the mass of reconstructed partner pairs, extending the cascade analyses to single-lepton and hadronic final states, and providing limits for broad resonances. None of them requires new detector capabilities, but all require a common definition of the signal for vectors and partners, which \HVTP{} provides.

\HVTPeta{} shows how the same framework accommodates a richer scalar sector. A composite pseudoscalar with couplings to the partners comparable to the composite Yukawa coupling can reduce the Higgs decays of the partners by up to a factor of two and replace them with final states that current searches do not cover. Below the $t\bar t$ threshold its couplings are strongly constrained by diphoton searches, unless local ultraviolet terms cancel the partner contributions to its couplings to gluons and photons. Above the threshold the pseudoscalar decays into top pairs, partner-pair production leads to final states with four or six top quarks, and the searches for $A\to t\bar t$ test the same couplings, although their reinterpretation requires signal templates that include the modification of the production amplitude by the partner loops.

At higher-energy colliders, vector-boson fusion, whose rate grows with the composite coupling, will become a complementary probe of the strongly coupled region. The framework can also be extended in several directions within the same structure: to further spin-one resonances of the coset and a nonlinear Higgs sector, to other representations of the top partners, to the non-minimal cosets and scalar sectors of models with an underlying fermionic description, and to higher-order corrections to the production of vectors and partners. Each extension can be released in the same form, as a tool for the reinterpretation of existing searches and for the design of new ones at the HL-LHC and at future colliders.

\clearpage\acknowledgments
I am grateful to all the colleagues with whom I have collaborated or exchanged ideas over the years on the topics explored in this paper.

I thank Takanori Kono, Quake Qin, Alvaro Lopez Solis, Philipp Gadow, Daniel Hayden, Tamara Vazquez Schroeder, Elin Bergeaas Kuutmann, Ana Maria Rodriguez Vera and the entire ATLAS Exotics group working on searches for vector-like quarks in Runs~2 and~3 of the LHC, with whom I have collaborated as a short-term associate of the ATLAS Collaboration. I have enjoyed working with them. The UFO model was originally developed for the ATLAS searches for vector-like quarks, and its public release and the present paper follow a suggestion from ATLAS. I hope that the models will be useful not only for these searches, but more generally for searches at the LHC and for phenomenological studies. I also thank Andrea Thamm and the colleagues of the Heavy Resonances Task Force for useful discussions, and I hope that this paper and the models will provide a useful input for the work of the Task Force.

The AI tools ChatGPT (OpenAI) and Claude (Anthropic) were used in a supporting role: to summarise the experimental searches considered in this work, to run validation tests of the model implementation, and for language editing. The conception of the work, the model and the physics interpretation are the author's own. All results, code and text were checked by the author, who takes full responsibility for the content of this work.

\clearpage\appendix
\section{Two-body decay widths}\label{app:widths}
This appendix collects the tree-level two-body widths used throughout the paper. They are evaluated with the mass-basis couplings of section~\ref{sec:model} and of the UFO, and agree with MadWidth for all channels of the partners and of the vectors, and for the $\eta$ channels implemented in \verb|analytic.py| (section~\ref{sec:validation}). For a decay $A\to B\,C$ with masses $M$, $m_B$, $m_C$ we use
\[
 \lambda(a,b,c)=a^2+b^2+c^2-2ab-2ac-2bc,\qquad x=\frac{m_B^2}{M^2},\quad y=\frac{m_C^2}{M^2},\quad D=\lambda(1,x,y),
\]
so that the daughter momentum in the rest frame of $A$ is $p=M\sqrt D/2$.

\subsection{Mass-basis couplings}
Let $O_C$ and $O_N$ be the orthogonal matrices that diagonalise the charged and neutral vector mass matrices of section~\ref{sec:vectormass}, with columns labelled by the mass eigenstates and rows by the site fields $(W_0,\rho_L)$ and $(W^3_0,B_0,\rho^3_L,\rho_Y)$. For the charged eigenstate $n$ the elementary and composite currents are
\[
 e_n=\frac{g}{\sqrt2}\Big(O_{C,0n}-\frac{O_{C,1n}}{k_2}\Big),\qquad
 c_n=\frac{g}{\sqrt2}\big(O_{C,0n}+k_2O_{C,1n}\big),
\]
and for the neutral eigenstate $n$ a field of isospin $T^3$, hypercharge $Y$ couples with
\[
 g\,T^3\big(O_{N,0n}-O_{N,2n}/k_2\big)+g'Y\big(O_{N,1n}-O_{N,3n}/k_1\big)\quad\text{(elementary)},
\]
\[
 g\,T^3\big(O_{N,0n}+k_2O_{N,2n}\big)+g'Y\big(O_{N,1n}+k_1O_{N,3n}\big)\quad\text{(composite)}.
\]
The colour octet couples with $-g_s/k_3$ to elementary and $g_sk_3$ to composite quarks. In the interaction basis these couplings form diagonal matrices $J_{L,R}$ in the space $(t,T,X_{2/3},\widetilde T)$ for charge $2/3$ and $(b,B)$ for charge $-1/3$; the unmixed $X_{5/3}$ is composite. The mass-basis couplings are
\[
 g_L=U_L^\dagger J_LU_L,\qquad g_R=U_R^\dagger J_RU_R,
\]
with $U_{L,R}$ from the biunitary diagonalisation $U_L^\dagger\mathcal M U_R={\rm diag}(m_i)$ of the fermion mass matrices; for charged currents the left and right rotations of the up and down sectors are used on the two sides. The Higgs couplings $\bar f_i(Y_{ij}P_R+Y^*_{ji}P_L)f_j\,h$ follow from the same rotations of the Yukawa matrix, and the pseudoscalar couplings of \HVTPeta{} from $K=U_L^T\,{\rm diag}(0,\kappa_Q,\kappa_Q,\kappa_S)\,U_R$.

\subsection{Heavy vectors}
For a vector $V$ of mass $M$ and an interaction $V_\mu\bar f_i\gamma^\mu(g_LP_L+g_RP_R)f_j$,
\begin{equation}
 \Gamma(V\to f_i\bar f_j)=\frac{C\,M\sqrt D}{48\pi}\Big[(|g_L|^2+|g_R|^2)\big(2-x-y-(x-y)^2\big)+12\,{\rm Re}(g_Lg_R^*)\sqrt{xy}\Big],
 \label{eq:Vff}
\end{equation}
with $x=m_i^2/M^2$, $y=m_j^2/M^2$. The colour factor is $C=1$ for leptons, $C=3$ for a colour-singlet vector decaying to quarks and $C=1/2$ for a fixed octet component, with ${\rm tr}(T^aT^b)=\delta^{ab}/2$. For a neutral vector and $i\ne j$ the conjugate final state $f_j\bar f_i$ contributes the same amount. For degenerate custodians, $m_{X_{5/3}}=m_{X_{2/3}}=M_X$ and $r=M_X^2/m_{W'}^2$, eq.~\eqref{eq:Vff} gives
\[
 \Gamma(W'^+\to X_{5/3}\bar X_{2/3})=\frac{g^2k_2^2m_{W'}}{8\pi}(1+2r)\sqrt{1-4r}.
\]
For the Yang--Mills vertex $g_{VV_1V_2}$ with the SM tensor structure, and for the vertex $i\,g_{VV_1h}\,g_{\mu\nu}$,
\begin{align}
 \Gamma(V\to V_1V_2)&=\frac{|g_{VV_1V_2}|^2M D^{3/2}}{192\pi\,xy}\big[1+10(x+y)+x^2+y^2+10xy\big],\label{eq:VVV}\\
 \Gamma(V\to V_1h)&=\frac{|g_{VV_1h}|^2\sqrt D}{48\pi M}\left[2+\frac{(1+x-y)^2}{4x}\right],\label{eq:VVh}
\end{align}
where $x=m_{V_1}^2/M^2$ and $y=m_{V_2}^2/M^2$ or $m_h^2/M^2$. The trilinear coupling $g_{VV_1V_2}$ is obtained by projecting each leg on the two sites, as given in section~\ref{sec:vectormass}, and the Higgs coupling of the charged vectors is $g_{W'Wh}=v\,(O_C^Tu_C)_W(O_C^Tu_C)_{W'}/2$ with $u_C=(g,gk_2)^T$; the neutral vectors are treated in the same way with $u_N$. For a neutral vector $N_n$ the bosonic channels are $W^+W^-$, $W^\pm W'^\mp$ (the two charge assignments contribute equally) and $N_mh$ for every lighter massive neutral state $N_m$ ($Z$, and $Z'$ or $B'$); the couplings are $g_{N_nW_aW_b}=\sum_r g_{r2}\,O^{(3)}_{rn}O^{(+)}_{ra}O^{(+)}_{rb}$ and $g_{N_nN_mh}=v\,(O_N^Tu_N)_m(O_N^Tu_N)_n/2$. The fermionic channels of $Z'$ and $B'$ follow from eq.~\eqref{eq:Vff} with the neutral currents given above, including the heavy--light and partner-pair final states.

\subsection{Top partners}
For a heavy fermion $F$ of mass $M$ decaying to a vector $V$ and a fermion $f$ through $V_\mu\bar f\gamma^\mu(g_LP_L+g_RP_R)F$,
\begin{align}
 \Gamma(F\to Vf)={}&\frac{\lambda^{1/2}(M^2,m_V^2,m^2)}{32\pi M^3}
 \bigg\{(|g_L|^2+|g_R|^2)\Big[M^2+m^2-2m_V^2+\frac{(M^2-m^2)^2}{m_V^2}\Big]\nonumber\\
 &\qquad\qquad-12\,mM\,{\rm Re}(g_Lg_R^*)\bigg\}.
 \label{eq:FVf}
\end{align}
For a scalar $\phi$ and the interaction $\phi\,\bar f(aP_R+bP_L)F$,
\begin{equation}
 \Gamma(F\to\phi f)=\frac{\lambda^{1/2}(M^2,m_\phi^2,m^2)}{32\pi M^3}
 \Big[(|a|^2+|b|^2)(M^2+m^2-m_\phi^2)+4\,mM\,{\rm Re}(ab^*)\Big].
 \label{eq:Fhf}
\end{equation}
For the Higgs $a=Y_{fF}$ and $b=Y_{Ff}^*$. For the pseudoscalar, $\mathcal L\supset-i\eta\,\bar\psi_i(K_{ij}P_R-K_{ji}P_L)\psi_j$ gives $a=K_{fF}$ and $b=-K_{Ff}$. The partner-to-partner transitions such as $X_{5/3}\to W^+T_{\rm light}$ and $T_i\to W^-X_{5/3}$ follow from eq.~\eqref{eq:FVf} with the corresponding composite currents. In the limit $M\gg m_V,m$ eq.~\eqref{eq:FVf} reduces to $\Gamma\simeq(|g_L|^2+|g_R|^2)M^3/(32\pi m_V^2)$, which is the Goldstone-equivalence form $[(\lambda^L)^2+(\lambda^R)^2]M/(32\pi)$ in terms of the transition couplings of eq.~\eqref{eq:partnerlambda}.

\subsection{The composite pseudoscalar}
The pseudoscalar decays to a top pair through the diagonal coupling $K_{tt}$,
\begin{equation}
 \Gamma(\eta\to t\bar t)=\frac{3\,K_{tt}^2}{8\pi}\,m_\eta\sqrt{1-\frac{4m_t^2}{m_\eta^2}},
\end{equation}
with a single power of the velocity, as appropriate for a CP-odd state. The loop-induced couplings are written as $\mathcal L\supset C_G\,\eta\,G^a_{\mu\nu}\widetilde G^{a\mu\nu}+C_A\,\eta\,F_{\mu\nu}\widetilde F^{\mu\nu}$, $\widetilde G^{\mu\nu}=\tfrac12\epsilon^{\mu\nu\rho\sigma}G_{\rho\sigma}$, with
\[
 C_G=\frac{\alpha_s}{8\pi}\Big[\sum_i\frac{K_{ii}}{m_i}\,\frac{f(\tau_i)}{\tau_i}+\frac{K_G}{f_\eta}\Big],\qquad
 C_A=\frac{\alpha}{8\pi}\Big[2\sum_iN_cQ_i^2\frac{K_{ii}}{m_i}\,\frac{f(\tau_i)}{\tau_i}+\frac{K_W+K_B}{f_\eta}\Big],
\]
where the sums run over all coloured, respectively charged, fermion mass eigenstates, $\tau_i=m_\eta^2/(4m_i^2)$, and
\[
 f(\tau)=\begin{cases}\arcsin^2\sqrt\tau, & \tau\le1,\\[2pt]
 -\frac14\Big[\ln\dfrac{1+\sqrt{1-1/\tau}}{1-\sqrt{1-1/\tau}}-i\pi\Big]^2, & \tau>1.\end{cases}
\]
For a heavy fermion $f(\tau)/\tau\to1$, so its contribution does not decouple as $K_{ii}/m_i$ when $K_{ii}$ grows with the mass. The widths are
\begin{equation}
 \Gamma(\eta\to gg)=\frac{2\,|C_G|^2m_\eta^3}{\pi},\qquad
 \Gamma(\eta\to\gamma\gamma)=\frac{|C_A|^2m_\eta^3}{4\pi}.
\end{equation}
In the UFO the corresponding vertices are $4i\,C_{G,A}\,\epsilon^{\mu\nu\rho\sigma}p_{1\rho}p_{2\sigma}$, with $\alpha_s$ evaluated at the scale chosen by the event generator; MadWidth evaluates it at $m_\eta$. For the electroweak bosons we write $\mathcal L\supset C_{ZZ}\,\eta Z_{\mu\nu}\widetilde Z^{\mu\nu}+C_{Z\gamma}\,\eta Z_{\mu\nu}\widetilde F^{\mu\nu}+C_{WW}\,\eta W^+_{\mu\nu}\widetilde W^{-\mu\nu}$. The ultraviolet terms $C_W\,\eta\,W^a_{\mu\nu}\widetilde W^{a\mu\nu}+C_B\,\eta\,B_{\mu\nu}\widetilde B^{\mu\nu}$, with $C_W=\alpha K_W/(8\pi s_W^2f_\eta)$ and $C_B=\alpha K_B/(8\pi c_W^2f_\eta)$, contribute $c_W^2C_W+s_W^2C_B$, $2s_Wc_W(C_W-C_B)$ and $2C_W$, respectively. The fermion loops are computed from the triangle diagrams with all mass eigenstates $i,j,k$ of equal charge sectors, an $\eta$ vertex $K_{ij}P_R-K_{ji}P_L$ and gauge vertices $\gamma^\mu(g_LP_L+g_RP_R)$ taken from the model, including the off-diagonal ones. For a CP-odd scalar the amplitude has the single structure $\epsilon^{\mu\nu\rho\sigma}p_{1\rho}p_{2\sigma}$, to which the terms depending on the loop momentum do not contribute, so that each diagram is finite; we integrate over one Feynman parameter analytically and over the other numerically. Above the $t\bar t$ threshold the triangles with an internal top quark that can decay into $Wb$ develop an anomalous threshold, which we regulate with the top width in the propagators ($m_t^2\to m_t^2-im_t\Gamma_t$, $\Gamma_t=1.42$ GeV); the second integration uses a composite rule clustered at the end points. Applied to photons, the same calculation reproduces $C_A$ above to $10^{-7}$ below the $t\bar t$ threshold and to 1\% above it, the difference being the effect of the top width. For a degenerate vector-like doublet of mass 4 TeV it reproduces the local operator coefficients $2N_c{\rm Tr}(T^2)K/m$ in all four channels to $5\times10^{-4}$, the expected size of the $m_\eta^2/m^2$ corrections, and the results are stable to $10^{-4}$ when the number of integration points is doubled (script \verb|validation/heavy_doublet_check.py|). The form factors are evaluated for on-shell bosons, or at zero virtuality when the decay is closed, and stored as constants, as for $C_G$. The widths are
\begin{gather}
 \Gamma(\eta\to ZZ)=\frac{|C_{ZZ}|^2m_\eta^3}{4\pi}\beta_Z^3,\qquad
 \Gamma(\eta\to W^+W^-)=\frac{|C_{WW}|^2m_\eta^3}{8\pi}\beta_W^3,\nonumber\\
 \Gamma(\eta\to Z\gamma)=\frac{|C_{Z\gamma}|^2m_\eta^3}{8\pi}\Big(1-\frac{m_Z^2}{m_\eta^2}\Big)^3,
\end{gather}
with $\beta_V=(1-4m_V^2/m_\eta^2)^{1/2}$. The contact vertices $\eta W^+W^-\gamma$ and $\eta W^+W^-Z$ from the non-Abelian part of $W^a_{\mu\nu}\widetilde W^{a\mu\nu}$ are included; their normalisation is fixed by the photon Ward identity for $u\bar d\to W^+\gamma\eta$, which MadGraph verifies. To this end the loop contribution to $C_{WW}$ is attributed to an effective coefficient of $W^a_{\mu\nu}\widetilde W^{a\mu\nu}$; this preserves the electromagnetic Ward identity, but is not a complete electroweak matching of the loop amplitudes off shell. These widths agree with MadWidth to $10^{-5}$.

\subsection{Production of the composite pseudoscalar}\label{app:etaprod}
The coupling of $\eta$ to gluons, eq.~\eqref{eq:eta} and $C_G$ above, receives contributions from the top quark and from all partners. The top quark couples to $\eta$ only through its composite components, with strength $0.11\,\kappa$ for the light spectrum $(m_Q,m_S)=(1.9,2.4)$ TeV, i.e.\ $0.16\,\kappa$ in units of $m_t/v$, whereas the partners couple with full strength; since the loop of a heavy fermion is suppressed only as $1/m$, the partners give a comparable or dominant part of the amplitude. The coupling of the top quark has the opposite sign to those of the partners, so that the two contributions interfere destructively. Below the $t\bar t$ threshold the rate exceeds that of the top loop alone by a factor of up to about three. Near the threshold the top loop is enhanced and comparable in size to the partner contribution, and at $m_\eta\simeq400$ GeV the two combine to a rate equal to that of the top loop alone; above the threshold the top loop decreases and the partners dominate, and the rate exceeds that of the top loop alone by factors of 3 and 15 at 500 and 800 GeV. At leading order and 13 TeV,
\begin{equation}
\sigma(pp\to\eta)\simeq\kappa^2\times\{1.48,\,0.41,\,0.27,\,0.27,\,0.13\}\ {\rm pb}\ \ {\rm for}\ \ m_\eta=\{200,300,400,500,700\}\ {\rm GeV},
\label{eq:etaxs}
\end{equation}
with branching fractions that do not depend on $\kappa$: below the $t\bar t$ threshold $\eta$ decays into $gg$ (75--88\%), $b\bar b$ (5--10\%), $\gamma\gamma$ (3--7\%) and, above about 180 GeV, $W^+W^-$ and $ZZ$, whose combined branching fraction grows from 1\% at 200 GeV to 9\% near the threshold, while $\mathcal B(Z\gamma)<0.4\%$; above it almost exclusively into $t\bar t$, with $\Gamma_\eta/m_\eta\simeq0.1\,\kappa^2\%$. In the fiducial region of the ATLAS diphoton search~\cite{ATLAS:2021uiz} (see refs.~\cite{Molinaro:2015cwg,Molinaro:2016oix,Molinaro:2017rpe} for the diphoton signatures of composite pseudoscalars) the diphoton rate amounts to about $50\,\kappa^2$ fb at 160 GeV, $33\,\kappa^2$ fb at 200 GeV and $13\,\kappa^2$ fb at 300 GeV.

The searches for $A\to t\bar t$ set limits on the coupling $g_{At\bar t}$, in units of $m_t/v$, of a pseudoscalar that couples only to the top quark, and include the interference with QCD $t\bar t$ production, which produces a peak--dip structure in $m_{t\bar t}$~\cite{ATLAS:2024vxm,CMS:2025kzt}. Relative to the top loop alone, the gluon-fusion amplitude of $\eta$ at $m_\eta=400$, 500 and 800 GeV is larger in modulus by factors 1.0, 1.8 and 3.9 and rotated by about $100^\circ$, $90^\circ$ and $80^\circ$. A real effective top coupling can reproduce the resonant rate, which depends only on the modulus, but not in general the interference, which depends on the phase. The green line of figure~\ref{fig:eta} uses the coupling $g_{A,\rm eff}=(|g_{\rm prod}|\,g_{\rm dec})^{1/2}$, which gives the same product of production and decay couplings, i.e.\ the same resonant rate at a common width, where $g_{\rm prod}$ reproduces the modulus of the full amplitude and $g_{\rm dec}=0.16\,\kappa$; it grows from $0.16\,\kappa$ at 400 GeV to $0.31\,\kappa$ at 800 GeV and reaches the CMS limit for $\Gamma/m=1\%$, obtained with the $t\bar t$ bound state $\eta_t$ in the background~\cite{CMS:2025kzt}, for $\kappa\simeq1.6$--3. This comparison is not a 95\% CL limit: a quantitative reinterpretation requires signal templates computed with the full complex production amplitude and the physical width, $\Gamma_\eta/m_\eta\simeq0.1\,\kappa^2\%$, i.e.\ between 0.3\% and 0.9\% for $\kappa=1.6$--3.

\section{Using the models: details}\label{app:usage}
This appendix complements section~\ref{sec:usage} with the content of the package, the details of the inputs, the particle names, the options of the scripts, practical recommendations for event generation and the details of the validation.

\subsection{Content of the package}
The package contains:
\begin{itemize}
\item ready-to-use models, \verb|HVTP_UFO| and \verb|HVTP_eta_UFO|, with their complete param cards, at the default point defined in \verb|point.json| and \verb|point_eta.json|;
\item \verb|prepare_point.py| and \verb|prepare_point_eta.py|, which generate a model for any point, with all two-body widths computed by MadWidth, and \verb|scan.py|, which does so for a grid of points;
\item \verb|physical_inputs.py|, which converts physical masses into Lagrangian inputs, and \verb|analytic.py|, an independent implementation of the spectrum and widths used for fast scans;
\item the UFO writers \verb|build.py| and \verb|build_eta.py|, the validation scripts of section~\ref{sec:validation}, example inputs in \verb|examples/| and a \verb|README.md| that summarises this section.
\end{itemize}

\subsection{Inputs}
\paragraph{Lagrangian inputs.} These are the parameters of the two-site Lagrangian of section~\ref{sec:model}, specified before EWSB, together with the SM inputs that fix the gauge couplings and the top Yukawa coupling. The vector masses $M_1,M_2,M_3$ are the masses of the heavy vectors before EWSB. The physical masses of $W'$, $Z'$ and $B'$ differ from them through mixing with the SM gauge bosons, by a relative amount of order $g^2k^2v^2/M^2$; the octet does not mix and $m_{G'}=M_3$. The fermion inputs are the Dirac masses $m_Q$ of the bidoublet and $m_S$ of the singlet and the degrees of compositeness $s_L,s_R$ of $q_L$ and $t_R$, eq.~\eqref{eq:superposition}. The mixing masses are $\Delta_L=m_Qs_L/\sqrt{1-s_L^2}$ and $\Delta_R=m_Ss_R/\sqrt{1-s_R^2}$, so that before EWSB the doublet $(T,B)$ has mass $m_Q/\sqrt{1-s_L^2}$, the singlet $\widetilde T$ has mass $m_S/\sqrt{1-s_R^2}$, and the doublet $(X_{5/3},X_{2/3})$, which does not mix, has mass $m_Q$. The composite Yukawa coupling $Y_*$ is not an input: it is determined by requiring that the lightest charge-$2/3$ eigenvalue equals $m_t$, searching in the interval $(0,Y_{\max})$. If no solution exists, typically because $s_L$ and $s_R$ are too small to reproduce the top mass, the scripts stop with an error. After EWSB the Yukawa interaction mixes the three heavy charge-$2/3$ states, so that their physical masses can differ substantially from $m_Q$ and $m_S$; only $m_{X_{5/3}}=m_Q$ is exact.

\paragraph{Physical inputs.} Because of these differences, the package also accepts the physical quantities in the lower part of table~\ref{tab:inputs}. Each of them replaces the Lagrangian input that it determines, which is then obtained by solving the tree-level mass relations numerically: for instance, \verb|mWp| fixes $M_2$ and \verb|mtp1|, the mass of the lightest heavy charge-$2/3$ state, fixes $m_S$ at given $m_Q$, $s_L$ and $s_R$. Giving both a physical input and the Lagrangian input that it replaces is an error. If the requested value cannot be reached with the other inputs fixed, the script reports the reachable range, and the solution is always checked against all requested values. Since the two neutral vectors are identified by their dominant component, nearly degenerate values of \verb|mWp| and \verb|mBp| cannot be requested together; such spectra are obtained from the Lagrangian inputs, e.g.\ $M_1=M_2$. The Lagrangian inputs actually used are always written to the output, so that every point can be reproduced.

\subsection{Particle content and names}
Table~\ref{tab:names} gives the MadGraph names and PDG codes of the new particles and relates each of them to the multiplets of section~\ref{sec:multiplets} and figure~\ref{fig:multiplets}. The neutral vector \verb|zp| is the mass eigenstate with the largest $\rho_L^3$ component and \verb|bp| the one with the largest $\rho_Y$ component. The three heavy charge-$2/3$ states are called \verb|tp1|, \verb|tp2| and \verb|tp3| in order of increasing mass, so that \verb|tp1| is always the lightest top partner, $T_{\rm light}$ in this paper. Their composition depends on the point. It is printed when the model is generated and stored in the param card, in the block \verb|HVTPCOMPOSITION|, whose entry $(i,j)$ gives the fraction of \verb|tp|$i$ in $T$, $X_{2/3}$ and $\widetilde T$ for $j=1,2,3$, averaged over chiralities; the remainder is the component along the elementary top fields, i.e.\ the elementary components of $q_L$ and $t_R$. The SM particles have their usual names; the Higgs is \verb|h|, the photon \verb|a|, and $b$, $\mu$ and $\tau$ are massive.

\begin{table}[htbp]\centering\footnotesize\setlength{\tabcolsep}{4pt}
\begin{tabular}{@{}lll>{\centering\arraybackslash}p{0.8cm}>{\raggedright\arraybackslash}p{3.3cm}>{\raggedright\arraybackslash}p{4.1cm}@{}}\toprule
State & MG5 name & PDG & $Q_{\rm em}$ & Multiplet (fig.~\ref{fig:multiplets}(a)) & Default point: mass [TeV] and content\\\midrule
$W'^\pm$ & \verb|wp+|, \verb|wp-| & $\pm9000011$ & $\pm1$ & $\rho_L^\pm$, $(\mathbf3,\mathbf1)_0$ & 4.407; mostly $\rho_L^\pm$\\
$Z'$ & \verb|zp| & 9000012 & 0 & $\rho_L^3$, $(\mathbf3,\mathbf1)_0$ & 4.409; mostly $\rho_L^3$\\
$B'$ & \verb|bp| & 9000013 & 0 & $\rho_Y$: hypercharge comb.\ of $\rho_R^3$, $\rho_X$ & 4.400; mostly $\rho_Y$\\
$G'$ & \verb|gp| & 9000014 & 0 & $\rho_c$, colour octet & 5.0; $\rho_c$\\\midrule
$T_{\rm light}$ & \verb|tp1| & 9000001 & 2/3 & mixture of $T$, $X_{2/3}$ (bidoublet) and $\widetilde T$ (singlet) & 1.73; $X_{2/3}$ 76\%, $\widetilde T$ 9\%, $T$ 5\%\\
$T_2$ & \verb|tp2| & 9000003 & 2/3 & as above & 2.17; $T$ 67\%, $X_{2/3}$ 14\%, $\widetilde T$ 4\%\\
$T_3$ & \verb|tp3| & 9000005 & 2/3 & as above & 3.33; $\widetilde T$ 67\%, $T$ 12\%, $X_{2/3}$ 8\%\\
$X_{5/3}$ & \verb|x53| & 9000002 & 5/3 & $X_{5/3}$ in $X$, bidoublet & 1.90 ($=m_Q$); unmixed\\
$B$ & \verb|bh| & 9000004 & $-1/3$ & $B$ in $Q$, bidoublet & 2.28 ($=m_Q/c_L$); $B$; its left-handed component has an elementary fraction $s_L^2$\\\midrule
$\eta$ & \verb|eta| & 9000020 & 0 & singlet completing the Higgs $\mathbf4$ to a $\mathbf5$ & input $m_\eta$; only in \HVTPeta{}\\\bottomrule
\end{tabular}
\caption{New particles in the UFO models, with the multiplets of figure~\ref{fig:multiplets} to which they belong and, as an example, their masses and composition at the default point. For the charge-$2/3$ states the percentages are the fractions in the composite fields, averaged over chiralities (block \texttt{HVTPCOMPOSITION} of the param card); the remainder is elementary top. The ordering by mass is fixed, whereas the composition depends on the point. The antiparticles carry a tilde, e.g.\ \texttt{x53\textasciitilde}.}\label{tab:names}
\end{table}

\subsection{Preparing a parameter point and scans}
The script \verb|prepare_point.py|, used as in section~\ref{sec:usage}, creates the output directory for the default point modified by the requested inputs. Each \verb|--set| overrides one key of table~\ref{tab:inputs}; alternatively, \verb|--point inputs.json| reads several inputs from a JSON file, with the command-line values taking precedence and missing keys taken from the defaults. The option \verb|--mg5| can be omitted if \verb|mg5_aMC| is in the path or the environment variable \verb|MG5| points to it. The script resolves the physical inputs, prints the Lagrangian inputs and the tree-level spectrum, writes the UFO, runs MadWidth for all two-body decays of the new states and of $t$, $W$, $Z$ and $h$, and writes the UFO again with the computed total widths as defaults. The output directory contains the model \verb|HVTP_UFO|, the complete \verb|param_card.dat| with all branching fractions, the Lagrangian inputs used (\verb|point.json|), the spectrum together with the requested physical inputs (\verb|spectrum.json|) and the MadWidth log. The param card also contains the block \verb|HVTPINPUTS|, which records the Lagrangian inputs and $Y_*$ (and, for \HVTPeta{}, the inputs of $\eta$). Since the param card is copied into the banner of the event files, every sample carries its parameter point. Like \verb|HVTPCOMPOSITION|, this block is informational: changing it has no effect on the model. The command
\begin{verbatim}
python3 prepare_point_eta.py my_eta_point --set m_eta=500 \
        --set kQ=2 --set kS=2
\end{verbatim}
does the same for \HVTPeta{}, which for $\kappa_Q=\kappa_S=K_G=K_W=K_B=0$ reduces to \HVTP{} plus a decoupled scalar.

For scans, \verb|scan.py| takes a JSON file with fixed and scanned inputs, for example
\begin{verbatim}
{"fixed": {"mQ": 1900, "mS": 2400},
 "scan":  {"mWp": [3500, 4000], "k2": [2, 3]}}
\end{verbatim}
and either generates a complete model for each point (option \verb|--mg5|) or, with \verb|--analytic|, computes only the spectrum and the two-body widths of the partners and vectors with \verb|analytic.py|, in a fraction of a second per point and without MadGraph. In both cases it writes a table \verb|summary.csv| with the inputs, masses, widths and main branching fractions.

\subsection{Generating events}
In a MadGraph session started from the same working directory, the model of a point is imported and used as any other model (section~\ref{sec:usage}).
The process directory inherits masses and widths, but its default param card contains no branching fractions; the complete \verb|param_card.dat| of the point must be used whenever the new states are decayed by the parton shower (see below). The following points help to avoid common mistakes.
\begin{itemize}
\item \textbf{Coupling orders.} Each coupling has the usual \verb|QCD| or \verb|QED| order and, if it involves heavy vectors, an additional order \verb|HV| equal to the number of heavy-vector legs ($W'$, $Z'$, $B'$, $G'$). Since the $G'$ couplings to quarks carry \verb|QCD| order, the default \texttt{generate p p \string> x53 x53\string~} includes both QCD production and $G'$ exchange. The contributions are best selected with explicit orders: \texttt{generate p p \string> x53 x53\string~ QCD=2 QED=0 HV=0} gives pure QCD production, \texttt{QCD=2 QED=2 HV=2} adds $G'$ and the neutral electroweak exchange ($\gamma$, $Z$, $Z'$, $B'$), and \verb|QCD=0| selects the colour-singlet electroweak part, which does not interfere with the colour-octet amplitudes in $q\bar q$ annihilation. Note that \verb|HV=0| removes the vertices with heavy vectors but keeps the mixing corrections to the couplings of the light states, so that at a finite \HVTP{} point it does not reproduce the SM exactly.
\item \textbf{Broad resonances.} The widths are on-shell, tree-level two-body widths. When a vector is broad (in practice $\Gamma/m\gtrsim10\%$) or a decay opens close to threshold, decay chains in the narrow-width approximation can misestimate the rate considerably (section~\ref{sec:fw}). One should then generate the full $2\to n$ process, e.g.\ \verb|p p > x53 tp1~| rather than \verb|p p > wp+, wp+ > x53 tp1~|, and increase \verb|bwcutoff| in the run card when the resonance is required on shell. The same holds for partners with large widths.
\item \textbf{Top partners alone.} For studies of the partners without vector effects, the vectors can be decoupled by setting $M_{1,2,3}$ to large values (e.g.\ 20 TeV), or excluded from the diagrams with \verb|HV=0|. Single production through $W$ exchange is then generated with \texttt{generate p p \string> tp1 j QCD=0 HV=0} (with \verb|define p = p b b~|, see below), and the decays use the branching fractions of the param card, which are computed at the same point as the production couplings.
\item \textbf{Parton shower and detector simulation.} When the new states are left undecayed in the hard process, Pythia~8 decays them with the \texttt{DECAY} tables of the param card stored in the event file. The complete param card of the point must therefore be placed in the \verb|Cards| directory of the process, or entered as a file path at the card-editing prompt of \verb|launch|. With the default card of the process directory, which has total widths but no branching fractions, Pythia leaves the partners stable and the hadronisation of these coloured states fails. The PDG codes of table~\ref{tab:names} do not coincide with any entry of the particle tables of Pythia~8 or ROOT. We have processed events of resonant $X_{5/3}\bar T_{\rm light}$ production, $G'\to T_{\rm light}\bar t$ and $\eta$ production with Pythia~8.312 and 8.318 and with Delphes~3.5.1 and its CMS card: all events are showered and simulated without errors, no new particle survives in the final state, and the decay fractions agree with the param card. In the Delphes truth record the new particles carry the charge $-999$, since they are not in the ROOT particle table; reconstructed objects are not affected.
\item \textbf{Flavour scheme.} Since $b$ is massive, the default proton and jet definitions follow the four-flavour scheme; processes with initial $b$ quarks, as in single production via $W$ exchange, require \verb|define p = p b b~| and a corresponding PDF choice.
\item \textbf{Running coupling.} All QCD-order couplings, including those of $G'$ and the $\eta gg$ and $\eta ggg$ vertices, scale with the appropriate power of $g_s$ and are rescaled event by event with the running $\alpha_s(\mu_R)$ used by MadGraph, which takes $\alpha_s(m_Z)$ from the PDF set; a PDF set with $\alpha_s(m_Z)$ close to the value in the param card keeps production and widths consistent. In this treatment $k_3$ does not depend on the scale: the elementary coupling $g_s/k_3$, which controls Drell--Yan production, runs as in QCD, and so does the composite coupling $g_sk_3$, whose actual running depends on the strong sector and is beyond the accuracy of the model. Since $\sigma(pp\to G')\propto\alpha_s(\mu_R)$, the rate of a multi-TeV octet depends on the scale choice; the default dynamical scale of MadGraph for $pp\to G'$ is $\mu_R=\mu_F=m_{G'}$. MadWidth evaluates $\alpha_s$ at the mass of the decaying particle, so that the $G'$ and $\eta\to gg$ widths in the param card correspond to $\alpha_s(m)$.
\item \textbf{Gauge and accuracy.} The models are written in unitary gauge and are meant for tree-level, leading-order calculations; the only loop-induced interactions are the couplings of $\eta$ to $gg$, $\gamma\gamma$, $Z\gamma$, $ZZ$ and $W^+W^-$ in \HVTPeta{}, with the associated contact vertices. They are form factors evaluated at fixed external virtualities and stored as constants: the UFO is not a one-loop generator, and applications to far off-shell bosons, such as vector-boson fusion at high energy, require additional validation.
\end{itemize}

\subsection{Validation}\label{app:validation}
The models have been validated in several ways. The two-body widths computed by MadWidth agree with the analytic formulas of appendix~\ref{app:widths}, evaluated with an independent implementation of the mass matrices (\verb|analytic.py|), for all channels of the partners and of the vectors $W'$, $Z'$, $B'$ and $G'$, including the decays of the neutral vectors into $W^+W^-$, $W^\pm W'^\mp$ and a lighter neutral vector plus a Higgs. At the default point and at P0 the $Z'$ and $B'$ widths agree within $4\times10^{-5}$ in every channel (the comparison is included in the package), and the total widths of all states to about $10^{-4}$ or better; the only larger differences, for $G'$ and $\eta\to gg$, come from the scale of $\alpha_s$. The structure of the couplings, the phases of the charged currents, the diagonality of the photon couplings and the Ward identities of amplitudes with external photons and gluons are tested by \verb|check_vertices.py| and \verb|check.mg5|. The script \verb|validation/symmetry_checks.py| verifies directly on the UFO the conservation of charge and colour at every vertex, the universality and diagonality of the photon and gluon couplings for all fermions, the hermiticity of the fermion couplings, the cancellation of the terms growing as $E^4$ and $E^2$ in $W^+W^-\to W^+W^-$, and the gauge sum rules that cancel the $E^2$ growth of $f_i\bar f_j\to W^+W^-$ with partners in the intermediate and external states; all are satisfied to numerical precision at the reference points. The script \verb|validation/oblique_T.py| computes the one-loop contribution of the partners to $\hat T$ discussed in section~\ref{sec:bprime}. For \HVTPeta{}, \verb|check_eta.py| verifies that the \HVTP{} vertices are unchanged and that the CP-odd couplings are hermitian, and \verb|check_eta.mg5| tests the Ward identities of $gg\to\eta g$ and $gg\to\eta gg$, which require the three-gluon contact term of the operator $\eta\,G\widetilde G$. The coefficient of this operator is evaluated at $q^2=m_\eta^2$ and stored as a constant; it is therefore a local approximation of the loop amplitude, adequate for single $\eta$ production and for partner decays but not for a far off-shell $\eta$ or for hard jets above the top threshold. We have also generated unweighted events with MadGraph5\_aMC@NLO 3.8.1 for single $G'$ production, resonant $X_{5/3}\bar T_{\rm light}$ production, partner pairs with and without vector exchange, and partner pairs with decay chains, and checked the masses, the event weights and the scale dependence. In $pp\to G'$ at $m_{G'}=5$ TeV the cross section decreases by a factor 1.661 between $\mu_R=m_Z$ and $\mu_R=m_{G'}$ at fixed factorisation scale, and by 1.068 between $m_{G'}$ and $2m_{G'}$, in agreement with the ratios of the values of $\alpha_s$ recorded in the events; the $G'$ couplings therefore follow the running of $\alpha_s$ exactly. The scripts that reproduce the figures and tables of this paper are distributed with the models.

\section{Inventory of experimental searches}\label{sec:inventory}
Table~\ref{tab:searches} inventories public searches with sensitivity to the TeV-scale spectrum and measurements relevant to its extensions. The energy of data taking defines the run. Several 2026 releases use the full Run~2 sample. The 13 TeV combinations collect multiple diboson and fermion analyses; the collaborations list the individual input channels and publications in their combination papers. The heavy--light cascade searches retain particular value because they publish results as a function of both resonance masses. At the time of writing the most recent of them are the full Run~2 $W'\to Tb,Bt$ search~\cite{expB2G20002} and the 2016 $Z'/G'\to Tt$ search~\cite{exp181206489}; no cascade search with the full Run~2 sample for the colour octet, nor any Run~3 cascade search, has been released.

{\small\setlength{\tabcolsep}{3pt}\renewcommand{\arraystretch}{1.16}
\begin{longtable}{@{}>{\raggedright\arraybackslash}p{.22\linewidth}>{\raggedright\arraybackslash}p{.11\linewidth}>{\raggedright\arraybackslash}p{.21\linewidth}>{\raggedright\arraybackslash}p{.39\linewidth}@{}}
\caption{Experimental inputs. Numerical luminosities are in fb$^{-1}$. Mass bounds are observed 95\% limits for the specified signal. PAS denotes a preliminary analysis summary. The inventory reflects the results public as of November 2026.}\label{tab:searches}\\
\toprule Analysis & Lumi. & Final state & Experimental result or recast input \\ \midrule\endfirsthead
\multicolumn{4}{l}{Table~\ref{tab:searches} continued}\\
\toprule Analysis & Lumi. & Final state & Experimental result or recast input \\ \midrule\endhead
\bottomrule\endfoot
\multicolumn{4}{@{}l}{\itshape Run 3 at 13.6 TeV and combined Run 2--3 data}\\*
ATLAS 2607.28334~\cite{exp260728334} & 140+165 & $ee,\mu\mu$ & 13+13.6 TeV. SSM $Z'$ 5.5 TeV. Fiducial spin-1 limits for several widths. \\
CMS PAS EXO-25-021~\cite{expEXO25021} & 283 & $ee,\mu\mu$ & Narrow peak normalised to the $Z$ yield. SSM and $\psi$ masses 5.55 and 4.95 TeV. \\
CMS PAS EXO-24-021~\cite{expEXO24021} & 62 & $e\nu,\mu\nu$ & SSM $W'$ 5.9 TeV. Fig.~4 gives rate bounds above transverse-mass thresholds. \\
CMS EXO-25-001~\cite{expEXO25001} & 207 total & Dilepton tail & 13+13.6 TeV. Oblique parameters and broad neutral vectors. Figs.~8--10 specify width assumptions. \\
ATLAS 2609.24950~\cite{exp260924950} & 140+56 & $t(t)Z'$, $Z'\to t\bar t$ & 13+13.6 TeV. Singlet vector 1.15 TeV for $c_t=1$, $\theta=0$. Match chirality and colour. \\
CMS SMP-25-013~\cite{expSMP25013} & 171 & Electroweak $WZjj$, $W^\pm W^\pm jj$ & Fiducial spectra for VBS. Test $WZjj$ with forward jets and coherent amplitudes. \\
\multicolumn{4}{@{}l}{\itshape Run 2 at 13 TeV: vectors and heavy--light cascades}\\*
ATLAS 2402.10607~\cite{atlas} & up to 139 & $VV,Vh$, fermions & HVT combination. Weak and strong benchmarks 5.8 and 4.4 TeV. \\
CMS B2G-25-003~\cite{cms} & 138 & $VV,Vh$, fermions & HVT weak and strong benchmarks 5.5 and 4.8 TeV. VBF benchmark reach up to 2.0 TeV. \\
ATLAS 2308.08521~\cite{exp230808521} & 139 & $tb$ & Leptonic and hadronic top categories. Use the published combination. \\
CMS 2310.19893~\cite{exp231019893} & 138 & $tb$, leptonic top & Width-dependent $W'$ limits and chirality hypotheses. Include interference for a left-handed current. \\
ATLAS 2512.17856~\cite{exp251217856} & 140 & $t\bar t$, one or two leptons & Singlet and KK-gluon templates. Masses tested from 0.4 to 5 TeV. \\
CMS PAS B2G-23-010~\cite{expB2G23010} & 138 & $t\bar t$ resonance & Cross-section limits for a colour octet and neutral singlet vectors. Use the appropriate width template. \\
ATLAS 1910.08447~\cite{exp191008447} & 139 & Dijet mass & Inclusive and flavour-tagged searches constrain light-quark currents and $G'$ production. \\
CMS 1911.03947~\cite{exp191103947} & 137 & Dijet mass & Use the $q\bar q$ template and width-dependent results for the octet. \\
CMS EXO-23-004~\cite{expEXO23004} & 117 & Scouting dijets & Narrow peaks at 0.6--1.8 TeV. Limits on $\sigma {\cal B} A$ cover lighter vectors. \\
CMS EXO-20-008~\cite{expEXO20008} & 138 & $b$-tagged dijets & Narrow-resonance limits. Evaluate the bottom fraction and tagging acceptance. \\
ATLAS 2402.16576~\cite{exp240216576} & 139 & $\tau\nu$ & High-$m_T$ limits for charged resonances with nonuniversal leptonic couplings. \\
ATLAS~\cite{atlasVV,atlasVH} & 139 & $WZ,WW,Wh,Zh$ & Separate charged and neutral Drell--Yan rate limits; superseded in sensitivity by the combinations above.\\
CMS~\cite{earlyCascade} & 35.9 & $W'\to Tb,Bt$ & First all-jets $thb$ result, superseded by the full Run~2 search below.\\
CMS B2G-20-002~\cite{expB2G20002} & 138 & $W'\to Tb,tB$ & All-jets $thb,tZb$. Compare both partner masses and decay fractions. \\
CMS 1812.06489~\cite{exp181206489} & 35.9 & $G',Z'\to Tt$, lepton+jets & Explicit heavy-gluon benchmark, $k_3=1/0.44$. Both resonance masses enter the interpretation. \\
CMS 1703.06352~\cite{exp170306352} & 2.6 & $Z'\to Tt$, all jets & $T\to Wb$. Neutral heavy--light benchmark for the reconstructed $Wtb$ system. \\
\multicolumn{4}{@{}l}{\itshape Run 2 at 13 TeV: pair production of fermionic partners}\\*
ATLAS 2212.05263~\cite{exp221205263} & 139 & VLQ pairs, missing momentum & Degenerate weak-doublet benchmark 1.59 TeV. Includes sensitivity to $X_{5/3}$. \\
ATLAS 2605.28538~\cite{exp260528538} & 139 & $T\bar T$, at least one $ht$ & Singlet 1.40 TeV, doublet 1.56 TeV, exclusive $ht$ 1.66 TeV. \\
ATLAS 2401.17165~\cite{exp240117165} & 140 & VLQ pairs, lepton+jets & Exclusive $Wb$ 1.70 TeV. Singlet 1.36 TeV. \\
CMS 2209.07327~\cite{exp220907327} & 138 & $T\bar T,B\bar B$, leptons & $T$ masses below 1.48 TeV excluded across the tested SM branching triangle. \\
CMS B2G-24-005~\cite{exp261114335} & 138 & $X_{5/3}\bar X_{5/3}$ & 1.63 TeV for unit $tW$ fraction. Exotic $tH^+$ reach 1.74 TeV for $m_{H^+}=0.8$ TeV. \\
ATLAS 2210.15413~\cite{exp221015413} & 139 & $T\bar T,B\bar B$ & At least one $Zt$ or $Zb$ transition. \\
ATLAS 1808.02343~\cite{exp180802343} & 36.1 & Pair combination & Earlier combination across SM branching fractions; overlaps the full Run 2 sample. \\
CMS 2402.13808~\cite{exp240213808} & 138 & $B\bar B$ & Dilepton and fully hadronic categories target neutral decays. \\
CMS 1906.11903~\cite{exp190611903} & 35.9 & $T\bar T,B\bar B$ & Fully hadronic topologies. \\
CMS 1810.03188~\cite{exp181003188} & 35.9 & $X_{5/3}\bar X_{5/3}$ & Same-sign dileptons and single leptons. The same-sign channel complements the later single-lepton analysis. \\
CMS 1710.01539~\cite{exp171001539} & 35.8 & $WbWb$ & Exclusive charged decays of pair-produced partners. \\
CMS 1812.09768~\cite{exp181209768} & 35.9 & $Z$ + jets & Opposite-sign lepton pair in VLQ pair decays. \\
CMS 1805.04758~\cite{exp180504758} & 35.9 & $T\bar T,B\bar B$ & Earlier multilepton and single-lepton selections. \\
CMS 2008.09835~\cite{exp200809835} & 137 & $B\bar B$ & Hadronic bottom-partner search preceding the dilepton plus jets update. \\
\multicolumn{4}{@{}l}{\itshape Run 2 at 13 TeV: electroweak single production}\\*
ATLAS single-$T$ combination~\cite{exp240808789} & 139 & $T\to ht,Zt$ & Limits on electroweak single production. Count the component searches once. \\
ATLAS 2506.15515~\cite{exp250615515} & 140 & Single $T\to Wb$ & Singlet $\kappa$ limits 0.22--0.52 at 1.15--2.30 TeV. Includes interference. \\
CMS B2G-22-004~\cite{cmsvlq} & 138 & Single $T\to Wb$ & Mass interval 0.8--2.0 TeV for $\kappa_W=0.2$, $\mathcal B(Wb)=0.5$. \\
CMS B2G-24-013~\cite{expB2G24013} & 138 & Single $B\to tW$ & Singlet produced with $bq$, $\Gamma/m=5\%$, 0.8--1.23 TeV. Match the doublet production current. \\
CMS B2G-19-001~\cite{expB2G19001} & 138 & Single $T\to ht,Zt$ & Hadronic decays and associated $b$ production. Compare transition currents and widths. \\
CMS 2605.13649~\cite{exp260513649} & up to 138 & Single $T\to ht$ & Opposite-sign dileptons. Rate limits from 2.0 pb at 0.6 TeV to 0.1 pb at 1.0 TeV. \\
ATLAS 2402.16561~\cite{monotopATLAS} & 139 & $T\to Zt$, invisible $Z$ & Monotop component of the ATLAS combination. Singlet reach 1.8 TeV for $\kappa=0.5$ and $\mathcal B(Zt)=0.25$. \\
ATLAS 2307.07584~\cite{exp230707584} & 139 & $T\to Zt$ & Multilepton input to the ATLAS single-$T$ combination. \\
ATLAS 2305.03401~\cite{exp230503401} & 139 & $T\to ht,Zt$ & Single-production search entering the same combination. \\
ATLAS 2201.07045~\cite{exp220107045} & 139 & $T\to ht$ & Fully hadronic boosted Higgs and top candidates. \\
ATLAS 2409.20273~\cite{exp240920273} & 139 & $T\to Wb$ & Fully hadronic charged-current selection; includes a $Y$ interpretation. \\
ATLAS 2308.02595~\cite{exp230802595} & 139 & $B\to hb$ & Three-bottom signature; match production current and Higgs fraction. \\
CMS 1708.01062~\cite{exp170801062} & 35.9 & $T\to Zt$ & Leptonic $Z$; also tests resonant production. \\
CMS 2201.02227~\cite{exp220102227} & 137 & $T\to Zt$ & Invisible $Z$ and hadronic top; widths up to 30 percent. \\
CMS 2302.12802~\cite{exp230212802} & 138 & $T\to ht$ & Diphoton Higgs tag gives complementary mass reconstruction. \\
CMS B2G-24-015~\cite{expB2G24015} & 138 & $T\to ht$, $h\to b\bar b$ & Anomaly-detection search for $X\to hY$; one benchmark is single $T\to ht$ with a hadronic top, $m_T=1.4$--3 TeV. \\
CMS 1909.04721~\cite{exp190904721} & 35.9 & $T\to ht,Zt$ & Earlier hadronic selection with associated top or bottom. \\
CMS 2510.25874~\cite{exp251025874} & 138 & $T\to t\phi$ & Lepton+jets and boosted scalar; overlaps the combined scalar result below. \\
CMS 1802.01486~\cite{exp180201486} & 35.9 & $B\to hb$ & Hadronic Higgs decay and associated jet. \\
CMS 1809.08597~\cite{exp180908597} & 35.9 & $B,X_{5/3}\to tW$ & Electroweak production of exotic or bottom-like partners. \\
CMS 1701.08328~\cite{exp170108328} & 2.3 & $Q\to Wb$ & Earlier single-production charged-current benchmark. \\
CMS 2104.12853~\cite{exp210412853} & 137 & $tW$ resonance & Hadronic resonance templates; transfer to a VLQ depends on production. \\
CMS 2111.10216~\cite{exp211110216} & 138 & $tW$ resonance & Lepton+jets templates; see the CMS review for VLQ interpretation. \\
\multicolumn{4}{@{}l}{\itshape Scalar extensions and invisible decays}\\*
ATLAS 2607.28381~\cite{exp260728381} & 140 & $T\bar T$, $T\to ts$ & Run 2. Fiducial diphoton rate versus $m_T,m_s$. \\
CMS B2G-24-016~\cite{expB2G24016} & 138 & $T\bar T$, $s\to\gamma\gamma$ & Run 2. Boosted hadronic tops. Efficiencies depend on both masses. \\
CMS B2G-22-001~\cite{expB2G22001} & 138 & Single $T\to ts$, $s\to b\bar b$ & Run 2. Scalar mass scan. SM Higgs and $\Gamma_T/m_T=5\%$, 0.85--1.3 TeV. \\
ATLAS 2401.13430~\cite{exp240113430} & 140 & $t\bar t$ + missing momentum & Run 2. Invisible daughter. A fermionic parent requires its own spin-dependent efficiency. \\
ATLAS 2607.26879~\cite{exp260726879} & 140+56 & $hh\to b\bar b\tau\tau$ & 13+13.6 TeV nonresonant di-Higgs input for the potential extension. \\
CMS HIG-24-010~\cite{expHIG24010} & 138+62 & $hh\to 4b$ & 13+13.6 TeV. Combined $\mu_{hh}<4.7$. Account for altered top-loop amplitudes. \\
\end{longtable}}
The direct-search inventory follows the third-generation channels catalogued by ATLAS~\cite{summaryATLAS}, and CMS~\cite{summaryCMS} (table 1 and sections 6.2--6.5), with subsequent releases listed individually. Early Run 2 inputs are retained for complementary final states. Dedicated light-flavour and exotic-charge $Y$ interpretations require couplings outside the benchmark. Shared events among successive releases or between inputs and their published combination preclude treating every row as an independent likelihood.

\section{Conversion of single-production limits to \HVTP{} couplings}\label{app:matching}
In the UFO, the coefficients of \texttt{FFV2} and \texttt{FFV3} in each vertex are $i g_L$ and $i g_R$, respectively. Table~\ref{tab:ufolambda} illustrates eq.~\eqref{eq:partnerlambda} at the default point of the package. To vary the transitions, regenerate the point with the desired masses and mixings as described in section~\ref{sec:usage}. The coupling names \texttt{GC\_n} depend on the point, whereas the field names identify the vertex.
\begin{table}[htbp]\centering
\begin{tabular}{@{}lllrr@{}}\toprule
UFO fields & $i g_L$ & $i g_R$ & $|\lambda^L|$ & $|\lambda^R|$\\\midrule
\verb|tp3~ b W+| & \verb|GC_435| & \verb|GC_436| & 1.160 & 0.008\\
\verb|Bh~ t W-| & \verb|GC_417| & \verb|GC_418| & 0.190 & 1.497\\
\verb|x53~ t W+| & \verb|GC_443| & \verb|GC_444| & 0.210 & 1.794\\
\verb|tp2~ t Z| & \verb|GC_117| & \verb|GC_118| & 0.016 & 1.433\\
\verb|tp1~ t Z| & \verb|GC_109| & \verb|GC_110| & 0.329 & 0.819\\
\bottomrule\end{tabular}
\caption{Chiral transition couplings at the default point of \texttt{point.json}, where \texttt{tp1}, \texttt{tp2} and \texttt{tp3} are mostly $X_{2/3}$, $T$ and $\widetilde T$. The middle columns identify the entries in \texttt{couplings.py} for this UFO.}\label{tab:ufolambda}
\end{table}
For the lightest state \texttt{tp1}, which is mostly $X_{2/3}$ but has a sizeable singlet and $T$ admixture, the left-handed neutral transition is not negligible. A direct application of a right-handed doublet contour would therefore require a check of the signal acceptance and branching fractions.

\paragraph{Charged singlet.}
We express the experimental sensitivities in the \HVTP{} coupling $\lambda$ defined in section~\ref{sec:lambda}. For a predominantly left-handed $Wb$ transition and negligible bottom mass,
\begin{equation}
 \lambda^2=\frac{32\pi}{F_W}\,\mathcal B(T\to Wb)\frac{\Gamma_T}{m_T},
 \qquad F_W=(1-x_W)^2(1+2x_W),\qquad x_W=m_W^2/m_T^2 .
 \label{eq:widthmatch}
\end{equation}
At the isolated-singlet point the rotated UFO vertex gives $|g_L|=0.05791$ and $|g_R|=2.02\times10^{-5}$. With $m_W=79.85$ GeV, the resulting $\lambda=1.241$ agrees to the quoted precision with eq.~\eqref{eq:widthmatch} evaluated using the MadWidth partial width $\Gamma(T\to Wb)=26.218$ GeV.

Experimental coupling symbols refer to specific signal parameterisations. For the chiral charged-current coefficient $\kappa_L^{Tb}$ multiplying $g/\sqrt{2}$ in eq.~(1) of ref.~\cite{exp240808789}, the translation is $\lambda=g m_T|\kappa_L^{Tb}|/(\sqrt{2}m_W)$ at $m_b=0$. The universal $\kappa$ in the ATLAS combination and the $\kappa_W$ used by CMS are template parameters. To match either convention to \HVTP{}, we use a published reference curve $\kappa_r(m_T)$ of known total width $r=\Gamma_T/m_T$ in the same singlet hypothesis. At fixed mass and branching fractions the partial width scales quadratically with the template coupling. Equation~\eqref{eq:widthmatch} then gives
\begin{equation}
 \lambda_{95}(m_T)=\sqrt{\frac{32\pi\,r\,\mathcal B(Wb)}{F_W}}\,
 \frac{\kappa_{95}(m_T)}{\kappa_r(m_T)} .
\end{equation}
For ATLAS we use the observed and median expected singlet contours in HEPData tables~\href{https://doi.org/10.17182/hepdata.153640.v1/t9}{9} and~\href{https://doi.org/10.17182/hepdata.153640.v1/t10}{10}, calibrated to the digitised 5\% width line in figure~6(a) of ref.~\cite{exp240808789}. For CMS, HEPData table~\href{https://doi.org/10.17182/hepdata.168402.v1/t8}{8} supplies the two limits and the singlet 10\% width line in the same left-axis units~\cite{cmsvlq}. The right-axis conversion $\kappa_W^T=\sqrt{2}\kappa_W^Y$ multiplies numerator and denominator equally. Both experiments are evaluated at $\mathcal B(Wb)=0.5$.

\paragraph{Doublet currents.}
For a purely chiral vector vertex and a daughter quark of mass $m_q$, eq.~\eqref{eq:partnerlambda} gives $|\lambda^a|=M|g_a|/m_V$ for the dominant chirality. Keeping the daughter masses in the two-body width gives
\begin{align}
 \frac{\Gamma(Q\to Vq)}{M}&=\frac{|\lambda^a|^2}{32\pi}F(x,y),\qquad
 x=\frac{m_V^2}{M^2},\quad y=\frac{m_q^2}{M^2},\\
 F(x,y)&=\sqrt{(1-x-y)^2-4xy}\big[(1-y)^2+x(1+y)-2x^2\big].
\end{align}
For $B$, the observed and expected CMS doublet limits and the 1\% reference coupling are supplied together in HEPData table~\href{https://doi.org/10.17182/hepdata.168733.v1/t19}{19}~\cite{expB2G24013}. We infer $\Gamma/M=0.01(\kappa_{95}/\kappa_{1\%})^2$ at each mass and retain the region below 10\%. For $T$ and $X_{2/3}$ the ATLAS doublet contours are in tables~\href{https://doi.org/10.17182/hepdata.153640.v1/t18}{18} and~\href{https://doi.org/10.17182/hepdata.153640.v1/t19}{19}~\cite{exp240808789}. We use the common 5\% width reference in figure~6 of that analysis, assign half the width to $Zt$ and retain $\Gamma/M<0.50$. These conversions assume the chiral and branching hypotheses of the corresponding templates.

For $X_{5/3}$ we use the right-handed 1\% limits in figure~8(b) of ref.~\cite{exp180908597} and the 13 TeV charge-summed production coefficients $\hat\sigma_1$ in table~2 of ref.~\cite{partnerRates}. In that convention the right-handed coupling is $g_R=g_2\hat c_R/2$, where $\hat c_R$ is the coupling parameter of ref.~\cite{partnerRates}, not to be confused with $c_R=\cos\phi_R$. With $v_{\rm EW}=2m_W/g_2$ this gives
\[
 \lambda_{X_{5/3},95}=\frac{M}{v_{\rm EW}}
 \sqrt{\frac{(\sigma{\cal B})_{95}}{\hat\sigma_1\,{\cal B}(X_{5/3}\to Wt)}} .
\]
This rate conversion assumes narrow-width factorisation and the acceptance of the published mass template. The width inferred from the coupling is recorded point by point in the accompanying data. The broader left-handed templates test a different top polarisation.

\section{Relative phase of the charged-vector vertex}
The relative sign of the bosonic and fermionic couplings of $W'$ does not affect its partial widths, but it enters every amplitude in which $W'$ exchange interferes with SM or other heavy-vector contributions, for instance in vector-boson fusion. A rephasing-invariant check is obtained by normalising the $W'Wh$ vertex to the SM $WWh$ vertex and dividing by the ratio of the $W'ud$ and $Wud$ couplings, taken with the same orientation of the charged fields. In \HVTP{} this quantity is real and equal to $-k_2^2$ at leading order in $v/M_2$, as expected from the opposite signs of the elementary and composite currents (section~\ref{sec:model}). Implementations that adopt a different phase convention for the charged bosonic terms can reproduce the same partial widths while differing at the amplitude level; the comparison with other mass-basis parameterisations therefore requires aligning the Higgs, SM and heavy charged currents together.

\section{Rate matching for the vector limits}\label{app:vectorrates}
We extract physical masses and open two-body widths from each generated UFO, on a grid of 150 GeV in the input mass and 0.2 in $k_i$. Interpretations based on narrow signal templates are restricted to $\Gamma/m\le5\%$, while the width-dependent CMS dilepton limits are used up to $\Gamma/m=10\%$; all of them assume Drell--Yan production in the narrow-width approximation; vector-boson fusion and interference with the SM require separate signal calculations.

\paragraph{CMS HVT combination.} For a triplet produced from light quarks the production rate scales with $g_F^2$ at fixed mass. If, within a final-state category, the relative weights of the individual channels are the same in HVT and in \HVTP{}, the limit can be transferred by rescaling the rate of the whole category. We adopt this approximation, whose limitations are discussed below: for a category $c$ whose limit $\sigma_{95}^{X}(c)$ on $\sigma(pp\to V')$ is quoted in the HVT benchmark $X$, we require
\[
 \frac{\sigma^X_{\rm th}(m)\,\big(g_F/g_F^X\big)^2\,\mathcal B_{\rm \HVTP{}}(c)}{\sigma_{95}^{X}(c)\,\mathcal B_X(c)}<1 ,
\]
with $g_F^2=16\pi\,\Gamma(W'\to u\bar d)/m_{W'}$ extracted from the generated widths. The lepton and quark categories are quoted for benchmark A and the boson category for benchmark B, with the theory cross sections of the same publication (HEPData tables for figures 4a and 4b of ref.~\cite{cms}). The reference branching fractions follow from $(g_F,g_H)=(0.556,0.556)$ for A and $(0.146,2.928)$ for B, using the high-mass HVT widths: $\Gamma_f/m=g_F^2/(4\pi)$ for the twelve fermion doublets and $\Gamma_{VV}+\Gamma_{Vh}=g_H^2m/(96\pi)$. This gives $\mathcal B_A=(0.24,0.72,0.04)$ and $\mathcal B_B=(0.014,0.042,0.944)$ for leptons, quarks and bosons. In \HVTP{} the lepton category includes $e,\mu,\tau$, the boson category $WZ,WW,Wh,Zh$, and the quark category all quark pairs including the third generation; the latter neglects the different acceptance of the $t\bar t$, $tb$ and dijet channels. As a check, the \HVTP{} line crosses benchmark B at $k_2\simeq4.5$, where our boson-category bound, 4.7 TeV, is consistent with the 4.8 TeV obtained by CMS from the full combination. The two-dimensional contours at 3 and 4 TeV (HEPData tables for figures 7a and 7b) exclude every point of the \HVTP{} line with $|g_H|\le3$, in agreement with the rescaled limits. The rescaling of a whole category is exact only if all its channels are rescaled by the same factor. This holds for the light fermions, but not necessarily for the third-generation quark channels, for the relative weight of $W'$ and $Z'$, whose partner thresholds differ, or for the bosonic channels when the partners are light. The resulting limits should therefore be regarded as approximate reinterpretations; an exact treatment requires the limits or likelihoods of the individual channels.

\paragraph{13.6 TeV lepton searches.} Let $P_V=\sum_{ij} L_{ij}\Gamma_{V\to ij}/m_V$, where $L_{ij}$ is the parton luminosity at $\mu_F=m_V$, including both beam orderings. In the narrow-width approximation the rate relative to a reference with the same spin and colour at the same energy is $(P_V/P_{\rm ref})\,\mathcal B(V\to f)/\mathcal B_{\rm ref}$. We use central CT18LO luminosities and its running $\alpha_s$~\cite{ct18lo}. The tree-level reference currents use $\sin^2\theta_W=0.2312$ and $\alpha^{-1}=127.9$. This procedure transfers the QCD normalisation of the sequential-model prediction and assumes comparable signal efficiencies. For $W'$ we take the observed, expected and SSM curves from figure~3(c) of ref.~\cite{expEXO24021}. For $Z'$ we take the ratios to the SSM prediction in figure~3(c) of ref.~\cite{expEXO25021}, interpolating limits logarithmically between widths and using the narrowest template below 0.6\%. The digitised SSM intersections reproduce the reported mass reaches to within 0.1 TeV. Since the SSM reference has a width of about 3\%, its off-shell tail is not identical to that of a narrower \HVTP{} signal. At the highest masses the off-shell component, enhanced by the parton luminosities, changes the signal efficiency, as noted by CMS; our on-shell rescaling does not account for this effect, so that the $\ell\nu$ and $\ell\ell$ reinterpretations are approximate above about 5 TeV.

\paragraph{Colour octet.} The reference is the axigluon/coloron column of HEPData table 4 of ref.~\cite{exp191103947}, with $\Gamma_q/m=\alpha_s/6$ per effectively massless flavour and $\mathcal B(jj)\simeq5/6$. We reweight all five initial flavours and use the calculated $u,d,s,c,b$ fractions in the final state. The \HVTP{} couplings and widths of $G'$ are computed at each point with $\alpha_s(m_{G'})$ of CT18LO, the same value used for the reference, so that the decrease of $\alpha_s$ at high mass is included in both. The reference includes the dijet rapidity acceptance; its transfer assumes the same parity-even spin-one decay distribution for the symmetric cuts.

\paragraph{Cascade search.} For the CMS $W'\to Tb,tB$ search we use the observed and expected limits on $\sigma\,\mathcal B$ for the three partner-mass hypotheses (HEPData tables for figure 7 of ref.~\cite{expB2G20002}). The \HVTP{} cross section is the HVT benchmark-B $W'$ cross section of ref.~\cite{cms} rescaled by $(g_F/g_F^B)^2$, so that the same QCD normalisation enters the triplet and cascade interpretations. The branching fraction sums $\mathcal B(W'\to T_i\bar b)\,[\mathcal B(T_i\to ht)+\mathcal B(T_i\to Zt)]$ over the charge-$2/3$ eigenstates and $\mathcal B(W'\to t\bar B)\,[\mathcal B(B\to hb)+\mathcal B(B\to Zb)]$, with all partner widths computed at the same point. The CMS limits refer to the sum of the $tB$ and $bT$ templates; the generalised limits of the same publication show that the $bT$ channel alone is somewhat more sensitive, so applying the summed limit is slightly conservative.

The scripts \texttt{hvt\_combination.py}, \texttt{cascade\_recast.py} and \texttt{vector\_limits.py} in the \texttt{figures} directory reproduce all three interpretations from the included HEPData tables.

\paragraph{Partner pair-production searches with resonant production.} The ATLAS $T\to ht$ search~\cite{exp260528538} selects events in which at least one partner decays into $ht$. Without a dedicated simulation, we take its efficiency for any pair to be proportional to the probability that at least one of the two partners decays into $ht$, normalised to the published doublet benchmark (probability 0.75, observed limit 1.26 fb at 1.73 TeV). Applied to the published singlet and $ht$-only benchmarks, this rule reproduces their limits to within 10\% and 20\%, respectively. As a cross-check we use the ATLAS search with one lepton and missing momentum~\cite{exp221205263}, which publishes limits for doublet $T\bar T$ and for $B\bar B\to WtWt$: each pair type is compared with the closest hypothesis, and the mixed pair $X_{5/3}\bar T_{\rm light}$ with the average per-leg efficiency of the two. The resonant rates include the finite-width correction of section~\ref{sec:fw} and the $Z'$- and $B'$-mediated pairs computed with MadGraph; the QCD rates are normalised to the NNLO predictions used by the experiments, which exceed the LO rates by about 15\%.

\paragraph{Same-sign dilepton analysis.} The analysis of $W'^+\to X_{5/3}\bar X_{2/3}$ in ref.~\cite{wp} was performed at 14 TeV for $s_L=0.9$, degenerate custodians at 0.9 TeV and $k_2=3$. It selects two positive leptons from $X_{5/3}\to W^+t\to W^+W^+b$ and at least four jets from $\bar X_{2/3}\to Z\bar t,h\bar t$, applies cuts on the lepton and jet transverse momenta, $H_T$ and $S_T$, and reconstructs the $X_{2/3}$ mass. We reproduce its parton-level input with \HVTP{}, with $(m_Q,m_S,s_L,s_R)=(0.9,6.0\,{\rm TeV},0.9,0.6)$, which gives $m_{X_{2/3}}=0.91$ TeV and $\mathcal B(X_{2/3}\to Zt,ht)=(0.46,0.52)$, close to the values of ref.~\cite{wp}. Dividing the published final cross sections by the LO rate times the leptonic $W$ and hadronic top, $Z$ and $h\to b\bar b$ decay fractions gives an overall efficiency that rises from 0.22 at $m_{W'}=1.9$ TeV to 0.41 at 3.5 TeV. We apply the latter to the light partner spectrum, with the mass window moved to the new partner mass, and include $\mathcal B(X_{5/3}\to W^+t)=0.84$ and the decay fractions of $T_{\rm light}$. As a check, we have simulated $10^4$ events of $pp\to W'^+\to X_{5/3}\bar T_{\rm light}$ at the default point with Pythia~8 and Delphes~3.5.1 (CMS card) and applied a comparable selection: two same-sign leptons with $p_T>30$ GeV, at least four jets with $p_T>30$ GeV, at least one $b$ tag, $H_T>1.2$ TeV and $S_T>1.8$ TeV. It retains $(0.63\pm0.08)\%$ of the events (statistical uncertainty), in agreement with the 0.61\% obtained from the rescaled efficiency and the branching fractions within this precision. The sample is generated as $pp\to X_{5/3}\bar T_{\rm light}$ with all contributing diagrams, dominated by the $W'^+$ resonance. In the same sample a Higgs-tagged and a top-tagged large-radius jet (the Higgs tag requires a $b$-tagged small-radius jet within $\Delta R<0.8$) reconstruct $T_{\rm light}$ with a peak shifted by $-5\%$, to 1.64 TeV, and a resolution of about 11\%; 4\% of the events fall in the window $1.45<m(ht)<2.0$ TeV, about seven times the same-sign yield. Without the corresponding backgrounds this does not by itself imply a better sensitivity. For the background we keep the final 14 TeV value of ref.~\cite{wp}, 0.028 fb (conservative); since harder cuts would reduce it for heavier partners, we also consider a background ten times smaller with both $W'$ charges included (optimistic). Signal rates at 13 and 14 TeV use CT18LO luminosities.

\paragraph{HL-LHC projections.} For the standard channels we take the expected category limits of the CMS combination~\cite{cms}, multiply the signal by the ratio of the LO $W'$ cross sections at 14 and 13 TeV and scale the limits by $\sqrt{L_0/L}$, appropriate for a background-dominated search, or by $L_0/L$ for a background-free one. The same scaling is applied to the expected limit of the ATLAS $T\to ht$ pair search, with the QCD rate rescaled by the LO ratio. Changes in detector performance and systematic uncertainties are ignored.

\bibliographystyle{JHEP}
\bibliography{references}
\end{document}